\documentclass[prd,amsmath,amssymb,cite,nofootinbib]{revtex4-2} 

\pdfoutput=1

\usepackage[utf8]{inputenc}
\usepackage{floatpag}
\usepackage{amsthm}
\usepackage{enumerate}
\usepackage{amssymb}
\usepackage{braket}
\usepackage{hyperref}
\hypersetup{
	colorlinks   = true, %Colours links instead of ugly boxes
	urlcolor     = blue, %Colour for external hyperlinks
	linkcolor    = blue, %Colour of internal links
	citecolor   = red %Colour of citations
}
\usepackage[capitalise]{cleveref}
\usepackage{graphicx}
\usepackage{caption}
\usepackage{subcaption}
\usepackage{epstopdf}
\usepackage{slashed}
\usepackage{color} % for inkscape
\graphicspath{{./Figures/}}

\usepackage{tikzit}
\tikzstyle{red dot}=[fill=black, draw=black, shape=circle, minimum size=0.5, scale=0.5]
\tikzstyle{green dot}=[fill=green, draw=black, shape=circle]
\tikzstyle{new style 0}=[fill=black, draw=black, shape=circle, minimum size=0.5, scale=0.5]

\tikzstyle{new edge style 0}=[-, draw=black, dashed=true]
\tikzstyle{new edge style 1}=[-, draw=black]
\tikzstyle{rojo}=[-, draw={rgb,255: red,137; green,0; blue,0}]

\usepackage{tikzexternal}

\definecolor{coloreqn}{HTML}{890000}

\usepackage{mathtools,leftidx}
\usepackage{array} % and/or
\allowdisplaybreaks

\usepackage{enumitem}

\begin{document}

\title{Multi-particle production in proton-nucleus collisions in the Color Glass Condensate}

\author{Pedro Agostini}
\email{pedro.agostini@usc.es}
\affiliation{Instituto Galego de F\'{\i}sica de Altas Enerx\'{\i}as IGFAE, Universidade de Santiago de Compostela, 15782 Santiago de Compostela, Galicia-Spain}

\author{Tolga Altinoluk}
\email{tolga.altinoluk@ncbj.gov.pl}
\affiliation{National Centre for Nuclear Research, 02-093 Warsaw, Poland}

\author{N\'estor Armesto}
\email{nestor.armesto@usc.es}
\affiliation{Instituto Galego de F\'{\i}sica de Altas Enerx\'{\i}as IGFAE, Universidade de Santiago de Compostela, 15782 Santiago de Compostela, Galicia--Spain}

%%\thankstext{t1}{Grants or other notes
%%about the article that should go on the front page should be
%%placed here. General acknowledgments should be placed at the end of the article.
%\thankstext{e1}{e-mail: pedro.agostini@usc.es}
%\thankstext{e2}{e-mail: Tolga.Altinoluk@ncbj.gov.pl}
%\thankstext{e3}{e-mail: nestor.armesto@usc.es}
%
%\institute{First Address, Street, City, Country\label{addr1}
%          \and
%          Second Address, Street, City, Country\label{addr2}
%          }

\vskip 0.5cm

%\date{\today}

\begin{abstract}

We compute multi-gluon production in the Color Glass Condensate approach in dilute-dense collisions, p$A$, extending previous calculations up to four gluons. We include the contributions that are leading in the overlap area of the collision but keep all orders in the expansion in the number of colors. We develop a diagrammatic technique to write the numerous color contractions and exploit the symmetries to group the diagrams and simplify the expressions. To proceed further, we use the McLerran-Venugopalan and Golec-Biernat-W\"usthoff models for the projectile and target averages, respectively. We use a form of the Lipatov vertices that leads to  the Wigner function approach for the projectile previously employed, that we generalise to take into account quantum correlations in the projectile wave function. We provide analytic expressions for integrated and differential two gluon cumulants and show a smooth dependence on the parameters defining the projectile and target Wigner function and dipole, respectively. For four gluon correlations we find that the second order four particle cumulant is negative, so a sensible second Fourier azimuthal coefficient can be defined. The effect of correlations in the projectile on this result results qualitatively and quantitatively large.

\end{abstract}

\maketitle

\tableofcontents

\section{Introduction}
\label{sec:intro}

Small collision systems, proton-proton ($pp$) and proton-nucleus ($p$A), studied at the Large Hadron Collider (LHC) show many of the characteristics~\cite{Schlichting:2016sqo,Loizides:2016tew,Schenke:2017bog,Nagle:2018nvi,Citron:2018lsq} that in heavy ion collisions are considered as signatures of the formation of hot deconfined partonic matter, the Quark Gluon Plasma. The most prominent example is the existence of azimuthal correlations in the two-particle inclusive distributions that are extended in pseudorapidity and show maxima when the particle transverse momenta are either parallel or antiparallel. This finding, named the ridge, was first observed in high multiplicity $pp$ collisions~\cite{Khachatryan:2010gv}, and then for smaller multiplicities~\cite{Aaboud:2016yar,Khachatryan:2015lva,Aad:2015gqa,Khachatryan:2016txc}, in $p$Pb collisions~\cite{CMS:2012qk,Abelev:2012ola,Aad:2012gla,Chatrchyan:2013nka,Abelev:2014mda,Aaij:2015qcq} and in association with $Z$ boson production~\cite{Aaboud:2019mcw}. It was also observed in $p$Au, $d$Au and $^3$HeAu collisions at the Relativistic Heavy Ion Collider (RHIC)~\cite{Adare:2014keg,Adamczyk:2015xjc,Adare:2015ctn,PHENIX:2018lia}. Azimuthal asymmetries in particle production have also been searched for in even smaller systems: in $\gamma$Pb through ultraperipheral collisions at the LHC~\cite{Aad:2021yhy} where they were found, and in $e^+e^-$ collisions~\cite{Badea:2019vey} at the Large Electron-Positron collider and deep inelastic scattering in $ep$ at the Hadron-Elektron-Ringanlage~\cite{ZEUS:2019jya} with inconclusive results.

The key open question nowadays is the clarification of the origin of such collective behaviour. In heavy ion collisions where the partonic density is very large, a natural explanation is that collectivity is built through the strong final state interactions of the created system. Such explanation looks justified by the success of viscous relativistic hydrodynamics~\cite{Jeon:2016uym,Romatschke:2017ejr} for describing the observed experimental features in soft particle production. The open questions at this moment are how the conditions for hydrodynamics to be applicable are reached from an initial state that is very far from equilibrium~\cite{Romatschke:2016hle} -- the emergence of the macroscopic description given by hydrodynamics from Quantum Chromodynamics (QCD) --, for which both strong and weak coupling explanations have been proposed (see, e.g.,~\cite{Keegan:2015avk}), and why hydrodynamics seems to work even for large anisotropies, outside its presumed range of applicability. Hydrodynamics appears as the effective theory for describing the soft modes of any field theory, see e.g.~\cite{Kurkela:2019set} and references therein.

The success of the application of hydrodynamics for describing azimuthal asymmetries in small systems, $pp$ and $p$Pb collisions~\cite{Romatschke:2017ejr,Schenke:2012wb}, while requiring careful choices of the initial conditions, pushes this description to small collision areas and low particle densities where non-hydrodynamic modes play a very important role~\cite{Romatschke:2017ejr,Kurkela:2019kip,Kirkpatrick:2021roq}. Therefore, it seems sensible to explore other alternatives. The Color Glass Condensate (CGC)~\cite{Gelis:2010nm,Kovchegov:2012mbw}, as weak coupling non-perturbative effective theory for QCD at high energies and partonic densities, offers a framework where azimuthal asymmetries can be calculated from first principles, see the review~\cite{Altinoluk:2020wpf} and references therein. Correlations in the final state reflect those found in the wave function of the projectile and target hadrons or nuclei, assuming that final state effects, including hadronisation, do not wash them out.

The initial versus final state origin of azimuthal correlations in small systems has been subject to intense scrutiny in recent years~\cite{Mace:2018yvl,Mace:2018vwq,Nagle:2018ybc}. At present, no CGC-based model is able to fully describe the existing experimental data. Still, the search for observables that may discriminate initial from final effect continues, e.g., the correlation of $v_2$ 
with the mean  transverse momentum of the particles produced in the collisions~\cite{Aad:2019fgl,Bozek:2016yoj,Giacalone:2020byk,Lim:2021auv} that has also been analysed in the CGC~\cite{Altinoluk:2020psk}. Also many particle cumulants are expected to be crucial. For example, four particle cumulants $c_2\{4\}$, with $v_2\{4\}=[-c_2\{4\}]^{1/4}$ (definitions of all these quantities will be provided below),
change sign from positive to negative with increasing particle multiplicity in the event, with a smooth behaviour from small to large systems and from smaller to larger energies.
This change of sign is associated with the onset of true collective flow of final state origin because higher order cumulants are less sensitive to non-flow contributions than those computed from two-particle correlations. In the CGC numerical implementation in~\cite{Dusling:2017dqg,Dusling:2017aot} the change of sign of $c_2\{4\}$ was interpreted as the transition from a dilute-dilute situation, described by the glasma graph approach~\cite{Dumitru:2008wn,Dumitru:2010iy} where azimuthal correlations correspond to the Bose enhancement of the gluons in the wave function of the colliding hadrons and to the Hanbury-Brown-Twiss (HBT) effect for the final gluons~\cite{Altinoluk:2015uaa,Altinoluk:2015eka,Kovchegov:2012nd,Kovchegov:2013ewa}, to a dilute-dense situation where multiple scattering dominates (for a discussion on density correlations to the dilute-dense situation, see~\cite{Schlichting:2019bvy} and references therein).

The goal of this work is the extension of the calculations of multiparticle production in the CGC in the dilute-dense situation (suitable for $p$A collisions) performed in~\cite{Altinoluk:2018ogz} to four gluon production (see~\cite{Martinez:2018tuf} for inclusive cross sections involving final state quark-antiquark pairs), and the computation of the two and four particle cumulants\footnote{As in standard CGC calculations, here odd azimuthal harmonics are absent,  see a discussion of the origin of the problem and proposed solutions in~\cite{Altinoluk:2020wpf} and references therein.}.
Note that up to four gluon production was previously computed in the glasma graph approach~\cite{Ozonder:2014sra}, and arguments in~\cite{Dumitru:2014yza} suggested that in such approximation $c_2\{4\}>0$ -- a result also found in~\cite{Dusling:2017aot} where only quark scattering is considered and partons in the projectile wave function are uncorrelated. In this work we use the argument in~\cite{Altinoluk:2018ogz,Altinoluk:2018hcu,Kovner:2017ssr,Kovner:2018vec} that captures those contributions of the ensembles of Wilson lines to multiparticle production that are leading in the overlap area of the collision (i.e., in the number of color domains or correlated particle sources), while keeping contributions to all orders in the number of colors. We use the Golec-Biernat-W\"usthoff (GBW) model~\cite{GolecBiernat:1998js,GolecBiernat:1999qd} for the target, and the generalised McLerran-Venugopalan (MV) model~\cite{McLerran:1993ni,McLerran:1993ka} for the projectile. In order to push the analytical calculations as far as possible, we employ the Wigner function ansatz used in~\cite{Lappi:2015vta,Dusling:2017aot,Davy:2018hsl} but extended to include quantum correlations in the projectile wave function. Due to the Gaussian forms that we employ for both the Wigner function and dipole, our results cannot be considered reliable for transverse momenta sizeably larger than the saturation scale.

This manuscript is organized as follows. In Section~\ref{sec1} we introduce the formalism to compute particle production in the CGC. Section~\ref{sec2} is devoted to the calculation of projectile and target averages required to obtain the final results. Then, in Section~\ref{sec3} we present our results and in Section~\ref{sec4} we give a summary and our conclusions. Appendices~\ref{app_AE}, \ref{app1}, \ref{app:wigner} and \ref{app:4gluon} contain a discussion on the validity of the area enhancement argument that we use for computing target ensembles of Wilson lines, useful integrals, a discussion of the Wigner function approach and details on the calculation of four gluon production, respectively.

\section{Theoretical background on multi-particle correlation}\label{sec1}

\subsection{Gluon production in dilute-dense collisions}

In this section we present a quick overview on multi-particle production in proton-nucleus collisions in the CGC framework. We  follow~\cite{Altinoluk:2014oxa,Altinoluk:2015gia} and references therein. The projectile is considered a highly boosted dilute system that is composed, mostly, by large-$x$ partons that act each as a color source with color charge density $\rho^a(\textbf{x})$, with superindex $a$ denoting color and $\textbf{x}$ the transverse position. The target is characterised by a strong field $\mathcal{A}^\mu (\textbf{x})=\mathcal{A}^{a\mu}(\textbf{x})T^a$, with $T^a$ the generators of the SU($N_c$) group in the adjoint representation. The nucleus ensemble is supposed to be much larger than the projectile in the transverse plane. In this picture, working in the light-cone gauge $\mathcal{A}^+=0$ and neglecting the transverse  components of the field, the amplitude for producing a gluon with transverse momentum $\textbf{k}$, pseudorapidity $\eta$, polarization $\lambda$ and color $a$ in the projectile-target collision is obtained by using the LSZ reduction formula (at leading order in the QCD coupling constant $g$) leading to
\begin{align}\label{factor}
	\mathcal{M}^a_\lambda(\eta,\textbf{k})=g \int \frac{d^2 \textbf{q}}{(2 \pi)^2} \overline{\mathcal{M}}^{ab}_\lambda(\eta,\textbf{k},\textbf{q}) \rho^b(\textbf{k}-\textbf{q}),
\end{align}
with $\rho^a(\textbf{q})$ the Fourier transform of the color charge density of the projectile  which is defined (see e.g.~\cite{Altinoluk:2011qy}) as 
\begin{align}\label{def:rho}
\rho^a(\textbf{x})=\int\frac{dp^+}{2\pi} a^{\dagger}_i( p^+,\textbf{x}) T^a a_i(p^+,\textbf{x}),
\end{align}
where $a^{\dagger}_i( p^+,\textbf{x})$ and $a_i(p^+,\textbf{x})$ are the creation and annihilation  operators for gluons with longitudinal momentum $p^+$ at transverse position $\textbf{x}$, respectively. The reduced matrix amplitude $\overline{\mathcal{M}}^{ab}_\lambda(\eta,\textbf{k},\textbf{q})$ is derived in \cite{Altinoluk:2014oxa,Gelis:2005pt,MehtarTani:2006xq} and it reads
\begin{align}\label{asda}
	\overline{\mathcal{M}}^{ab}_\lambda(\eta,\textbf{k},\textbf{q})&=\epsilon^{i*}_\lambda(\textbf{k}) i e^{i k^- L^+} \Bigg\{ 2\frac{\textbf{k}^i}{\textbf{k}^2} \int_\textbf{y} e^{-i\textbf{q}\textbf{y}}   \mathcal{U}_{\textbf{y}}^{ab}(L^+,0)
	-2\frac{(\textbf{k}-\textbf{q})^i}{(\textbf{k}-\textbf{q})^2} \int_{\textbf{y},\textbf{x}} e^{i (\textbf{k}-\textbf{q}) \textbf{y}-i \textbf{k} \textbf{x}}  \mathcal{G}^{ab}_{k^+}(L^+,\textbf{x};0,\textbf{y})
	\Bigg. \nonumber \\
	\Bigg. &+ \int_{\textbf{x},\textbf{y}} e^{i (\textbf{k}-\textbf{q}) \textbf{y}} \frac{1}{k^+} \int_0^{L^+} dy^+ e^{-i\textbf{k}\textbf{x}} [\partial_{\textbf{y}^i} \mathcal{G}^{ac}_{k^+}(L^+,\textbf{x};y^+,\textbf{y})] \mathcal{U}_{\textbf{y}}^{cb}(y^+,0) \Bigg\}.
\end{align}
In this equation $\int_\textbf{x}\equiv \int d^2\textbf{x}$, $\epsilon^{i*}_\lambda (\textbf{k})$ is the polarisation vector, $k^-=\textbf{k}^2/(2k^+)$,  $k^+=e^\eta/\sqrt{2}$, $L^+$ is the longitudinal length of the target, $\textbf{q}$ is the transverse momentum transferred from the target during the interaction and $\textbf{k-q}$ is the transverse momenta of the projectile color charge density, and
\begin{equation}
	\mathcal{G}^{ab}_{k^+}(x^+,\textbf{x};y^+,\textbf{y})= \Theta(x^+-y^+) \int_{\textbf{y}}^{\textbf{x}} \mathcal{D} \textbf{z} \exp \left[  \frac{i k^+}{2} \int_{y^+}^{x^+} d z^+ \dot{\textbf{z}}^2(z^+) \right] \mathcal{U}^{ab}_{\textbf{z}} \left( x^+,y^+ \right)
\end{equation}
is the scalar gluon propagator, with the path integral taking into account the Brownian motion of the gluon in transverse plane, for fixed ends of the trajectory $\textbf{z}(x^+)=\textbf{x}$, $\textbf{z}(y^+)=\textbf{y}$. We use light-cone coordinates $x^\pm=(x^0\pm x^3)/\sqrt{2}$.
\begin{equation}
	\mathcal{U}_{\textbf{x}}^{ab}( x^+,y^+)=\mathcal{P} \exp\left\{ i g \int_{y^+}^{x^+} dz^+ \mathcal{A}^-(z^+,\textbf{x}) \right\}^{ab}
\end{equation}
is the Wilson line that accounts for the multiple gluon exchanges with the target. 

 It is necessary to mention that the reduced matrix amplitude $\overline{\mathcal{M}}^{ab}_\lambda(\eta,\textbf{k},\textbf{q})$ given in Eq. \eqref{asda} is written for a target with a longitudinal width $L^+$ and therefore goes beyond the standard eikonal approximation commonly adopted in CGC calculations. In the eikonal approximation 
%As commonly done in the CGC framework, we use the eikonal approximation, see~\cite{Altinoluk:2020wpf}. In this approximation 
(see~\cite{Altinoluk:2020wpf} and references there) the target and  projectile are taken as very highly boosted systems without longitudinal extent because of Lorentz contraction. This is equivalent to taking the limit $L^+ \rightarrow 0$, $k^+ \rightarrow \infty$ and assuming that the target field is a local shock-wave, $\mathcal{A}^-(z^+,\textbf{x}) \propto \delta(z^+)$. In  the present work, we restrict ourselves to the standard CGC framework and adopt the eikonal approximation. Within this approximation, the scalar gluon propagator simplifies and can be written as
%Thus, in the eikonal approximation we can write the scalar gluon propagator as
\begin{equation}
	\mathcal{G}^{ab}_{k^+}(\underline{x},\underline{y})\to\mathcal{U}^{ab}_{\textbf{x}} \left( x^+,y^+ \right) \delta^{(2)}(\textbf{x}-\textbf{y}).
\end{equation}
Consequently, the reduced matrix amplitude given in \cref{asda} simplifies as well and it reads
\begin{align}\label{amplitude}
	\overline{\mathcal{M}}^{ab}_\lambda(\textbf{k},\textbf{q})&
%=2 i \epsilon^{i*}_\lambda L^i(\textbf{k,\textbf{q}}) \int_y e^{-i \textbf{q}\textbf{y}}   \mathcal{U}_{\textbf{y}}^{ab}(L^+,0)
%	\nonumber \\
%	&
=2 i \epsilon^{i*}_\lambda (\textbf{k}) L^i(\textbf{k,\textbf{q}}) \int_y e^{-i \textbf{q}\textbf{y}}   U^{ab}(\textbf{y}), 
\end{align}
where we have introduced the Lipatov vertex
\begin{align}\label{lipatovv}
	L^i(\textbf{k,\textbf{q}})=\frac{\textbf{k}^i}{\textbf{k}^2}-\frac{(\textbf{k}-\textbf{q})^i}{(\textbf{k}-\textbf{q})^2}
\end{align}
and changed the notation of the Wilson lines, $U^{ab}(\textbf{y})=\mathcal{U}_{\textbf{y}}^{ab}(L^+,0)$.
%and the Eikonal Wilson line
%\begin{align}
%	U^{ab}(\textbf{y})=\mathcal{P}^+ e^{ig \int dz^+ (\mathcal{A}(z^+,\textbf{x})T)^{ab}}.
%\end{align}

\begin{figure}
	\centering
	\includegraphics*[scale=0.9]{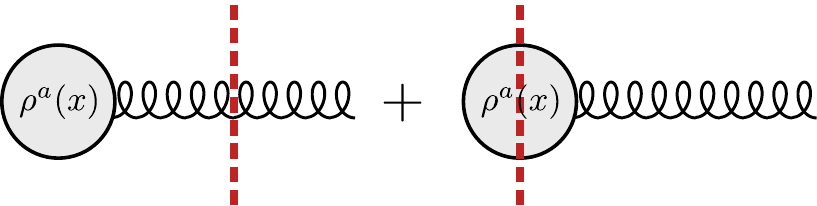}
	\caption{Physical interpretation of the Lipatov vertex, with the vertical dashed line denoting the interaction with the target.}
	\label{lipatov_fig}
\end{figure}

The physical interpretation of the Lipatov vertex, see Fig.~\ref{lipatov_fig}, is such that the first element in the sum in \cref{lipatovv} accounts for interaction of the color source $\rho^a(\textbf{x})$ with the target before emitting the gluon and the second element accounts for a gluon being emitted from the source and then interacting with the target.

In~\cite{Agostini:2019avp,Agostini:2019hkj} it was shown that the corrections with respect to the eikonal approximation stemming from the target having a finite length can be important for collision energies below a few hundred GeV. Corrections coming from the inclusion of transverse components of the background field have also been considered in~\cite{Balitsky:2015qba,Balitsky:2016dgz,Chirilli:2018kkw,Altinoluk:2020oyd,Chirilli:2021lif}, but until now no estimation is available of their quantitative impact on particle production. In the remainder of this work we restrict to the eikonal approximation.

The multiplicity for producing $n$-gluons with transverse momentum $\textbf{k}_i$, pseudorapidity $\eta_i$, color $a_i$ and polarization $\lambda_i$ is given, in terms of the amplitude matrix that is leading for $g\rho^a (\textbf{q})\sim 1$, as
\begin{align}\label{multiplicity}
	2^n (2 \pi)^{3n} \frac{d^n N}{\prod_{i=1}^n d \eta_i d^2 \textbf{k}_i}= \Big \langle \mathcal{M}_{\lambda_1}^{a_1}(\eta_1,\textbf{k}_1)\cdots \mathcal{M}_{\lambda_n}^{a_n}(\eta_n,\textbf{k}_n) \left(\mathcal{M}_{\lambda_n}^{a_n}(\eta_n,\textbf{k}_n)\right)^{\dagger} \cdots \left(\mathcal{M}_{\lambda_1}^{a_1}(\eta_1,\textbf{k}_1)\right)^{\dagger} \Big \rangle_{p,T},
\end{align}
where $ \braket{\cdots}_{p,T} $ denotes the average over the color charge density configurations of the projectile and target. The factor of $2^n$ on the right hand side of Eq.~\eqref{multiplicity}  originates from the Lorentz invariant phase space written in terms of rapidity $\eta_i$.

\begin{figure}[htb]
	\centering
	\vskip -1.5cm
	\includegraphics*[scale=0.9]{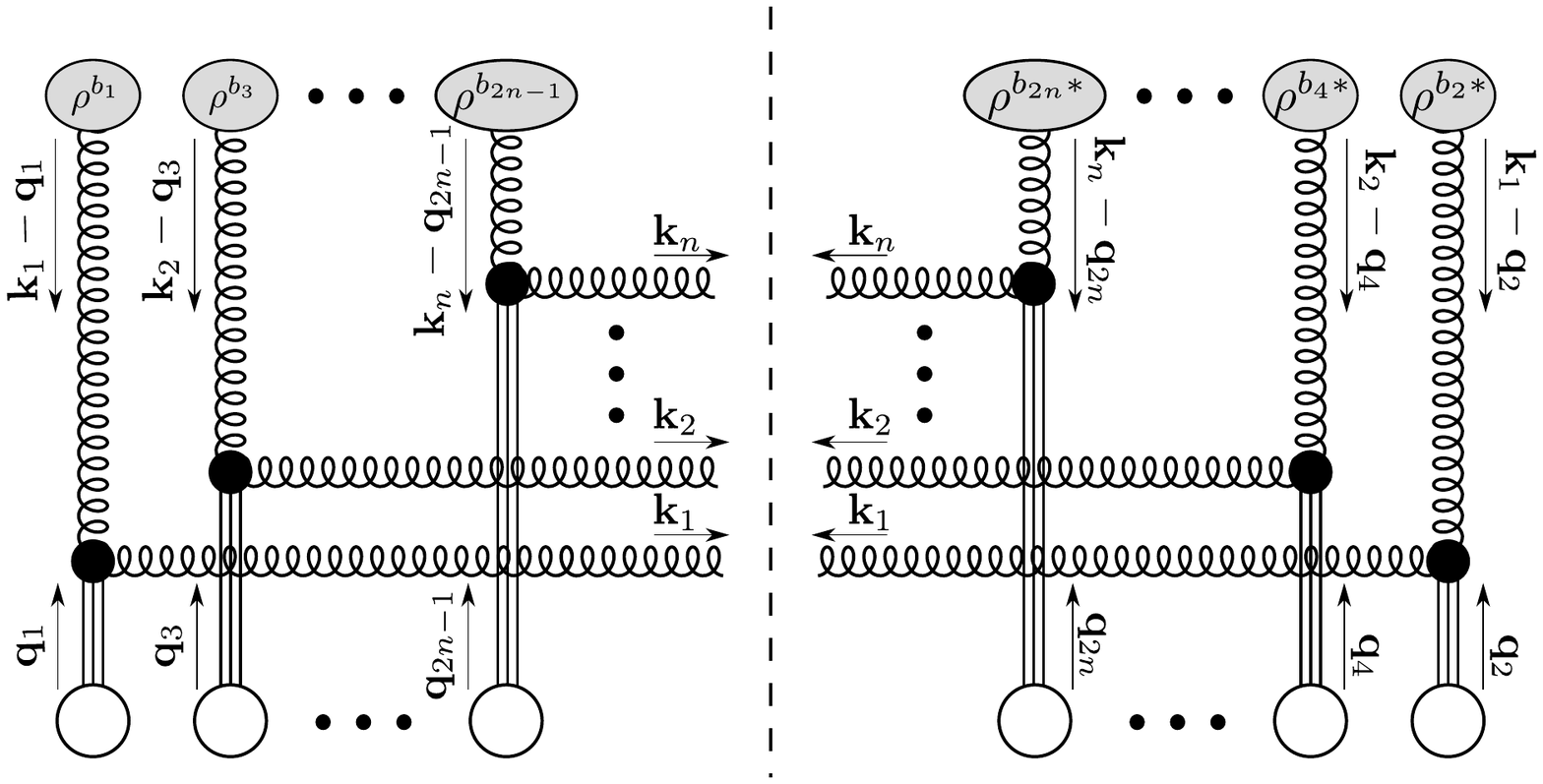}
	\vskip -16.2cm
	\caption{Diagram showing Eq.~\eqref{mul_kfac} with its momentum assignments. Each group of three vertical straight lines represents the rescattering with the target, the black blobs Lipatov vertices and the vertical dashed line the cut.}
	\label{mastereq_fig}
\end{figure}

Using \cref{factor} and dropping the dependence on $\eta$ due to the eikonal approximation, we can write this expression as (see Fig.~\ref{mastereq_fig})
\begin{align}\label{mul_kfac}
	2^n(2 \pi)^{3n} \frac{d^n N}{\prod_{i=1}^{n}d^2 \textbf{k}_i}=&g^{2n} \int \left(\prod_{i=1}^{2n} \frac{d^2 \textbf{q}_i}{(2 \pi)^2}\right) \Big \langle \rho^{b_1}(\textbf{k}_1-\textbf{q}_1)\rho^{b_2 \dagger}(\textbf{k}_1-\textbf{q}_2) \cdots \rho^{b_{2n-1}}(\textbf{k}_n-\textbf{q}_{2n-1})\rho^{b_{2n} \dagger}(\textbf{k}_n-\textbf{q}_{2n}) \Big \rangle_{p}
	\nonumber 
	\\
	&\times
	\Big \langle \overline{\mathcal{M}}^{a_1 b_1}_{\lambda_1}(\textbf{k}_1,\textbf{q}_1) \overline{\mathcal{M}}^{b_2 a_1 \dagger}_{\lambda_1}(\textbf{k}_1,\textbf{q}_2) \cdots \overline{\mathcal{M}}^{a_{n} b_{2n-1}}_{\lambda_{n}}(\textbf{k}_n,\textbf{q}_{2n-1}) \overline{\mathcal{M}}^{b_{2n} a_{n} \dagger}_{\lambda_{n}}(\textbf{k}_n,\textbf{q}_{2n}) \Big \rangle_{T}.
\end{align}
Solving this equation is the main point of this work and will be the focus of the discussion in the next sections.

\cref{mul_kfac} involves the $2n$-point correlation functions of the color charge densities of the projectile and the target. Solving these objects is a highly non trivial task. It is known that using the MV Gaussian weight~\cite{McLerran:1993ni,McLerran:1993ka} it is possible to find a closed-form solution for these correlators~\cite{Gelis:2001da,Fujii:2002vh,Blaizot:2004wv,Dominguez:2008aa,Dominguez:2012ad,Apolinario:2014csa,Shi:2017gcq,Dusling:2017aot}. However, the solution can be extremely complicated for $n>2$ if the large-$N_c$ limit is not taken. Corrections to the Gaussian weight for the 2-point correlator were considered in~\cite{Dumitru:2011ax}, and solutions based on high-energy evolution at large $N_c$ in~\cite{Iancu:2011ns}, but in this work we restrict ourselves to the MV model.

When $|\textbf{k}|/Q_s \gg 1$ with $Q_s$ the saturation momentum of the gluons in the target~\cite{Gelis:2010nm,Kovchegov:2012mbw}, or equivalently $g A^- \ll 1$, we can expand the product of $2n$ Wilson lines up to order $g^{2n}$. Thus, if a Gaussian weight is chosen for the target color charge density we can apply  Wick's theorem and write the $2n$-point functions as the sum of $(2n-1)!!$ products of $n$ 2-point functions, thus being able to solve the correlator exactly. This  is the  glasma graph approach previously mentioned and has been used in several works~\cite{Dumitru:2008wn,Dusling:2009ar,Dusling:2009ni,Dusling:2013oia,Ozonder:2014sra,Gelis:2009wh} to produce phenomenological results. The main disadvantage of this approximation is that it works in a small kinematic range, being only suitable for dilute-dilute collisions.

Another approach for evaluating the $n$-point functions that keeps the simplicity of the glasma graph approximation but without having to restrict ourselves to the dilute limit is the so called area enhancement argument~\cite{Altinoluk:2018ogz,Altinoluk:2018hcu,Kovner:2017ssr,Kovner:2018vec}. We will explain the this argument in the next section.

\subsection{The area enhancement argument}\label{secarea}

One of the key points to evaluate the multiplicity for multi-particle production is the calculation of the average over charge color densities of $2n$ matrix amplitudes:
\begin{align}
	&\Big \langle \overline{\mathcal{M}}^{a_1 b_1}_{\lambda_1}(\textbf{k}_1,\textbf{q}_1) \overline{\mathcal{M}}^{b_2 a_1 \dagger}_{\lambda_1}(\textbf{k}_1,\textbf{q}_2) \cdots \overline{\mathcal{M}}^{a_{n} b_{2n-1}}_{\lambda_{n}}(\textbf{k}_n,\textbf{q}_{2n-1}) \overline{\mathcal{M}}^{b_{2n} a_{n} \dagger}_{\lambda_{n}}(\textbf{k}_n,\textbf{q}_{2n}) \Big \rangle_{T}
	\nonumber \\
	\propto 
	& \int_{\textbf{y}_1 \cdots \textbf{y}_{2n}} e^{-i \textbf{q}_1 \cdot \textbf{y}_1+i \textbf{q}_2 \cdot \textbf{y}_2 \cdots +i \textbf{q}_{2n} \cdot \textbf{y}_{2n} }
	\Big \langle U(\textbf{y}_1)^{a_1 b_1}U^\dagger(\textbf{y}_2)^{b_2 a_2} \cdots U^\dagger(\textbf{y}_{2n})^{b_{2n} a_{2n}}\Big \rangle_{T},
\end{align}
where we have used the fact that the only part of the reduced amplitude that depends on the target charge density are the Wilson lines.

For the sake of simplicity, let us just consider the case where we just have 4 Wilson lines in such a way that the color indices are contracted forming a single trace. In this case the object that we have to evaluate is the quadrupole operator
\begin{align}\label{area1}
	\tilde{Q}(\textbf{q}_1,\textbf{q}_2,\textbf{q}_3,\textbf{q}_4)&=\int_{\textbf{y}_1 \textbf{y}_2 \textbf{y}_3 \textbf{y}_4} e^{-i \textbf{q}_1 \cdot \textbf{y}_1+i \textbf{q}_2 \cdot \textbf{y}_2 -i \textbf{q}_3 \cdot \textbf{y}_3 +i \textbf{q}_{4} \cdot \textbf{y}_{4} }
	\frac{1}{N_c^2-1}\Big \langle Tr \left[U(\textbf{y}_1) U^\dagger(\textbf{y}_2) U(\textbf{y}_3) U^\dagger(\textbf{y}_{4})\right]\Big \rangle_{T}
	\nonumber \\ 
	&\equiv\int_{\textbf{y}_1 \textbf{y}_2 \textbf{y}_3 \textbf{y}_4} e^{-i \textbf{q}_1 \cdot \textbf{y}_1+i \textbf{q}_2 \cdot \textbf{y}_2 -i \textbf{q}_3 \cdot \textbf{y}_3 +i \textbf{q}_{4} \cdot \textbf{y}_{4} } 
	Q(\textbf{y}_1,\textbf{y}_2,\textbf{y}_3,\textbf{y}_4).
\end{align}

Following the arguments in~\cite{Kovner:2017ssr,Kovner:2018vec}, the configuration of the transverse coordinates $\textbf{y}_i$ that maximises the integral is such that these legs are as far away as possible between them. On the other hand, in the CGC picture the target ensemble is composed by domains of chromoelectric field with a typical correlation length, $Q_s^{-1}$, that is fixed by the saturation scale, where color neutralises. Therefore two objects that only depend on the target color charge density, sitting at two different points $\textbf{y}_i$ and $\textbf{y}_j$, will have a vanishing correlation when $|\textbf{y}_i-\textbf{y}_j| \gg Q_s^{-1}$. This implies that the only way of obtaining a non vanishing correlator is by grouping the legs in, at least, pairs where the distance between the transverse points is smaller than the correlation length. Thus, for the case of the quadrupole, this is equivalent to write
\begin{align}\label{area2}
	Q(\textbf{y}_1,\textbf{y}_2,\textbf{y}_3,\textbf{y}_4) \approx 
	&\lim_{ {|\textbf{y}_1-\textbf{y}_2| \lesssim Q_s^{-1} \atop |\textbf{y}_3-\textbf{y}_4| \lesssim Q_s^{-1}} \atop |\textbf{y}_1-\textbf{y}_3| \lesssim Q_s^{-1} } Q(\textbf{y}_1,\textbf{y}_2,\textbf{y}_3,\textbf{y}_4)
	+
	\lim_{ {|\textbf{y}_1-\textbf{y}_2| \lesssim Q_s^{-1} \atop |\textbf{y}_3-\textbf{y}_4| \lesssim Q_s^{-1}} \atop |\textbf{y}_1-\textbf{y}_3| \gg Q_s^{-1} } Q(\textbf{y}_1,\textbf{y}_2,\textbf{y}_3,\textbf{y}_4)
	\nonumber \\
	+
	&\lim_{ {|\textbf{y}_1-\textbf{y}_3| \lesssim Q_s^{-1} \atop |\textbf{y}_2-\textbf{y}_4| \lesssim Q_s^{-1}} \atop |\textbf{y}_1-\textbf{y}_2| \gg Q_s^{-1} } Q(\textbf{y}_1,\textbf{y}_2,\textbf{y}_3,\textbf{y}_4)
	+
	\lim_{ {|\textbf{y}_1-\textbf{y}_4| \lesssim Q_s^{-1} \atop |\textbf{y}_2-\textbf{y}_3| \lesssim Q_s^{-1}} \atop |\textbf{y}_1-\textbf{y}_2| \gg Q_s^{-1} } Q(\textbf{y}_1,\textbf{y}_2,\textbf{y}_3,\textbf{y}_4).
\end{align}

The first term of this equation, although it gives a non vanishing contribution to the 4-point correlator, is constrained to a smaller region of  phase space than the other 3 terms. This will imply that, after performing the integration in \cref{area1}, it will be suppressed by the area of the target with respect to the other ones. On the other hand, the other three terms can be just written as a product of dipoles, that is
\begin{align}
	\lim_{ {|\textbf{y}_1-\textbf{y}_2| \lesssim Q_s^{-1} \atop |\textbf{y}_3-\textbf{y}_4| \lesssim Q_s^{-1}} \atop |\textbf{y}_1-\textbf{y}_3| \gg Q_s^{-1}}
	\Big \langle U(\textbf{y}_1)^{a_1 b_1}U^\dagger(\textbf{y}_2)^{b_1 a_2} U(\textbf{y}_3)^{a_2 b_2} U^\dagger(\textbf{y}_{4})^{b_{2} a_{1}}\Big \rangle_{T} 
	\approx \Big \langle U(\textbf{y}_1)^{a_1 b_1}U^\dagger(\textbf{y}_2)^{b_1 a_2} 
	\Big \rangle_{T} \Big \langle
	U(\textbf{y}_3)^{a_2 b_2} U^\dagger(\textbf{y}_{4})^{b_{2} a_{1}}\Big \rangle_{T}
\end{align}
and analogously to the other terms. Thus we can write the quadrupole operator as a sum of products of 2-point functions,
\begin{align}
	\Big \langle U(\textbf{y}_1)^{a_1 b_1}U^\dagger(\textbf{y}_2)^{b_1 a_2} 
	U(\textbf{y}_3)^{a_2 b_2} U^\dagger(\textbf{y}_{4})^{b_{2} a_{1}}\Big \rangle_{T}
	\approx  &\Big \langle U(\textbf{y}_1)^{a_1 b_1}U^\dagger(\textbf{y}_2)^{b_1 a_2} 
	\Big \rangle_{T} \Big \langle
	U(\textbf{y}_3)^{a_2 b_2} U^\dagger(\textbf{y}_{4})^{b_{2} a_{1}}\Big \rangle_{T}
	\nonumber \\
	+&\Big \langle U(\textbf{y}_1)^{a_1 b_1} U(\textbf{y}_3)^{a_2 b_2} 
	\Big \rangle_{T} \Big \langle
	U^\dagger(\textbf{y}_2)^{b_1 a_2}  U^\dagger(\textbf{y}_{4})^{b_{2} a_{1}}\Big \rangle_{T}
	\nonumber \\
	+&\Big \langle U(\textbf{y}_1)^{a_1 b_1}U^\dagger(\textbf{y}_{4})^{b_{2} a_{1}} 
	\Big \rangle_{T} \Big \langle
	U^\dagger(\textbf{y}_2)^{b_1 a_2} U(\textbf{y}_3)^{a_2 b_2} \Big \rangle_{T},
\end{align}
keeping in mind that this approximation is only good after performing the phase space integral since, otherwise the first term in \cref{area2} is non-negligible. In Appendix~\ref{app_AE} we discuss the validity of this argument, that we call {\it area enhancement argument}.

This result can be generalised to the case of any number or configuration of the Wilson lines by noting that the contribution of the multipole that is enhanced by the area of the target, i.e., that is leading in $S_\perp Q_s^{-2}$ with $S_\perp$ the area of the projectile (or the overlap area in a dilute-dense collision), is always a sum over all possible combinations of 2-point functions. This is analogous to assume that the target averages of Wilson lines follow a Gaussian statistics and thus we are able to apply  Wick's theorem to them:
\begin{align}\label{wickarea}
	\Big \langle U(\textbf{y}_1)^{a_1 b_1}U(\textbf{y}_2)^{a_2 b_2} \cdots U(\textbf{y}_{2n})^{a_{2n} b_{2n}}\Big \rangle_{T}=\sum_{\sigma \in \Pi(\chi)} \prod_{\{\alpha,\beta\}\in \sigma}  
	\Big \langle U(\textbf{y}_\alpha)^{a_\alpha b_\alpha} U(\textbf{y}_\beta)^{a_\beta b_\beta} \Big \rangle_{T}\, ,
\end{align}
being $\chi=\{1,2,\dots,2 n\}$ and $\Pi(\chi)$ the set of partitions of $\chi$ with disjoint pairs. \cref{wickarea} simplifies enormously the evaluation of  multipoles and shares its simplicity with the glasma graph approach through  Wick's theorem. The main difference between them is that the first does not rely on the dilute limit and thus is applicable to dilute-dense scattering.	
This approach has been used recently~\cite{Altinoluk:2018ogz,Altinoluk:2020psk} in order to evaluate the phase space integral of 4-point and 6-point functions. We will use it in order to evaluate \cref{mul_kfac}.

\subsection{Particle correlations}\label{sec1b}

In this section we summarise the main ideas behind particle correlations within the CGC effective theory and provide the general formulae that we will employ to study azimuthal correlations. For a complete review of the former aspect, we refer to~\cite{Altinoluk:2020wpf} and references therein. 

\begin{figure}[h!]
	\centering
	\includegraphics*[scale=0.3]{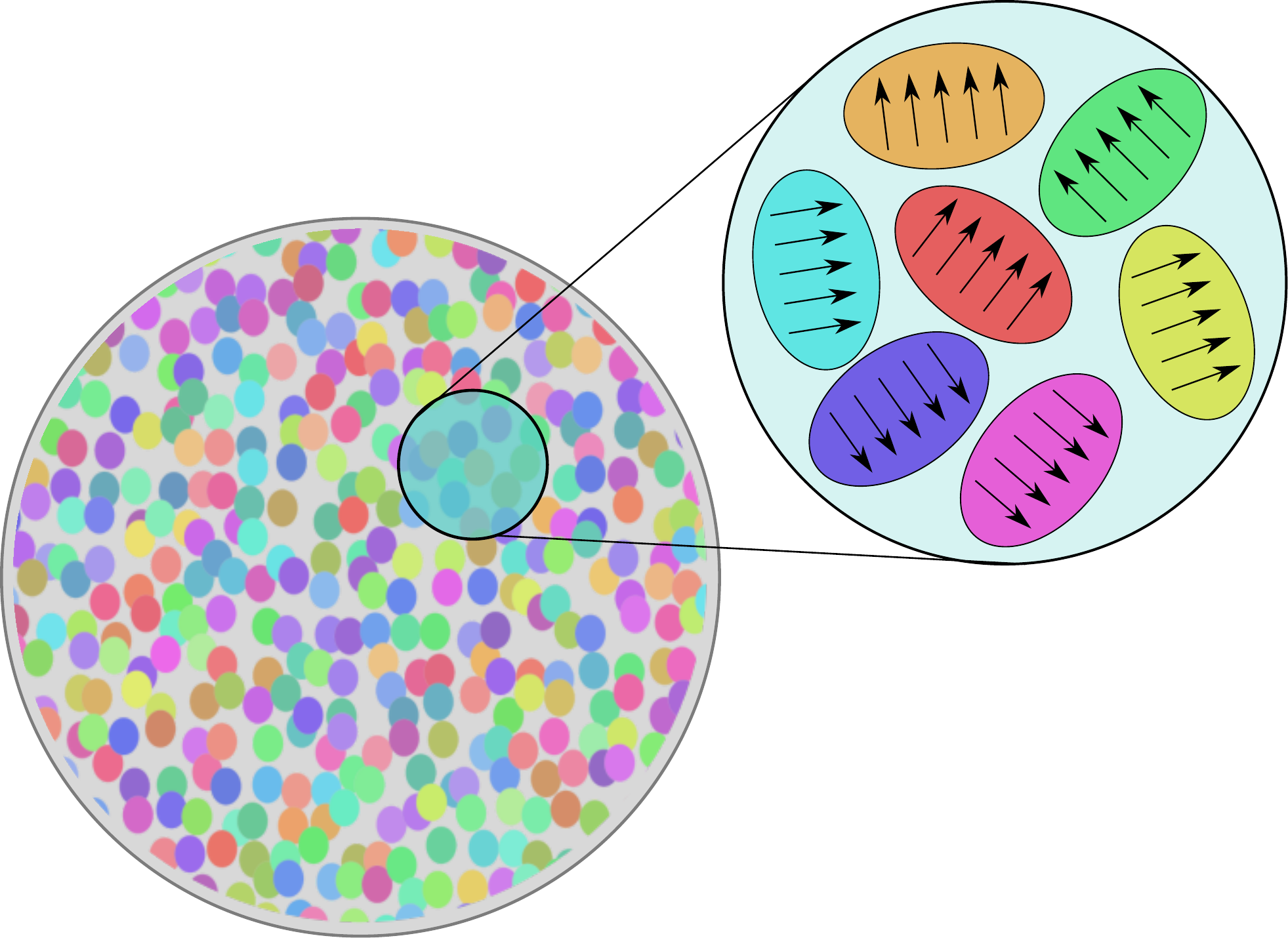}
	\caption{Picture of the chromoelectric fields inside the target.}
	\label{target_domains}
\end{figure}

Following the argument in~\cite{Kovner:2010xk,Kovner:2011pe}, we make the picture of angular correlations in dilute-dense scatterings through the interactions of the projectile partons with the strong chromoelectric fields generated by the target. The strength of the chromoelectric field in the target wave function is characterised by the saturation momentum, $Q_s$, which is also the typical momentum of partons in the wave function, $k_\perp \sim Q_s$. The correlation length of the fields is roughly $Q_s^{-1}$ and the target ensemble can be modelled as a compound of domains with different chromoelectric fields that change from event to event as illustrated in~\cref{target_domains}. When a parton coming from the projectile wave function hits the target it will scatter in one of these domains and will pick a momentum that is proportional to the chromoelectric field inside this domain. Thus angular correlation appears when two partons scatter in the same chromoelectric domain. As gluons belong to a real representation of SU(3), scattering with parallel and antiparallel momenta is identical. Thus, this picture is also able to explain the absence of odd azimuthal correlations in gluon production.

In order to study the azimuthal harmonics appearing in multi-particle correlation it is convenient to use the cumulant method~\cite{Borghini:2001vi}. This method aims to reduce the contribution of the so-called "non-flow" correlation, i.e., contributions to the correlation function that come from other processes other than true collective flow, such as resonance decays or jet correlations, to the definition of the azimuthal harmonics. In this method we define the 2- and 4-particle cumulants of order $n$ as
\begin{align}
	c_n\{2\} &= \Big \langle e^{i n ( \phi_1 -\phi_2)} \Big \rangle,
	\label{cn2_def}
	\\
	c_n\{4\} &= \Big \langle e^{i n ( \phi_1 +\phi_2-\phi_3-\phi_4)} \Big \rangle - 2 \Big \langle e^{i n ( \phi_1 -\phi_2)} \Big \rangle^2,
	\label{cn4_def}
\end{align}
where $\langle \cdots \rangle$ denotes the average over all events and particles. For convenience we define the $n^{\text{th}}$-order $\kappa$-function
\begin{equation}\label{kappa_def}
	\kappa_n\{m\}= 
	\int \left(\prod_{i=1}^m \frac{d^2 \textbf{k}_i}{(2\pi)^2}\right) 
	\frac{d^m N}{\prod_{i=1}^m d^2 \textbf{k}_i} 
	e^{i n (\phi_1+\cdots+\phi_{m/2}-\phi_{m/2+1}-\cdots-\phi_m)},
\end{equation}
in such a way that the event average can be written as
\begin{equation}
	\Big\langle e^{i n (\phi_1+\cdots+\phi_{m/2}-\phi_{m/2+1}-\cdots-\phi_m)} \Big\rangle = \frac{\kappa_n \{m\}}{\kappa_0 \{m\}}\ .
\end{equation}

Given this definition of the cumulants we can write the 2- and 4-particle Fourier harmonics of order $n$ as
\begin{align}
	v_n\{2\}&=(c_n\{2\})^{1/2},\label{vn2}
	\\
	v_n\{4\}&=(-c_n\{4\})^{1/4}.\label{vn4}
\end{align}

Similarly, the harmonics $v_n$ can also be defined as a function of the transverse momentum. In order to do that we also define the so-called "differential" cumulants:
\begin{align}
	d_n\{2\}(p_\perp)&=\frac{\tilde{\kappa}_n \{2\}(p_\perp)}{\tilde{\kappa}_0 \{2\}(p_\perp)}\ ,
	\label{dn2_def}
	\\
	d_n\{4\}(p_\perp)&=\frac{\tilde{\kappa}_n \{4\}(p_\perp)}{\tilde{\kappa}_0 \{4\}(p_\perp)}-2 \frac{\tilde{\kappa}_n \{2\}(p_\perp)}{\tilde{\kappa}_0 \{2\}(p_\perp)} \ \frac{\kappa_n\{2\}}{\kappa_0\{2\}},
	\label{dn4_def}
\end{align}
where $p_\perp=|\textbf{p}|$ and we have defined the differential $\kappa$-functions as
\begin{equation}\label{kappadif}
	\tilde{\kappa}_n \{m\}(p_\perp) \equiv \frac{d\kappa_n \{m\}}{p_\perp d p_\perp}=\int_0^{2 \pi} d\phi_1 
	\int \left(\prod_{i=2}^m \frac{d^2 \textbf{k}_i}{(2\pi)^2}\right) 
	\frac{d^m N}{\prod_{i=1}^m d^2 \textbf{k}_i} \Bigg\rvert_{\textbf{k}_1=\textbf{p}} 
	e^{i n (\phi_1+\cdots+\phi_{m/2}-\phi_{m/2+1}-\cdots-\phi_m)}.
\end{equation}
With this prescription, the differential azimuthal harmonics are given by\footnote{In experimental analysis, e.g.~\cite{Chatrchyan:2013nka}, these harmonics are usually normalised as
	\begin{align*}
		v_n\{2\}(p_\perp) &= \frac{d_n\{2\}(p_\perp)}{(c_n\{2\})^{1/2}},
		\\
		v_n\{4\}(p_\perp) &= \frac{-d_n\{4\}(p_\perp)}{(-c_n\{4\})^{3/4}}.
	\end{align*}
}
\begin{align}
	v_n\{2\}(p_\perp) &= [d_n\{2\}(p_\perp)]^{1/2},
	\\
	v_n\{4\}(p_\perp) &= [-d_n\{4\}(p_\perp)]^{1/4}.\label{vn4pt}
\end{align}

\section{Evaluating the target and projectile correlation functions}\label{sec2}

In this section we will introduce the notation and the arguments followed in order to evaluate the $2n$-point correlation functions for both  projectile and  target ensembles.	

\subsection{Setting up the notation}

We write the reduced matrix amplitude  as
\begin{align}
	\Lambda_i \equiv \overline{\mathcal{M}}^{a_i b_i}_{\lambda_i}(\textbf{k}_i,\textbf{q}_i),
\end{align}
where $i={1,\dots,2 n}$. We should note, however, that in this notation when $i$ is even the reduced matrix element is conjugate (i.e., to the right of the cut) and when it is odd it is not conjugate (i.e., to the left of the cut). Furthermore, since the produced gluon has the same momentum, polarisation and color both in the real and conjugate spaces we have to apply the following constraints:
\begin{align}
	\textbf{k}_{2m}=\textbf{k}_{2m-1}\ , \label{eqk}
	\\
	\lambda_{2m}=\lambda_{2m-1}\ , \label{eql}
	\\
	a_{2m}=a_{2m-1}\ , \label{eqa}
\end{align}
with $m=1,\dots,n$. Thus this notation also changes the usual labelling of the gluon final momenta, $\textbf{k}_i$, since now they are labeled by only odd numbers ($1,3,5,\dots$) instead of  $1,2,3,\dots$. With these convention we can write the $2n$-point function of the reduced matrix amplitudes in the simplified form
\begin{align}\label{lambdacorr}
	\Big \langle \overline{\mathcal{M}}^{a_1 b_1}_{\lambda_1}(\textbf{k}_1,\textbf{q}_1) \overline{\mathcal{M}}^{b_2 a_1 \dagger}_{\lambda_1}(\textbf{k}_1,\textbf{q}_2) \cdots \overline{\mathcal{M}}^{a_{n} b_{2n-1}}_{\lambda_{n}}(\textbf{k}_n,\textbf{q}_{2n-1}) \overline{\mathcal{M}}^{b_{2n} a_{n} \dagger}_{\lambda_{n}}(\textbf{k}_n,\textbf{q}_{2n}) \Big \rangle_{T} =
	\Big \langle \Lambda_1 \Lambda_2 \cdots \Lambda_{2n-1} \Lambda_{2n} \Big \rangle_{T}.
\end{align}

As we have pointed out in \cref{secarea}, in this work we will use the area enhancement argument in order to evaluate the multipole correlators. Thus, using the same arguments that we have used for obtaining \cref{wickarea}, we apply Wick's theorem and  \cref{lambdacorr} reads
\begin{align}\label{Wick1}
	\Big \langle \Lambda_1 \Lambda_2 \cdots \Lambda_{2n-1} \Lambda_{2n} \Big \rangle_{T}=\sum_{\sigma \in \Pi(\chi)} \prod_{\{\alpha,\beta\}\in \sigma}  \Big \langle \Lambda_\alpha \Lambda_\beta \Big \rangle_{T},
\end{align}
with $\chi=\{1,2,\dots,2 n\}$ and $\Pi(\chi)$ the set of partitions of $\chi$ with disjoint pairs. On the other hand, in order to evaluate the 2-point function we use \cref{amplitude},
\begin{align}\label{2point}
	\Big \langle \Lambda_\alpha \Lambda_\beta \Big \rangle_{T}= 4  
	\epsilon^{i*}_{\lambda_\alpha}(\textbf{k}_\alpha) L^i(\textbf{k}_\alpha,\textbf{q}_\alpha) \epsilon^{j*}_{\lambda_\beta} (\textbf{k}_\beta) L^j(\textbf{k}_\beta,\textbf{q}_\beta) 
	\int_{\textbf{y}_\alpha} e^{i(-1)^\alpha \textbf{q}_\alpha \textbf{y}_\alpha + i(-1)^\beta \textbf{q}_\beta \textbf{y}_\beta}   
	\Big \langle U^{a_\alpha b_\alpha}(\textbf{y}_\alpha) U^{a_\beta b_\beta}(\textbf{y}_\beta)  \Big \rangle_{T},
\end{align}
where we do not write the overall sign of the equation which should be $-(-1)^{\alpha+\beta}$ since the number of real and complex conjugate matrix elements are the same and therefore the net sign of \cref{Wick1} will always be positive. Another simplification that we can make is by noting that the Lipatov vertices will always be contracted with the one that is evaluated at the same momentum $\textbf{k}_i$. This follows from the fact that two gluons with the same transverse momentum will also have the same polarisation and thus the polarisation vectors fulfill
\begin{align}
	\epsilon^{i*}_{\lambda}(\textbf{k}) \epsilon^{j}_{\lambda}(\textbf{k}) = \delta^{ij},
\end{align}
which  implies a contraction of the two Lipatov vertices with the same $\textbf{k}$-momentum. Thus we will write directly $L^\lambda(\textbf{k},\textbf{q})$ in \cref{2point} instead of $ \epsilon^{i*}_{\lambda}(\textbf{k}) L^i(\textbf{k},\textbf{q})$ because both expressions lead to the same result. 

On the other hand, we should also evaluate the average of two Wilson lines. In order to do so we follow \cite{Kovner:2018vec} and  use the fact that the target ensemble is globally color invariant, which implies that the average of any tensor in this ensemble has to be proportional to a linear combination of invariant tensors. Thus
\begin{align}
	\Big \langle U^{a_\alpha b_\alpha}(\textbf{y}_\alpha) U^{a_\beta b_\beta}(\textbf{y}_\beta)  \Big \rangle_{T}
	=\frac{\delta^{a_\alpha a_\beta}\delta^{b_\alpha b_\beta}}{(N_c^2-1)^2} \Big \langle Tr \left[U(\textbf{y}_\alpha) U(\textbf{y}_\beta)\right]  \Big \rangle_{T}
	\equiv \frac{\delta^{a_\alpha a_\beta}\delta^{b_\alpha b_\beta}}{N_c^2-1} D(\textbf{y}_\alpha,\textbf{y}_\beta),
\end{align}
where we have introduced the dipole operator $D(\textbf{x},\textbf{y})$.

Therefore, making the change of variables $\textbf{y}_{\alpha,\beta}=\textbf{b} \pm \textbf{r}/2$ we can write \cref{2point} as
\begin{align}
	\Big \langle \Lambda_\alpha \Lambda_\beta \Big \rangle_{T}=
	4 L^{\lambda_\alpha}(\textbf{k}_\alpha,\textbf{q}_\alpha)L^{\lambda_\beta}(\textbf{k}_\beta,\textbf{q}_\beta)
	\frac{\delta^{a_\alpha a_\beta}\delta^{b_\alpha b_\beta}}{N_c^2-1} 
	\int_{\textbf{r},\textbf{b}} 
	e^{i\textbf{b}[(-1)^\alpha \textbf{q}_\alpha+(-1)^\beta \textbf{q}_\beta]+i\textbf{r}/2[(-1)^\alpha \textbf{q}_\alpha-(-1)^\beta \textbf{q}_\beta]}D(\textbf{r},\textbf{b}).
\end{align}

This equation can be simplified further if we exploit the fact that the target ensemble has a much larger extension in the transverse plane than the projectile one and then we assume translational invariance of the dipole operator, that is, $D(\textbf{r},\textbf{b}) = D(|\textbf{r}|)$. Thus, defining the Fourier transform of the dipole operator
\begin{align}
	d(\textbf{q})=\int_{\textbf{r}} e^{-i \textbf{q} \cdot \textbf{r}} D(|\textbf{r}|),
\end{align}
we obtain our final expression for the 2-point function of the reduced matrix amplitude:
\begin{align}\label{target}
	\Big \langle \Lambda_\alpha \Lambda_\beta \Big \rangle_{T}=4\frac{\delta^{a_\alpha a_\beta}\delta^{b_\alpha b_\beta}}{N_c^2-1} (2 \pi)^2 \delta^{(2)}[\textbf{q}_\alpha+(-1)^{\alpha+\beta}\textbf{q}_\beta] L^{\lambda_\alpha}(\textbf{k}_\alpha,\textbf{q}_\alpha)L^{\lambda_\beta}(\textbf{k}_\beta,\textbf{q}_\beta) d(\textbf{q}_\alpha).
\end{align}

In order to obtain a final expression for \cref{mul_kfac} we should also evaluate the $2n$-point function of the projectile color charge densities. In this case we will use the generalised MV model and also use the Wick's theorem. Introducing again the simplified notation
\begin{align}
	g \rho^{b_i}(\textbf{k}_i-\textbf{q}_i) \equiv \Omega_i\ ,
\end{align}
we can write the $2n$-point function as
\begin{align}
\label{omega_corr}
	g^{2n} \Big \langle \rho^{b_1}(\textbf{k}_1-\textbf{q}_1)\rho^{b_2 \dagger}(\textbf{k}_1-\textbf{q}_2) \cdots \rho^{b_{2n-1}}(\textbf{k}_n-\textbf{q}_{2n-1})\rho^{b_{2n} \dagger}(\textbf{k}_n-\textbf{q}_{2n}) \Big \rangle_{p}=
	\Big \langle \Omega_1 \Omega_2 \cdots \Omega_{2n-1} \Omega_{2n} \Big \rangle_{p}.
\end{align}
Here, as in \cref{lambdacorr}, even indices correspond to complex conjugates.

This correlator, \cref{omega_corr}, has the following Wick expansion:
\begin{align}\label{corrho}
	\Big \langle \Omega_1 \Omega_2 \cdots \Omega_{2n-1} \Omega_{2n} \Big \rangle_{p}=\sum_{\omega \in \Pi(\chi)} \prod_{\{i,j\}\in \omega} \Big \langle \Omega_i \Omega_j \Big \rangle_{p}.
\end{align}

In the generalised MV model this 2-point function can be written as
\begin{align}\label{rho}
	\Big \langle \Omega_i \Omega_j \Big \rangle_{p}=\frac{\delta^{b_i b_j}}{N_c^2-1} \mu^2\left[\textbf{k}_i-\textbf{q}_i,(-1)^{i+j}(\textbf{k}_j-\textbf{q}_j)\right],
\end{align}
where $\mu^2(\textbf{k},\textbf{q})$ is a function peaked around $\textbf{k}+\textbf{q}=0$. In the strict MV model we have that $\mu^2(\textbf{k},\textbf{q})\propto  \delta^{(2)}(\textbf{k}+\textbf{q})$.

All in all, using the area enhancement argument for computing the target correlator and the MV model for computing the projectile one, we arrive at the following general result for the multiplicity of $n$-gluon production:
\begin{align}\label{multiplicity3}
	2^n (2 \pi)^{3n} \frac{d^n N}{\prod_{i=1}^{n}d^2 \textbf{k}_i}=&\int \left(\prod_{i=1}^{2n} \frac{d^2 \textbf{q}_i}{(2 \pi)^2}\right)
	\left(\sum_{\sigma \in \Pi(\chi)} \prod_{\{i,j\}\in \sigma}  \Big \langle \Omega_i \Omega_j \Big \rangle_{p}\right)
	\left(\sum_{\omega \in \Pi(\chi)} \prod_{\{\alpha,\beta\}\in \omega}  \Big \langle \Lambda_\alpha \Lambda_\beta \Big \rangle_{T}\right),
\end{align}
that, together with \cref{target,rho}, provides the full expression that will be used along this work.

\subsection{Wick diagrams}\label{sec2c}

Since the expression of \cref{multiplicity3} involves the product of two Wick expansions it includes the sum of $(2n-1)!!^2$ products of $2n$ 2-point functions. Thus, when $n>2$ we will have to deal with a large number of terms and, for this reason, it is convenient to introduce a shorthand notation for each of these objects involved in the sum. 
Therefore we introduce in this work a diagrammatic notation for each term inside the sum of \cref{multiplicity3} analogous to the diagrams introduced in \cite{Gelis:2009wh} within the glasma graph approach. In our case, the diagrams consist of two parts that are separated by a vertical dashed line. In both parts we draw 2 rows and $n$ columns of dots where the dots of the upper row are labelled by odd numbers and the ones of the lower row are labelled by even numbers, and the labels are the same in both sides:
\begin{equation}
	\input{intro_1.tikz}.
\end{equation}
Each column of both parts of the diagram corresponds to a produced gluon. The columns defined by (1,2) corresponds to gluon 1, the ones defined by (3,4) to gluon 2 and so on. The upper row (odd indices) will represent the real space and the lower row (even indices) will represent the conjugate space. As we have said, each term of the sum of \cref{multiplicity3} will have a product of $n$ 2-point functions coming from the projectile average and $n$ 2-point functions coming from the target average that are labelled by 2 indices that goes from 1 to $2n$. We will draw these 2-point functions as lines that connect the dots in the diagram. We choose the left part of the diagram to represent the 2-point correlators of the projectile and the right part to represent the 2-point correlators of the target and schematically what we will draw is the following\footnote{This diagrammatic approach is also very similar to the notation used in \cite{Kovner:2018vec} where they wrote the terms of the Wick expansion of the target as $[i_1,i_2][i_3,i_4]\cdots[i_{2n-1},i_{2n}]$, being the indices inside the brackets the ones that define the 2-point functions in the expansion. For the projectile they used the same notation changing the brackets by curly brackets.}:
\begin{align}
	\Big \langle \Omega_i \Omega_j \Big \rangle_{p}
	\Big \langle \Lambda_\alpha \Lambda_\beta \Big \rangle_{T} 
	=\input{definicion.tikz}.
\end{align}

As an illustrative example let us select one of the $5!!^2=225$ terms that appear in \cref{multiplicity3} when $n=3$:
\begin{align}\label{example}
	\int \left(\prod_{i=1}^{6} \frac{d^2 \textbf{q}_i}{(2 \pi)^2}\right)
	\Big \langle \Omega_1 \Omega_6 \Big \rangle_{p} \Big \langle \Omega_3 \Omega_4 \Big \rangle_{p} \Big \langle \Omega_2 \Omega_5 \Big \rangle_{p}
	\Big \langle \Lambda_1 \Lambda_5 \Big \rangle_{T} \Big \langle \Lambda_3 \Lambda_4 \Big \rangle_{T} \Big \langle \Lambda_2 \Lambda_6 \Big \rangle_{T}.
\end{align}
This term will be represented by the following diagram:
\begin{align}\label{ejemplo}
	\input{ejemplo1_n3.tikz}.
\end{align}
Note that the integration over the $\textbf{q}'s$ is implicit.

If we want to write the diagram in an equation form we just have to use \cref{target,rho}. For example, the diagram in \cref{ejemplo} reads (remember that we are labeling $\textbf{k}_i$ with odd indices, and \cref{eqk,eql,eqa})
\begin{align}\label{example2}
	\input{ejemplo1_n3.tikz}&=\int \left(\prod_{i=1}^{6} \frac{d^2 \textbf{q}_i}{(2 \pi)^2}\right) \frac{1}{(N_c^2-1)^6} \delta^{b_1 b_6} \delta^{b_2 b_5} \delta^{b_3 b_4} \mu^2[\textbf{k}_1-\textbf{q}_1,-(\textbf{k}_5-\textbf{q}_6)] \mu^2[\textbf{k}_1-\textbf{q}_2,-(\textbf{k}_5-\textbf{q}_5)] \nonumber
	\\
	&\times
	\mu^2[\textbf{k}_3-\textbf{q}_3,-(\textbf{k}_3-\textbf{q}_4)] \delta^{a_1 a_5}\delta^{b_1 b_5}\delta^{a_1 a_5}\delta^{b_2 b_6}\delta^{a_3 a_3}\delta^{b_3 b_4} 4^3 L^{\lambda_1}(\textbf{k}_1,\textbf{q}_1)  L^{\lambda_1}(\textbf{k}_1,\textbf{q}_2) L^{\lambda_5}(\textbf{k}_5,\textbf{q}_6) 
	\nonumber \\ &\times
	L^{\lambda_5}(\textbf{k}_5,\textbf{q}_5)L^{\lambda_3}(\textbf{k}_3,\textbf{q}_3) L^{\lambda_3}(\textbf{k}_3,\textbf{q}_4)
	(2 \pi)^6 \delta^{(2)}[\textbf{q}_1+\textbf{q}_5] \delta^{(2)}[\textbf{q}_2+\textbf{q}_6] \delta^{(2)}[\textbf{q}_3-\textbf{q}_4]
	d(\textbf{q}_1) d(\textbf{q}_2) d(\textbf{q}_3)
	\nonumber \\ &=4^3
	\frac{1}{(N_c^2-1)^2}\int_{\textbf{q}_1, \textbf{q}_2, \textbf{q}_3} d(\textbf{q}_1) d(\textbf{q}_2) d(\textbf{q}_3) L^{i}(\textbf{k}_1,\textbf{q}_1)  L^{i}(\textbf{k}_1,\textbf{q}_2) L^{j}(\textbf{k}_3,\textbf{q}_3)  L^{j}(\textbf{k}_3,\textbf{q}_3)
	L^{k}(\textbf{k}_5,-\textbf{q}_1) 
	\nonumber \\ &\times
	L^{k}(\textbf{k}_5,-\textbf{q}_2)
	\mu^2[\textbf{k}_1-\textbf{q}_1,-(\textbf{k}_5+\textbf{q}_2)] \mu^2[\textbf{k}_1-\textbf{q}_2,-(\textbf{k}_5+\textbf{q}_1)] \mu^2[\textbf{k}_3-\textbf{q}_3,-(\textbf{k}_3-\textbf{q}_3)],
\end{align}
with $\int_\textbf{q}\equiv\int d^2\textbf{q}/(2\pi)^2$. 

Besides making the notation more compact we can also exploit the structure of the diagrams in order to find symmetries between them,  the associated power in $(N_c^2-1)$ for each diagram and which kind of quantum correlations (Bose enhancement or HBT) it includes, by making use of the following properties:
\begin{enumerate}[label=\roman*)]
	
	\item\label{prop1} \textit{Interchanging two dots within a column, $2 m \leftrightarrow 2m-1$, of a given diagram is equivalent to make the change of variables $\textbf{k}_{2m-1} \rightarrow - \textbf{k}_{2m-1}$}. 
	
	For example,
	\begin{align}
		\input{ejemplo2_n3.tikz}(\textbf{k}_5 \rightarrow-\textbf{k}_5)=\input{ejemplo1_n3.tikz}.
	\end{align}
	
	In order to prove this property it is enough to evaluate $ \Big \langle \Omega_{2m} \Omega_i \Big \rangle_{p} \Big \langle \Omega_{2m-1} \Omega_j \Big \rangle_{p}
	\Big \langle \Lambda_{2m} \Lambda_\gamma \Big \rangle_{T} \Big \langle \Lambda_{2m-1} \Lambda_\beta \Big \rangle_{T} $, being $i,j,\gamma$ and $\beta$ arbitrary indices, since it is the only piece of \cref{multiplicity3} that depends on the dots $2m$ and $2m-1$. This expression can be computed using \cref{target,rho}. Then, if one makes the change of variables with unit Jacobian $\textbf{q}_{2m} \rightarrow - \textbf{q}_{2m-1}$ and $\textbf{q}_{2m-1} \rightarrow - \textbf{q}_{2m}$ and uses the fact that $L^\lambda(\textbf{k},-\textbf{q})=-L^\lambda(-\textbf{k},\textbf{q})$ and $\mu^2(-\textbf{k},-\textbf{q})=\mu^2(\textbf{k},\textbf{q})$, we see that
	\begin{align}
		&\Big \langle \Omega_{2m} \Omega_i \Big \rangle_{p} \Big \langle \Omega_{2m-1} \Omega_j \Big \rangle_{p}
		\Big \langle \Lambda_{2m} \Lambda_\gamma \Big \rangle_{T} \Big \langle \Lambda_{2m-1} \Lambda_\beta \Big \rangle_{T}
		\nonumber \\
		=
		&\Big \langle \Omega_{2m-1} \Omega_i \Big \rangle_{p} \Big \langle \Omega_{2m} \Omega_j \Big \rangle_{p}
		\Big \langle \Lambda_{2m-1} \Lambda_\gamma \Big \rangle_{T} \Big \langle \Lambda_{2m} \Lambda_\beta \Big \rangle_{T}
		(\textbf{k}_{2m-1} \rightarrow - \textbf{k}_{2m-1}).
	\end{align}
	
	\item\label{prop2} \textit{Interchanging two columns $(2 m,2 m-1)$ and $(2 k, 2 k-1)$ of a given diagram is equivalent to make the change of variables $\textbf{k}_{2m-1} \leftrightarrow \textbf{k}_{2k-1}$}. 
	
	For example,
	\begin{align}
		\input{ejemplo3_n3.tikz}(\textbf{k}_3 \leftrightarrow \textbf{k}_5)=\input{ejemplo1_n3.tikz}.
	\end{align}
	
	The proof of this property is trivial since, by definition, each column of dots corresponds to a momentum $ \textbf{k}_i $ and, therefore, interchanging two columns in both sides is equivalent to interchange the label of two momenta.
	
	\item\label{prop3} We can extract the powers of $(N_c^2-1)$ by looking at the structure of each side of the diagram. Before making a statement of the property we will start by using \cref{ejemplo} as an illustrative example. Using \cref{target,rho} we can extract the counting in powers of $(N_c^2-1)^{-1}$ of this diagram by writing the Kronecker deltas
	\begin{align}
		\input{ejemplo1_n3.tikz} \propto \frac{1}{(N_c^2-1)^6} 
		&\delta^{a_1 a_5}\delta^{a_3 a_4}\delta^{a_2 a_6} \times \delta^{a_1 a_2}\delta^{a_3 a_4}\delta^{a_5 a_6}
		\nonumber \\
		\times 
		&\delta^{b_1 b_5}\delta^{b_3 b_4}\delta^{b_2 b_6} \times \delta^{b_1 b_6}\delta^{b_2 b_5}\delta^{b_3 b_4},
	\end{align}
	where the second group of deltas of the first line is introduced to preserve \cref{eqa}, that is, that the color of the produced gluons is the same in the real and the conjugate spaces. The first group of deltas of both lines accounts to the target configuration (right side of the diagram) and the last group of deltas accounts to the projectile configuration. If we organize this equation in such a way that all the indices in the deltas are closed we have that
	\begin{align}\label{ejemploNc}
		\input{ejemplo1_n3.tikz} \propto \frac{1}{(N_c^2-1)^6} 
		&\left(\textcolor{coloreqn}{\delta^{a_5 a_1}} \delta^{a_1 a_2} \textcolor{coloreqn}{\delta^{a_2 a_6}} \delta^{a_6 a_5}\right) \times \left(\textcolor{coloreqn}{\delta^{a_4 a_3}}  \delta^{a_3 a_4}\right)
		\nonumber \\
		\times 
		&\left(\textcolor{coloreqn}{\delta^{b_5 b_1}} \delta^{b_1 b_6} \textcolor{coloreqn}{\delta^{b_6 b_2}} \delta^{b_2 b_5}\right) \times \left(\textcolor{coloreqn}{\delta^{b_4 b_3}} \delta^{b_3 b_4}\right)
		\nonumber \\
		&=\frac{(N_c^2-1)^2(N_c^2-1)^2}{(N_c^2-1)^6}=(N_c^2-1)^{-2},
	\end{align}
	where we have written the deltas that come from the target side of the diagram in a different color by convenience. We can do the same procedure that we did in the last equation in a diagrammatic and faster way by just drawing the target (right) side of the diagram on top of the left side and counting the number of closed lines that we obtain (which is equivalent to the second line of \cref{ejemploNc}) and drawing vertical lines in the right side of the diagram and counting the number of closed lines that we obtain (which is equivalent to the first line of \cref{ejemploNc}),
	\begin{align}
		\input{ejemplo1_n3.tikz} \longrightarrow \input{ejemplonc.tikz} \longrightarrow \frac{(N_c^2-1)^2(N_c^2-1)^2}{(N_c^2-1)^6}=(N_c^2-1)^{-2},
	\end{align}
	where we can identify the red lines in the second diagram as the red Kronecker deltas of \cref{ejemploNc}. In general, \textit{if we call $n_p$ the number of closed lines that we obtain by projecting the right side of the diagram on top of the left side and $n_T$ the number of closed lines that we obtain by projecting vertical lines on top of the right side of the diagrams the counting in powers of $(N_c^2-1)$ of a given diagram for general $n$ is
		\begin{align}
			\frac{(N_c^2-1)^{n_p}(N_c^2-1)^{n_T}}{(N_c^2-1)^{2n}}=(N_c^2-1)^{n_p+n_T-2n}.
	\end{align}}
	
	As we will see through this work, this property is useful for organising the terms of \cref{multiplicity3} in powers of $(N_c^2-1)^{-1}$ in a systematic way, especially when $n$ is large.
	
	\item\label{prop4} The types of quantum correlation that we have in a given diagram can be obtained as follows. \textit{If the same two dots are linked in both sides of the diagram we have two possibilities: if the dots belong to the same column labelled by $(2k,2k-1)$ it means that the gluon $k$ is uncorrelated (disconnected piece) and if the dots belong to different columns it means that the gluons that define these columns have an HBT correlation. By exclusion, all the gluons involved in other kind of links have a Bose enhancement correlation either in the projectile or in the target wave function}.
	
	For example,
	\begin{align}
		\input{correlation.tikz}
	\end{align}
	in this diagram the 3$^{rd}$ produced gluon is uncorrelated, gluons 1-2 and 2-4 have an HBT correlation and gluons 1-4-5 have a Bose enhancement correlation.
	
	In order to check this property it is enough to evaluate the terms in \cref{multiplicity3} that contain the same links in both the projectile and target sides. That is, we are interested in terms that contain
	\begin{align}
		\Big \langle \Omega_{a} \Omega_b \Big \rangle_{p}
		\Big \langle \Lambda_{a} \Lambda_b \Big \rangle_{T},
	\end{align}
	where $a,b=1,...,2n$ are generic dots. The objects of \cref{target} and \cref{rho} that contain the information of the quantum interference correlations are the Dirac deltas and the functions $\mu^2(\textbf{k},\textbf{q})$ respectively (the Lipatov vertices and the dipole functions give a different kind of correlation). Thus, we can write
	\begin{align}
		\Big \langle \Omega_{a} \Omega_b \Big \rangle_{p}
		\Big \langle \Lambda_{a} \Lambda_b \Big \rangle_{T}
		&\propto
		\mu^2\left[ \textbf{k}_a-\textbf{q}_a,(-1)^{a+b} (\textbf{k}_b-\textbf{q}_b) \right] \delta^{(2)}\left[ \textbf{q}_a+(-1)^{a+b}\textbf{q}_b \right]
		\nonumber \\
		&= \mu^2\left[ \textbf{k}_a-\textbf{q}_a,(-1)^{a+b} \textbf{k}_b+\textbf{q}_a \right] \delta^{(2)}\left[ \textbf{q}_a+(-1)^{a+b}\textbf{q}_b \right].
	\end{align}
	Since $\mu^2(\textbf{k},\textbf{q})$ is peaked around $\textbf{k}=-\textbf{q}$ this implies that we have a peak around $\textbf{k}_a=-(-1)^{a+b} \textbf{k}_b$ which is an HBT correlation. In the case in which $a$ and $b$ belong to the same column, that is, $a=2k-1$ and $b=2k$ (or vice-versa), it is clear that we loose the correlation in function $\mu^2$ -- in fact we loose any kind of correlation since in this case the Lipatov vertices and the dipole function can be factorized. 
\end{enumerate}

\section{Results}\label{sec3}

In this section we  present the calculation of \cref{multiplicity3} for $n=2$, 3 and 4. Larger values of $n$ can be also considered in the same fashion, contingent upon sufficient computation power. In order to compute \cref{multiplicity3} we need \cref{target,rho} which contain two functions that need to be modelled, $\mu^2(\textbf{k},\textbf{q})$ and $d(\textbf{q})$. 

As indicated before, in the strict MV model $\mu^2(\textbf{k},\textbf{q})$ is proportional to a Dirac delta. However, in order to be more realistic, we choose a smoother function that is also peaked around $\textbf{k}+\textbf{q}=0$, such as a Gaussian\footnote{In order to preserve dimensions, this function should be multiplied by $2 \pi B_p$. However, since the multiplicity that we are evaluating will be normalised in such a way that the integrated multiplicity for single inclusive gluon production is dimensionless we do not need to care about this factor. In fact, any constant factor in the multiplicity is irrelevant for studying correlations because of the normalisation of the cumulants.}:
\begin{align}\label{mu2}
	\mu^2(\textbf{k},\textbf{q})=e^{-\frac{(\textbf{k}+\textbf{q})^2}{4 B_p^{-1}}},
\end{align}
where $B_p$ is the gluonic transverse area of the projectile.

For the dipole  we use the Fourier transform of the  GBW saturation model~\cite{GolecBiernat:1998js,GolecBiernat:1999qd}:
\begin{align}\label{dipole}
	d(\textbf{q})=\frac{4 \pi}{Q_s^2} e^{-\frac{\textbf{q}^2}{Q_s^2}}.
\end{align}

We should also account for the infrared divergences of the Lipatov vertices. The product of two Lipatov vertices is 
\begin{align}
	L^i(\textbf{k},\textbf{q}_1)L^i(\textbf{k},\textbf{q}_2)=\left[ \frac{\textbf{k}^i}{\textbf{k}^2}-\frac{(\textbf{k}-\textbf{q}_1)^i}{(\textbf{k}-\textbf{q}_1)^2}\right] \left[ \frac{\textbf{k}^i}{\textbf{k}^2}-\frac{(\textbf{k}-\textbf{q}_2)^i}{(\textbf{k}-\textbf{q}_2)^2}\right].
\end{align}
Usually these divergences are regulated by introducing an infrared cutoff both in all the integration over the momenta. However, in this work we use the following expression for the product of two Lipatov vertices:
\begin{align}\label{lipatov}
	L^i(\textbf{k},\textbf{q}_1)L^i(\textbf{k},\textbf{q}_2)=\frac{(2 \pi)^2}{\xi^2} \exp\left\{-\frac{[\textbf{k}-(\textbf{q}_1+\textbf{q}_2)/2]^2}{\xi^2}\right\},
\end{align}
where $\xi^2$ is a parameter with dimensions of momentum squared. This choice, although it does not maintain some important properties of the Lipatov vertices, it is much simpler to deal with and, as we show in \cref{app:wigner}, it is equivalent to using the Wigner function approach~\cite{Lappi:2015vha,Lappi:2015vta,Dusling:2017dqg,Dusling:2017aot,Davy:2018hsl} but including quantum correlations in the projectile wave function.
Thus, for two partons in the projectile the joint Wigner function that we use reads
\begin{align}\label{wigner2}
	W^{b_1 b_2 b_3 b_4}(\textbf{b}_1,\textbf{p}_1,\textbf{b}_2,\textbf{p}_2)&=\frac{1}{(N_c^2-1)^2} \frac{1}{\pi^4 \xi^4 B_p^2} e^{-(\textbf{p}_1^2+\textbf{p}_2^2)/\xi^2} e^{-(\textbf{b}_1^2+\textbf{b}_2^2)/B_p} \Big[ \delta^{b_1 b_2} \delta^{b_3 b_4}
	\nonumber \\
	&
	+\delta^{b_1 b_3} \delta^{b_2 b_4} 2 \pi B_p \delta^{(2)}(\textbf{b}_1-\textbf{b}_2) e^{-(\textbf{p}_1+\textbf{p}_2)^2/(2 B_p^{-1})}
	\nonumber \\
	&
	+\delta^{b_1 b_4} \delta^{b_2 b_3} 2 \pi B_p \delta^{(2)}(\textbf{b}_1-\textbf{b}_2) e^{-(\textbf{p}_1-\textbf{p}_2)^2/(2 B_p^{-1})}
	\Big],
\end{align}
where one sees the uncorrelated term as the first one in the sum on the right hand side and the four color indices correspond to the four $\rho$'s in the projectile average for the double inclusive gluon cross section.

%We note that the main problem of \cref{lipatov} is that it only depends on the momentum of the parent parton, $\textbf{k}_i-\textbf{q}_i$, and not on the final momentum, $\textbf{k}_i$. Therefore, \cref{lipatov} only includes the contribution in which the gluon is emitted from the source and then interacts with the target, thus missing part of the physics. The final momentum is acquired by the interaction with the target which is suitable for the projectile collinear limit in which the Wigner function approach is usually employed. Note also that, due to the assumed Gaussian forms, our final expressions cannot be considered reliable for transverse momenta sizeably larger than the saturation scale.

We note that the main problem of \cref{lipatov} is that it only depends on the momentum of the parent parton, $\textbf{k}_i-\textbf{q}_i$, and not on the final momentum, $\textbf{k}_i$. Therefore, \cref{lipatov} only includes the contribution in which the gluon is emitted from the source and then interacts with the target, thus missing part of the physics. The final momentum is acquired by the interaction with the target which is suitable for the projectile collinear limit. In principle, in this limit the so-called "hybrid factorization" is employed and it corresponds to forward production of partons near the proton fragmentation region~\cite{Dumitru:2005gt}. The approach that we adopt in this manuscript is suitable for central production even though the approximation used for the Lipatov vertices in \cref{lipatov} is more appropriate for considering the forward limit. Therefore, admittedly the validity of our approach is reduced to the forward region but not yet near the proton fragmentation one. In this region, the projectile partons are defined in terms of Wigner functions (see~\cite{Lappi:2015vha,Lappi:2015vta,Dusling:2017dqg,Dusling:2017aot,Davy:2018hsl}). However, we would like to emphasize that the Wigner functions adopted in these references are factorized for two partons and do not include quantum correlations in the projectile. The two parton joint Wigner function (given in \cref{wigner2}) that we use in our approach indeed encodes the correlations in the projectile which is one of the novelties of the present manuscript\footnote{Quantum correlations in the projectile have been taken into account in~\cite{Kovner:2017ssr,Kovner:2018vec} but not for more than two partons.}. Moreover, adopting \cref{lipatov} for the Lipatov vertices and \cref{wigner2} for the joint Wigner function to describe the projectile partons, allows us to perform the computation analytically until the very end, even though they restrict the validity region of our results. In our approach, one can generalise the computation to the production of any number of particles and can perform the study analytically within its limits of the validity. Other approaches that are strictly valid for central production, such as the study performed in~\cite{Altinoluk:2020psk} or the one in \cite{Dusling:2017aot}, rely on final numerical integrations which would be extremely difficult in the case of four particle correlations, or the computation is performed numerically from the very beginning making it difficult to control, respectively. Finally, due to the assumed Gaussian forms, our final expressions cannot be considered reliable for transverse momenta sizeably larger than the saturation scale.

\subsection{Double inclusive gluon production} \label{doubleinclusive}

The case $n=2$, that is, the spectrum for double inclusive gluon production is the most studied case. It was well described using the exact solution for the dipole correlators in the MV model~\cite{Lappi:2015vta,Davy:2018hsl}, in the glasma graph approximation~\cite{Dumitru:2008wn} and using the area enhancement argument~\cite{Altinoluk:2018hcu,Altinoluk:2018ogz}. The result that we present in this section is the same obtained in~\cite{Altinoluk:2018ogz} but now, with the help of \cref{mu2,dipole,lipatov}, we are able to obtain a closed-form solution for both the multiplicity and the azimuthal harmonics.

In this case, the expansion of \cref{multiplicity3} in terms of the Wick diagrams is
\begin{align}\label{n2}
	\frac{d^2 N}{d^2 \textbf{k}_1 d^2 \textbf{k}_3}=
	\input{n2_1.tikz}
	&+\left(\input{n2_3.tikz}+\input{n2_4.tikz}+\textbf{k}_3 \rightarrow -\textbf{k}_3 \right)
	\nonumber \\
	&+\left( \input{n2_2.tikz}+\input{n2_5.tikz}+\textbf{k}_3 \rightarrow -\textbf{k}_3 \right),
\end{align}
where we have grouped the 9 diagrams by their powers in $(N_c^2-1)^{-1}$. 

The Wick diagrams of \cref{n2} can be computed using \cref{target,rho,mu2,dipole,lipatov} in a straightforward way since all the arguments of the $\textbf{q}_i$ integrals are Gaussian functions and, therefore, they can be trivially solved. The result is
\begin{align}
	&\input{n2_1.tikz}=\frac{1}{\pi^2(\xi^2+Q_s^2)^2} \exp \left[ - \frac{\textbf{k}_1^2+\textbf{k}_3^2}{\xi^2+Q_s^2} \right],
	\\ &
	\input{n2_3.tikz}=\frac{1}{(N_c^2-1)}\frac{1}{ \pi^2(\xi^2+Q_s^2)[Q_s^2+\xi^2(1+B_p Q_s^2)]}
	\nonumber \\
	&\qquad \qquad \qquad \qquad \qquad \qquad \qquad \qquad \times
	\exp \left[ - \frac{2\xi^2 B_p (\textbf{k}_1+\textbf{k}_3)^2+[Q_s^2+\xi^2(1+B_pQ_s^2)](\textbf{k}_1^2+\textbf{k}_3^2)}{(\xi^2+Q_s^2)[Q_s^2+\xi^2(1+B_pQ_s^2)]} \right],
	\\ &
	\input{n2_4.tikz}=\frac{1}{(N_c^2-1)}\frac{1}{ \pi^2 \xi^2 (\xi^2+Q_s^2)} 
	\exp \left[ - \frac{2 \xi^2(\textbf{k}_1^2+\textbf{k}_3^2)+(B_p Q_s^2 \xi^2+B_p \xi^4+Q_s^2)(\textbf{k}_1+\textbf{k}_3)^2}{2 \xi^2(\xi^2+Q_s^2)} \right],
	\\ &
	\input{n2_2.tikz}=\frac{1}{(N_c^2-1)^2}\frac{1}{\xi^2 \pi^2 (\xi^2+Q_s^2)(1+B_p Q_s^2)} \exp \left[ - \frac{2 \xi^2 (\textbf{k}_1^2+\textbf{k}_3^2)+Q_s^2(\textbf{k}_1+\textbf{k}_3)^2}{2 \xi^2(\xi^2+Q_s^2)} \right],
	\\ &
	\input{n2_5.tikz}=\frac{1}{(N_c^2-1)^2}\frac{1}{ \pi^2 \xi^2[Q_s^2+\xi^2(1+B_p Q_s^2)]} 
	\nonumber \\
	&\qquad \qquad \qquad \qquad \qquad \qquad \times 
	\exp \left[ - \frac{B_p \xi^4 (\textbf{k}_1-\textbf{k}_3)^2+(Q_s^2+B_p Q_s^2 \xi^2)(\textbf{k}_1+\textbf{k}_3)^2+2\xi^2(\textbf{k}_1^2+\textbf{k}_3^2)}{2 \xi^2[Q_s^2+\xi^2(1+B_p Q_s^2)] } \right].
\end{align}

With these 5 equations we have fully determined the differential multiplicity in \cref{n2}. In order to obtain the value of the integrated spectrum we just have to perform again Gaussian integrations over $\textbf{k}_i$ obtaining
\begin{align}\label{N2}
	N=\kappa_0\{2\}=1+\frac{2}{N_c^2-1}\left[\frac{2}{1+B_p \xi^2} \right] +\frac{2}{(N_c^2-1)^2}\left[\frac{1}{1+B_p Q_s^2}+\frac{1}{1+B_p \xi^2} \right].
\end{align}
We can see from this equation that, apart from the suppression in powers of $(N_c^2-1)^{-1}$, the correlated terms contain suppression factors $(1+B_p Q_s^2)^{-1}$ and $(1+B_p \xi^2)^{-1}$. Following the domain picture that we have discussed in \cref{sec1b}, $B_p Q_s^2 \equiv n_D$ is the number of color domains in the overlap area of the projectile with the target in the transverse plane. We should expect decorrelation of the produced gluons in the limit of $n_D \rightarrow \infty$ since the probability of two gluons scattering off the same domain vanishes in this limit. Therefore, to fix $\xi^2$ it makes sense to choose a value that is proportional to $Q_s^2$ in order to preserve decorrelation in the limit $n_D \rightarrow \infty$. For this reason we will choose $\xi^2 = \alpha Q_s^2$, being $\alpha$ a real number, in the rest of this work\footnote{Since $\xi^2$ is a momentum scale of the projectile wave function it should be related with $B_p^{-1}$ and not with $Q_s^2$ which is a momentum scale of the target wave function. However, the choice $\xi^2=Q_s^2$ is the one that has given more consistent phenomenological results and for this reason we use it through all this work. In~\cite{Lappi:2015vta,Lappi:2015vha} and~\cite{Davy:2018hsl} the choices $\xi^2=B_p^{-1}$ and $\xi=Q_s/4$ have been made, respectively, and the sensitivity of the results to variations of these choices has been examined.}.

\begin{figure}[h!]
	\centering
	\includegraphics[scale=0.8]{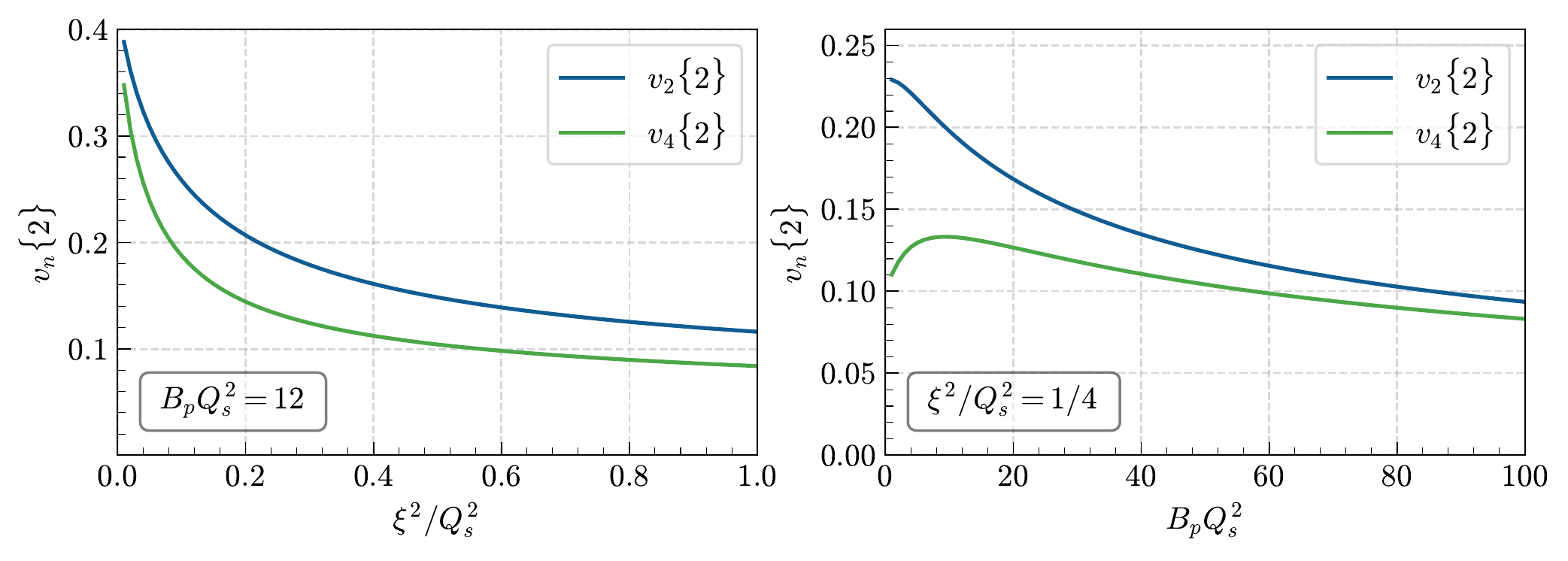}
	\caption{Dependence of the even 2-particle azimuthal harmonics, $v_{2n}\{2\}$, on $\alpha \equiv \xi^2/Q_s^2$ (left, for $B_p Q_s^2=12$) and $n_D \equiv B_p Q_s^2$ (right, for $\xi^2/Q_s^2=1/4$).}
	\label{vn2_qs2}
\end{figure}

The 2-particle azimuthal harmonics, \cref{cn2_def}, can be obtained by performing the integration over $\textbf{k}_i$ with the help of \cref{doublebesselintegr}. The result for the second order $\kappa$-function is
\begin{align}\label{kapan2}
	\kappa_{2n} \{ 2 \}=
	&\frac{8 \Gamma \left(n+1\right)^2}{ \Gamma (2n+1)} \Bigg\{\frac{\alpha (1+\alpha)}{N_c^2-1} \Bigg[
	\frac{(1+\alpha+\alpha n_D)}{\alpha^4 n_D^2} \left(\frac{\alpha^2 n_D}{2+\alpha^2 n_D+2\alpha(1+n_D)}\right)^{2(n+1)} 
	\nonumber \\ & \qquad \qquad \qquad \qquad \qquad \qquad \qquad \qquad \qquad \times
	\, _2F_1\left(n+1,n+1;2n+1;\left(\frac{\alpha^2 n_D}{2+\alpha^2 n_D+2\alpha(1+n_D)}\right)^2 \right)
	\nonumber  \\ & \qquad \qquad \qquad \qquad \qquad
	+\frac{1}{(1+\alpha^2 n_D+\alpha(2+n_D))^2} \left(\frac{1+\alpha n_D+\alpha^2 n_D}{1+\alpha^2 n_D+\alpha(2+n_D)}\right)^{2n} 
	\nonumber \\ & \qquad \qquad \qquad \qquad \qquad \qquad \qquad \qquad \qquad \times
	\, _2F_1\left(n+1,n+1;2n+1;\left(\frac{1+\alpha n_D+\alpha^2 n_D}{1+\alpha^2 n_D+\alpha(2+n_D)}\right)^2 \right)
	\Bigg]
	\nonumber \\  & \qquad \qquad \qquad
	+\frac{1}{(N_c^2-1)^2}\Bigg[ \frac{\alpha(1+\alpha+\alpha n_D)}{(1+\alpha^2 n_D+\alpha(2+n_D))^2} \left(\frac{1+\alpha n_D-\alpha^2 n_D}{1+\alpha^2 n_D+ \alpha(2+n_D)}\right)^{2 n} 
	\nonumber \\ & \qquad \qquad \qquad \qquad \qquad \qquad \qquad \qquad \qquad \times
	\, _2F_1\left(n+1,n+1;2n+1;\left( \frac{1+\alpha n_D-\alpha^2 n_D}{1+\alpha^2 n_D+ \alpha(2+n_D)} \right)^2 \right)
	\nonumber \\ & \qquad \qquad \qquad \qquad \qquad \qquad
	+\frac{\alpha(1+\alpha)}{1+n_D} (1+2\alpha)^{-2(n+1)} \, _2F_1\left(n+1,n+1;2n+1;\frac{1}{(1+2\alpha)^2}\right) \Bigg] \Bigg\},
\end{align}
where we have defined $\alpha=\xi^2/Q_s^2$, we have taken $n>0$ and due to the symmetry $\textbf{k}_3 \rightarrow -\textbf{k}_3$ of \cref{n2} all odd harmonics vanish. Using \cref{N2,kapan2} we can evaluate the 2-particle azimuthal harmonics as 
\begin{align}\label{vn2p}
	v_{2n} \{2\} = \sqrt{\frac{\kappa_{2n} \{2\}}{\kappa_{0} \{2\}}}.
\end{align}
In \cref{vn2_qs2} we plot the dependence of $v_{2n}\{2\}$ with respect to $n_D$ and $\alpha$ by fixing $N_c=3$. The value of the even azimuthal harmonics grows rapidly as both $n_D$ and $\alpha$ approach zero and it decreases slowly when these parameters are large. This decrease with $n_D$ is what we should expect in the color domain picture of particle correlation since as $n_D$ gets larger the probability of two gluons scattering in the same domain is smaller and thus the overall correlation. On the other hand, the decrease with $\alpha$ must be taken with care because $\alpha$ gives the ratio between the momentum transfers from projectile and target. The dilute-dense approximation that we are using makes sense only for $\alpha$ sizeably smaller than 1.

\begin{figure}[h!]
	\centering
	\includegraphics[scale=0.8]{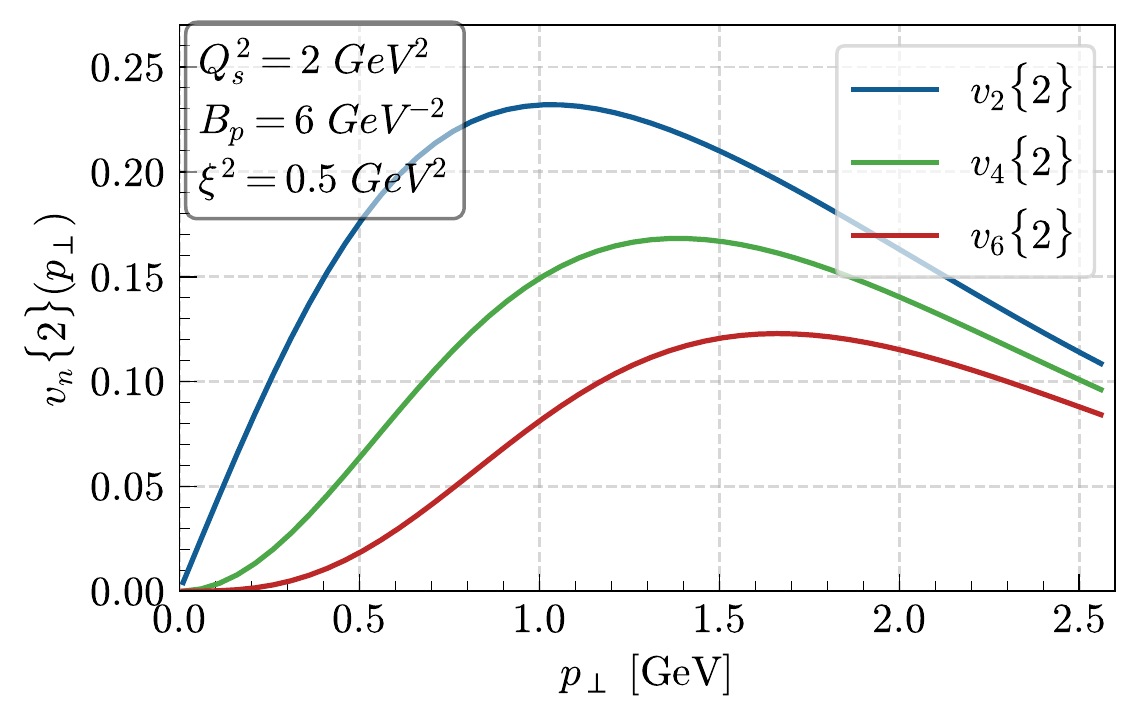}
	\caption{Dependence of the differential 2-particle even azimuthal harmonics, $v_{2n}\{2\}$, on transverse momentum $p_\perp$. In this graph we have used $B_p= 6$ GeV$^{-2}$, $\xi=Q_s/2$, $Q_s^2=2$ GeV$^2$ and $N_c=3$.}
	\label{vn2_pt}
\end{figure}

As we have seen in \cref{sec1b}, we can also compute the azimuthal harmonics as a function of transverse momentum by using the differential $\kappa$-function defined in \cref{kappadif}. The $\textbf{k}_i$ integral can be solved with the help of \cref{besselintegr} and the result is
\begin{align}
	\tilde{\kappa}_0\{2\}(p_\perp)=
	\frac{e^{-\frac{p_\perp^2}{\alpha Q_s^2}}}{Q_s^2}
	\Bigg\{ \frac{2 e^{\frac{p_\perp^2}{\alpha(1+\alpha) Q_s^2}}}{1+\alpha}  &+ \frac{8}{N_c^2-1} \left[\frac{
		e^{ \frac{(1+\alpha n_D-\alpha^2 n_D) p_\perp^2}{\alpha[1+\alpha^2 n_D+\alpha(2+n_D)]Q_s^2} }}{1+\alpha^2 n_D+\alpha(2+n_D)} 
	+\frac{e^{\frac{(2+2\alpha n_D-\alpha^2 n_D)p_\perp^2}{\alpha[2+\alpha^2 n_D+2 \alpha(1+n_D)] Q_s^2} }}{2+\alpha^2 n_D+2 \alpha(1+n_D)} 
	\right]    
	\nonumber \\
	&+
	\frac{8}{(N_c^2-1)^2} \left[  \frac{e^{ \frac{p_\perp^2}{\alpha(1+2\alpha)Q_s^2} }}{(1+2 \alpha)(1+n_D)}  
	+ \frac{
		e^{ \frac{(1+\alpha n_D-\alpha^2 n_D) p_\perp^2}{\alpha[1+\alpha^2 n_D+\alpha(2+n_D)]Q_s^2} }}{1+\alpha^2 n_D+\alpha(2+n_D)}   \right]
	\Bigg\},
\end{align}
and
\begin{align}
	\tilde{\kappa}_{2n}\{2\}(p_\perp)&=
	\frac{8  \Gamma(1+n)}{Q_s^2\Gamma(1+2n)} \left(\frac{p_\perp^2}{2 Q_s^2}\right)^{n} e^{-\frac{p_\perp^2}{\alpha Q_s^2}} \Bigg\{ \frac{1}{N_c^2-1} \Bigg[   
	\frac{e^{-\frac{(-1+\alpha n_D +\alpha^2 n_D)p_\perp^2}{2 \alpha(1+\alpha) Q_s^2}}}{1+\alpha^2 n_D+\alpha(2+n_D)} \left(\frac{ (1+\alpha n_D+\alpha^2 n_D)^2}{\alpha(1+\alpha)(1+\alpha^2 n_D+\alpha(2+n_D))}\right)^{n} 
	\nonumber \\  & \qquad  \qquad \qquad \qquad \qquad \qquad \qquad \qquad  \times
	\, _1F_1\left(n+1;2n+1;\frac{ (1+\alpha n_D+\alpha^2 n_D)^2 p_\perp^2}{2 \alpha(1+\alpha)[1+\alpha^2 n_D+\alpha(2+n_D)] Q_s^2}\right)
	\nonumber \\
	&\qquad \qquad \qquad \qquad
	+\frac{e^{-\frac{ [-2+\alpha^3 n_D-2\alpha(1+n_D)]p_\perp^2}{2 \alpha (1+\alpha)(1+\alpha+\alpha n_D) Q_s^2}}}{2+\alpha^2 n_D+2\alpha(1+n_D)}  \left(\frac{\alpha^4 n_D^2}{(1+\alpha)(1+\alpha+\alpha n_D)[2+\alpha^2 n_D+2 \alpha (1+n_D)]}\right)^{n} 
	\nonumber \\ & \qquad \qquad \qquad \qquad \qquad \times
	\, _1F_1\left(n+1;2n+1;\frac{\alpha^4 n_D^2 p_\perp^2}{2(1+\alpha)(1+\alpha+\alpha n_D)[2+\alpha^2 n_D+2 \alpha (1+n_D)]Q_s^2}\right)
	\Bigg] \nonumber \\ &+
	\frac{1}{(N_c^2-1)^2} \Bigg[ 
	\frac{e^{-\frac{(-1-\alpha n_D+\alpha^2 n_D) p_\perp^2}{2 \alpha (1+\alpha+\alpha n_D) Q_s^2}}}{1+\alpha^2 n_D+\alpha(2+n_D)}  
	\left(\frac{(1+\alpha n_D-\alpha^2 n_D)^2}{\alpha (1+\alpha+\alpha n_D)(1+\alpha^2 n_D+\alpha(2+n_D))}\right)^{n} \, 
	\nonumber \\ & \qquad \qquad \qquad \qquad \qquad \times
	\, _1F_1\left(n+1;2n+1;\frac{(1+\alpha n_D-\alpha^2 n_D)^2 p_\perp^2}{2\alpha (1+\alpha+\alpha n_D)(1+\alpha^2 n_D+\alpha(2+n_D))Q_s^2}\right)
	\nonumber \\ 
	&\qquad \qquad   +
	\frac{e^{-\frac{p_\perp^2}{2 \alpha (1+\alpha) Q_s^2}}}{(1+2\alpha)(1+n_D)}   
	\left( \frac{1}{\alpha (1+\alpha)(1+2\alpha)} \right)^n
	\, _1F_1\left(n+1;2n+1;\frac{p_\perp^2}{2\alpha (1+\alpha)(1+2\alpha)Q_s^2}\right)
	\Bigg] \Bigg\},
\end{align}
where $n>0$. The fact that $\tilde{\kappa}_{2n}(p_\perp)$ is proportional to $(p_\perp^2/Q_s^2)^n$ was also obtained in~\cite{Davy:2018hsl} although there a different model for the target average was employed. The differential 2-particle even azimuthal harmonics can be obtained by evaluating
\begin{align}
	v_{2n}\{2\}(p_\perp)=\sqrt{\frac{\tilde{\kappa}_{2n}\{2\}(p_\perp)}{\tilde{\kappa}_0\{2\}(p_\perp)}}
\end{align}
and the result is plotted in \cref{vn2_pt} for $n=1,2,3$ and $B_p= 6$ GeV$^{-2}$, $\xi=Q_s/2$, $Q_s^2=2$ GeV$^2$ and $N_c=3$. Although we do not aim for a comparison with experimental data, the obtained values are in the ballpark of them. Note that due to the Gaussian forms that we employ, our results cannot be considered reliable for $p_\perp$  sizeably larger than $Q_s$.

\subsection{Triple inclusive gluon production}\label{sec:triple}

In this section we show the result for \cref{multiplicity3} when $n=3$, that is, the triple inclusive gluon spectrum. Since in this work we are mainly interested in computing azimuthal harmonics we will just show the expansion of the spectrum in terms of the Wick diagrams. However, it has be shown recently \cite{Altinoluk:2020psk} that this result is useful for studying the correlation between the 2-particle azimuthal harmonics and multiplicity and average transverse momentum.

As we did for $n=2$, we can group the Wick diagrams in the expression for the $n=3$ gluon spectrum in powers of $(N_c^2-1)^{-1}$ as
\begin{align}\label{n3}
	8(2 \pi)^{9} \frac{d^3 N}{d^2 \textbf{k}_1 d^2 \textbf{k}_3d^2 \textbf{k}_5}=N_3^{(0)}+N_3^{(1)}+N_3^{(2)}+N_3^{(3)}+N_3^{(4)}.
\end{align}

In order to obtain each one of these terms we have to use the property \ref{prop3} of \cref{sec2c}. In this case the suppression of each diagram is given by $(N_c^2-1)^{n_p+n_T-6}$ and we can have three kind of configurations on each side of the diagram
\begin{align}
	&\input{n3_ex1.tikz},
	\label{conf_n3_1}
	\\
	\Bigg( &\input{n3_ex2.tikz}+\textbf{k}_5\rightarrow -\textbf{k}_5 \Bigg)+\textbf{k}_1 \leftrightarrow \textbf{k}_3+\textbf{k}_1 \leftrightarrow \textbf{k}_5,
	\label{conf_n3_2}
	\\
	\Bigg( &\input{n3_ex3.tikz}+\textbf{k}_1\rightarrow -\textbf{k}_1
	+\textbf{k}_3\rightarrow -\textbf{k}_3
	+\textbf{k}_5\rightarrow -\textbf{k}_5 \Bigg)+\textbf{k}_1 \leftrightarrow \textbf{k}_5.
	\label{conf_n3_3}
\end{align}
It is easy to see that the only way of obtaining $n_T=3$ ,2 and 1 is having the first, second and third configuration on the right side of the diagram, \cref{conf_n3_1,conf_n3_2,conf_n3_3} respectively. On the other hand, the only way of obtaining $n_p=3$, 2 and 1 is  having the same configuration on the left side of the diagram as the one on the right side, a configuration on the left side that has one link equal to the configuration on the right side and the other 2 links different, and at configuration on the left side that has all the links different than the configuration on the right side, respectively. One can also check that for a given configuration on the right side of the diagram the number of possibilities for $n_p=3$ is 1, for $n_p=2$ is 6 and for $n_p=1$ is 8.

With this taken into account, let us show as an example how to find all the Wick diagrams suppressed by $(N_c^2-1)^{-2}$. In this case $n_p+n_T=4$. There are three possibilities:
\begin{enumerate}[label=\roman*)]
	\item $n_p=1$ and $n_T=3$.
	
	This implies that we have to have the configuration of \cref{conf_n3_1} on the right side and configurations on the left side that has all the links different than the one on the right side. As we have said, there are 8 possibilities for this case:
	\begin{align}
		\input{n3_ex1_1.tikz} \ , \ \input{n3_ex1_2.tikz} \ , \ \input{n3_ex1_3.tikz} \ , \ \input{n3_ex1_4.tikz}  \ ,
		\nonumber \\
		\input{n3_ex1_5.tikz} \ , \ \input{n3_ex1_6.tikz} \ , \ \input{n3_ex1_7.tikz} \ , \ \input{n3_ex1_8.tikz} \ .
		\nonumber		
	\end{align}
		
	\item $n_p=2$ and $n_T=2$.
	
	This implies that we have to have the configuration of \cref{conf_n3_2} on the right side of the diagram and configurations on the left side that have one link in common with the right side and the other ones different. There are $6\times 6$ possibilities in this case:
	\begin{align}
		\input{n3_ex2_1.tikz} \ , \ \input{n3_ex2_2.tikz} \ , \ \input{n3_ex2_3.tikz} \ , \ \input{n3_ex2_4.tikz}  \ ,
		\input{n3_ex2_5.tikz} \ , \ \input{n3_ex2_6.tikz}
		\nonumber		
	\end{align}
	and the $5$ permutations of \cref{conf_n3_2} for each diagram.
	
	\item $n_p=3$ and $n_T=1$.
	
	This implies that we have to have the configuration of \cref{conf_n3_3} on the right side of the diagram and configurations on the left side that have all the links in common with the right side. There are $1 \times 8$ possibilities in this case:
	\begin{align}
		\input{n3_35.tikz} 
		\nonumber
	\end{align}
	and the 7 permutations of \cref{conf_n3_3}.
\end{enumerate}

All in all, we can write all the 52 Wick diagrams that have a suppression of $(N_c^2-1)^{-2}$ as
\begin{flalign}\label{n32}
	N_3^{(2)}&=\Bigg\{  \Bigg[ \Bigg(  \input{n3_ex1_3.tikz}+ \input{n3_35.tikz} \Bigg) +\textbf{k}_1\rightarrow -\textbf{k}_1
	+\textbf{k}_3\rightarrow -\textbf{k}_3
	+\textbf{k}_5\rightarrow -\textbf{k}_5 \Bigg]+\textbf{k}_1 \leftrightarrow \textbf{k}_5  \Bigg\} &&
	\nonumber \\
	&+\Bigg\{  \Bigg[ \Bigg(  \input{n3_ex2_1.tikz} +\input{n3_ex2_2.tikz} + \input{n3_ex2_3.tikz} + \input{n3_ex2_4.tikz}  +
	\input{n3_ex2_5.tikz} + \input{n3_ex2_6.tikz} \Bigg) &&
	\nonumber \\
	&+\textbf{k}_5\rightarrow -\textbf{k}_5  \Bigg]
	 +\textbf{k}_1 \leftrightarrow \textbf{k}_3+\textbf{k}_1 \leftrightarrow \textbf{k}_5 \Bigg\}.&&
\end{flalign}

This procedure, although tedious, is straightforward to implement on a computer.
Repeating it we find that there is 1 diagram suppressed by $(N_c^2-1)^{0}$:
\begin{flalign}\label{n30}
	&N_3^{(0)}=\input{n3_11.tikz},&
\end{flalign}
12 diagrams suppressed by $(N_c^2-1)^{-1}$:
\begin{flalign}\label{n31}
	&N_3^{(1)}=\Bigg\{ \Bigg[ \Bigg( \input{n3_21.tikz}+\input{n3_22.tikz}\Bigg) +\textbf{k}_5\rightarrow -\textbf{k}_5 \Bigg]+\textbf{k}_1 \leftrightarrow \textbf{k}_3+\textbf{k}_1 \leftrightarrow \textbf{k}_5 \Bigg\},&
\end{flalign}
96 diagrams suppressed by $(N_c^2-1)^{-3}$:
\begin{flalign}\label{n33}
	N_3^{(3)}=\Bigg\{  \Bigg[ \Bigg(  &\input{n3_3_1.tikz} +\input{n3_3_2.tikz} + \input{n3_3_3.tikz} + \input{n3_3_4.tikz} +\input{n3_3_5.tikz} +\input{n3_3_6.tikz}&&
	\nonumber \\
	+ &\input{n3_3_7.tikz} + \input{n3_3_8.tikz} \Bigg)
	+\textbf{k}_5\rightarrow -\textbf{k}_5  \Bigg]
	+\textbf{k}_1 \leftrightarrow \textbf{k}_3+\textbf{k}_1 \leftrightarrow \textbf{k}_5 \Bigg\}&&
	\nonumber \\
	+\Bigg\{  \Bigg[ \Bigg(  &\input{n3_32_1.tikz} +\input{n3_32_2.tikz} + \input{n3_32_3.tikz} + \input{n3_32_4.tikz} +\input{n3_32_5.tikz} +\input{n3_32_6.tikz} \Bigg) &&
	\nonumber \\ &+\textbf{k}_1\rightarrow -\textbf{k}_1
	+\textbf{k}_3\rightarrow -\textbf{k}_3
	+\textbf{k}_5\rightarrow -\textbf{k}_5 \Bigg]+\textbf{k}_1 \leftrightarrow \textbf{k}_5  \Bigg\}, &&
\end{flalign}
and 64 diagrams suppressed by $(N_c^2-1)^{-4}$:
\begin{flalign}\label{n34}
	N_3^{(4)}=\Bigg\{  \Bigg[ \Bigg(  &\input{n3_4_1.tikz} +\input{n3_4_2.tikz} + \input{n3_4_3.tikz} + \input{n3_4_4.tikz} +\input{n3_4_5.tikz} +\input{n3_4_6.tikz} &&
	\nonumber \\ 
	+ &\input{n3_4_7.tikz}+\input{n3_4_8.tikz} \Bigg)+\textbf{k}_1\rightarrow -\textbf{k}_1
	+\textbf{k}_3\rightarrow -\textbf{k}_3
	+\textbf{k}_5\rightarrow -\textbf{k}_5 \Bigg]+\textbf{k}_1 \leftrightarrow \textbf{k}_5  \Bigg\}. &&
\end{flalign}

All in all, we get the $225=(5!!)^2$. \cref{n32,n30,n31,n33,n34} give the full Wick expansion of the triple inclusive gluon spectrum to all orders in $(N_c^2-1)^{-1}$. These equations were computed in~\cite{Altinoluk:2018ogz} up to order $(N_c^2-1)^{-2}$. In order to write \cref{n3} just as a function of $\textbf{k}_i$ one just has to employ \cref{target,rho,mu2,dipole,lipatov} and then perform the $\textbf{q}_i$ integrals. These integrals are straightforward if the arguments of the integrals are Gaussian functions, as we assumed before.

\subsection{Four gluon inclusive production}

In this section we evaluate \cref{multiplicity3} for $n=4$, that is, the four gluon inclusive spectrum. Since in this case the number of diagrams involved is very large ($[7!!]^2=11025$) and thus their writing is not viable, we will start by discussing the simpler case in which the partons in the wave function of the projectile are initially not correlated. Then we will consider to the more general case that is explained in detail in \cref{app:4gluon}.  

The case in which the partons in the projectile wave function are initially uncorrelated was discussed for scattering quarks~\cite{Dusling:2017aot}, and for gluons~\cite{Ozonder:2014sra} within the glasma graph approach. In our case, this implies writing \cref{multiplicity3} as
\begin{align}\label{multiplicity_uncorr}
	2^n(2 \pi)^{3n} \frac{d^n N}{\prod_{i=1}^{n}d^2 \textbf{k}_i}=&\int \left(\prod_{i=1}^{2n} \frac{d^2 \textbf{q}_i}{(2 \pi)^2}\right)
	\prod_{i=1}^{n}  \Big \langle \Omega_{2i-1} \Omega_{2i} \Big \rangle_{p}
	\left(\sum_{\sigma \in \Pi(\chi)} \prod_{\{\alpha,\beta\}\in \sigma}  \Big \langle \Lambda_\alpha \Lambda_\beta \Big \rangle_{T}\right),
\end{align}
and, instead of having $[(2n-1)!!]^2$ terms in the sum, we just have $(2n-1)!!$. Using \cref{multiplicity_uncorr} we can write the 4-gluon inclusive production as the sum of 105 diagrams (also known as rainbow diagrams~\cite{Gelis:2009wh}) in the following form:
\begin{align}\label{mul4}
	&16(2 \pi)^{12} \frac{d^4 N}{d^2 \textbf{k}_1 d^2 \textbf{k}_3 d^2 \textbf{k}_5 d^2\textbf{k}_7}=
	\input{n4_11.tikz}
	\nonumber \\ &+
	\left[\left( \input{n4_12.tikz}+\textbf{k}_7 \rightarrow -\textbf{k}_7   \right)+\textbf{k}_1 \leftrightarrow \textbf{k}_5+\textbf{k}_1 \leftrightarrow \textbf{k}_7+\textbf{k}_3 \leftrightarrow \textbf{k}_5 +\textbf{k}_3 \leftrightarrow \textbf{k}_7  +(\textbf{k}_3 \leftrightarrow \textbf{k}_5)(\textbf{k}_1 \leftrightarrow \textbf{k}_7)\right]
	\nonumber \\ &+
	\Bigg\{\left[\left( \input{n4_13.tikz}
	+\textbf{k}_3 \rightarrow -\textbf{k}_3
	+\textbf{k}_5 \rightarrow -\textbf{k}_5
	+\textbf{k}_7 \rightarrow -\textbf{k}_7  \right)
	+\textbf{k}_3 \leftrightarrow \textbf{k}_7
	\right]
	+\textbf{k}_1 \leftrightarrow \textbf{k}_3
	+\textbf{k}_1 \leftrightarrow \textbf{k}_5
	+\textbf{k}_1 \leftrightarrow \textbf{k}_7 \Bigg\}
	\nonumber \\ &+
	\left[\left( \input{n4_15.tikz}
	+\textbf{k}_1 \rightarrow -\textbf{k}_1
	+\textbf{k}_5 \rightarrow -\textbf{k}_5
	+(\textbf{k}_1 \rightarrow -\textbf{k}_1)(\textbf{k}_5 \rightarrow -\textbf{k}_5)
	\right)
	+\textbf{k}_3 \leftrightarrow \textbf{k}_5
	+\textbf{k}_3 \leftrightarrow \textbf{k}_7
	\right]
	\nonumber \\ &+
	\Bigg\{ \Bigg[ \input{n4_14.tikz}+\textbf{k}_1 \rightarrow -\textbf{k}_1+\textbf{k}_3 \rightarrow -\textbf{k}_3 +\textbf{k}_5 \rightarrow -\textbf{k}_5 +\textbf{k}_7 \rightarrow -\textbf{k}_7
	 \\ &
	+ \frac{1}{2} \Bigg(  
	(\textbf{k}_1 \rightarrow -\textbf{k}_1)(\textbf{k}_3 \rightarrow -\textbf{k}_3)
	+(\textbf{k}_1 \rightarrow -\textbf{k}_1)(\textbf{k}_5 \rightarrow -\textbf{k}_5)
	+(\textbf{k}_1 \rightarrow -\textbf{k}_1)(\textbf{k}_7 \rightarrow -\textbf{k}_7)
	+(\textbf{k}_3 \rightarrow -\textbf{k}_3)(\textbf{k}_5 \rightarrow -\textbf{k}_5)
	\nonumber \\ &
	+(\textbf{k}_3 \rightarrow -\textbf{k}_3)(\textbf{k}_7 \rightarrow -\textbf{k}_7)
	+(\textbf{k}_5 \rightarrow -\textbf{k}_5)(\textbf{k}_7 \rightarrow -\textbf{k}_7)
	\Bigg) \Bigg] 
	+\textbf{k}_1 \leftrightarrow \textbf{k}_3
	+\textbf{k}_1 \leftrightarrow \textbf{k}_7
	+\textbf{k}_3 \leftrightarrow \textbf{k}_5
	+\textbf{k}_3 \leftrightarrow \textbf{k}_7
	+\textbf{k}_5 \leftrightarrow \textbf{k}_7  \Bigg\}.\nonumber
\end{align}
In this expression, the term in the first line corresponds to the case in which all the generated gluons are uncorrelated . The 12 terms in the second line correspond to the case in which 2 gluons are uncorrelated and 2 gluons are correlated. The 32 terms in the third line correspond to the case in which 1 gluon is uncorrelated and the remaining 3 ones are correlated. The 12 terms of the fourth line correspond to the case in which two pair of gluons are correlated independently, i.e.,  factorisable connected diagrams. Finally, the 48 terms of the last lines (the factor 1/2 avoids double counting of the diagrams) correspond to the case in which all the gluons are correlated between them, i.e., fully connected diagrams. Note that the first, second, third and fourth, and fifth terms in the sum on the right hand side correspond to terms with increasing powers in $(N_c^2-1)^{-2}$.

The Wick diagrams of \cref{mul4} can be computed in the same fashion as in \cref{doubleinclusive}. However, since we are only interested in computing the 4-particle cumulants, \cref{cn4_def}, we will exploit the $\textbf{k}_i \leftrightarrow \textbf{k}_j$ and $\textbf{k}_i \rightarrow -\textbf{k}_i$ symmetries in order to simplify the calculation. When evaluating the 4-particle $\kappa$-function in \cref{kappa_def} all the terms that contain at least one disconnected piece, i.e., two vertical lines in both sides of the diagram, will vanish trivially due to rotational invariance. For this reason the diagrams of the first three lines of \cref{mul4} will not contribute to the 4-particle $\kappa$-function when $n>0$ and therefore we can write
\begin{align}\label{kn4ex}
	\kappa_{2n}\{4\}=\int d^2\textbf{k}_1 d^2\textbf{k}_3 d^2\textbf{k}_5 d^2\textbf{k}_7 e^{i2n(\phi_1+\phi_3-\phi_5-\phi_7)} 
	\Bigg[ &\left(\input{n4_15.tikz} + \text{perm}_4 \right)
	\nonumber \\
	+ &\left( \input{n4_14.tikz} + \text{perm}_5 \right)  \Bigg],
\end{align}
where we have written schematically $\text{perm}_4$, which encodes all the factorizable connected diagrams, and $\text{perm}_5$, which includes all the fully connected diagrams, as all the permutations of the fourth and fifth lines of \cref{mul4} respectively. Note that we have also dropped the factors 2 and $2\pi$ as they cancel in \cref{cn2_def,cn4_def,dn2_def,dn4_def}.

On the other hand, we can read from the permutations for the fully connected diagrams $\text{perm}_5$, that all the diagrams that are related by a change of variables $\textbf{k}_i \rightarrow -\textbf{k}_i$ will give the same result for the integral in \cref{kn4ex} since this transformation is equivalent to making $\phi_i \rightarrow \phi_i + \pi$ in the argument of the exponential and, thus, leaves the integral invariant. Furthermore, it is easy to check that all the diagrams of the last three lines of \cref{mul4} that are related by the change of variables $\textbf{k}_1 \leftrightarrow \textbf{k}_3$, $\textbf{k}_5 \leftrightarrow \textbf{k}_7$ and $\textbf{k}_3 \leftrightarrow \textbf{k}_7$ also give the same value for the integral. By inspection of the symmetries, which is detailed in \cref{app:4gluon}, we can write the 48 integrals defined by the permutations of the last three lines of \cref{mul4} as
\begin{align}
	\int d^2\textbf{k}_1 d^2\textbf{k}_3 d^2\textbf{k}_5 d^2\textbf{k}_7 e^{i2n(\phi_1+\phi_3-\phi_5-\phi_7)} &
	\left(\input{n4_14.tikz} + \text{perm}_5 \right)
	\nonumber \\ 
	=\int d^2\textbf{k}_1 d^2\textbf{k}_3 d^2\textbf{k}_5 d^2\textbf{k}_7 e^{i2n(\phi_1+\phi_3-\phi_5-\phi_7)} &
	\left( 32 \input{n4_14.tikz} + 16 \input{fig11.tikz} \right),
\end{align}
where the last diagram can be seen as the first one with the change of variables $\textbf{k}_1 \leftrightarrow \textbf{k}_7$.

Furthermore, 4 out of the 12 diagrams of the fourth term of \cref{mul4}
\begin{align}
	\input{n4_15.tikz} +\textbf{k}_1 \rightarrow -\textbf{k}_1
	+\textbf{k}_3 \rightarrow -\textbf{k}_3+(\textbf{k}_1 \rightarrow -\textbf{k}_1)(\textbf{k}_3 \rightarrow -\textbf{k}_3)
\end{align}
only depend on $\phi_1-\phi_3$ and $\phi_5-\phi_7$ and therefore vanish due to rotational invariance. Having this into account we can write \cref{kn4ex} as
\begin{align}\label{kn4_example}
	\kappa_{2n}\{4\}%=\int d^2\textbf{k}_1 d^2\textbf{k}_3 d^2\textbf{k}_5 d^2\textbf{k}_7 \frac{d^4 N}{d^2 \textbf{k}_1 d^2 \textbf{k}_3d^2 \textbf{k}_5 d^2\textbf{k}_7} e^{i2n(\phi_1+\phi_3-\phi_5-\phi_7)}&&
%	\nonumber \\
	=\int d^2\textbf{k}_1 d^2\textbf{k}_3 d^2\textbf{k}_5 d^2\textbf{k}_7 e^{i2n(\phi_1+\phi_3-\phi_5-\phi_7)} 
	\Bigg( 8 \input{kn4_1.tikz}  &+ 32 \input{n4_14.tikz} 
	\nonumber \\
	&+ 16 \input{fig11.tikz}   \Bigg).
\end{align}

On the other hand, the 2-particle $\kappa$-function in the case in which the partons are initially uncorrelated in the projectile wave function is
\begin{align}
	2 \kappa_{2n}\{2\}^2=2 \left(2 \int d^2\textbf{k}_1 d^2\textbf{k}_3 e^{i2n(\phi_1-\phi_3)}  \input{n2_2.tikz}  \right)^2=8 \int d^2\textbf{k}_1 d^2\textbf{k}_3 d^2\textbf{k}_5 d^2\textbf{k}_7 e^{i2n(\phi_1+\phi_3-\phi_5-\phi_7)} \input{kn4_1.tikz}. 
\end{align}

Therefore we can write \cref{kn4_example} as
\begin{align}\label{kn4ex2}
	\kappa_{2n}\{4\} = \int d^2\textbf{k}_1 d^2\textbf{k}_3 d^2\textbf{k}_5 d^2\textbf{k}_7 e^{i2n(\phi_1+\phi_3-\phi_5-\phi_7)} 
	\Bigg(  32 \input{n4_14.tikz} + 16 \input{fig11.tikz}   \Bigg)+2\kappa_n\{2\}^2.
\end{align}

In order to compute $\kappa_0 \{4\}$ we have to have into account all the diagrams of \cref{mul4}. However, since all the permutations are related by the change of variables $\textbf{k}_i \rightarrow -\textbf{k}_i$ or $\textbf{k}_i \leftrightarrow
\textbf{k}_j$ ($i \ne j$) that leave the integral invariant we can write
\begin{align}\label{kn0ex}
	\kappa_{0}\{4\} = \int d^2\textbf{k}_1 d^2\textbf{k}_3 d^2\textbf{k}_5 d^2\textbf{k}_7 \Bigg( \input{n4_11.tikz} &+12 \input{n4_12.tikz} +32 \input{n4_13.tikz} 
	\nonumber \\
	&+12 \input{n4_15.tikz} +48 \input{n4_14.tikz}  \Bigg).
\end{align}
This integral can be easily performed since all the terms are just Gaussian functions.

All in all, the 4-particle cumulant can be computed by using \cref{cn4_def} and \cref{kn4ex2,kn0ex}
\begin{align}\label{cn42}
	c_{2n}\{4\} &= \frac{1}{\kappa_0\{4\}}\int d^2\textbf{k}_1 d^2\textbf{k}_3 d^2\textbf{k}_5 d^2\textbf{k}_7 e^{i2n(\phi_1+\phi_3-\phi_5-\phi_7)} 
	\Bigg(  32 \input{n4_14.tikz} + 16 \input{fig11.tikz}   \Bigg)
	\nonumber \\
	&+2 \kappa_{2n}\{2\}^2 \left( \frac{1}{\kappa_0\{4\}}-\frac{1}{\kappa_0\{2\}^2} \right)
\end{align}
and the four particle even azimuthal harmonics is obtained as
%using \cref{vn4}.
\begin{align}\label{vn4uncorr}
	v_{2n}\{4\}=(-c_{2n}\{4\})^{1/4}.
\end{align}

\begin{figure}[h!]
	\centering
	\includegraphics[scale=0.8]{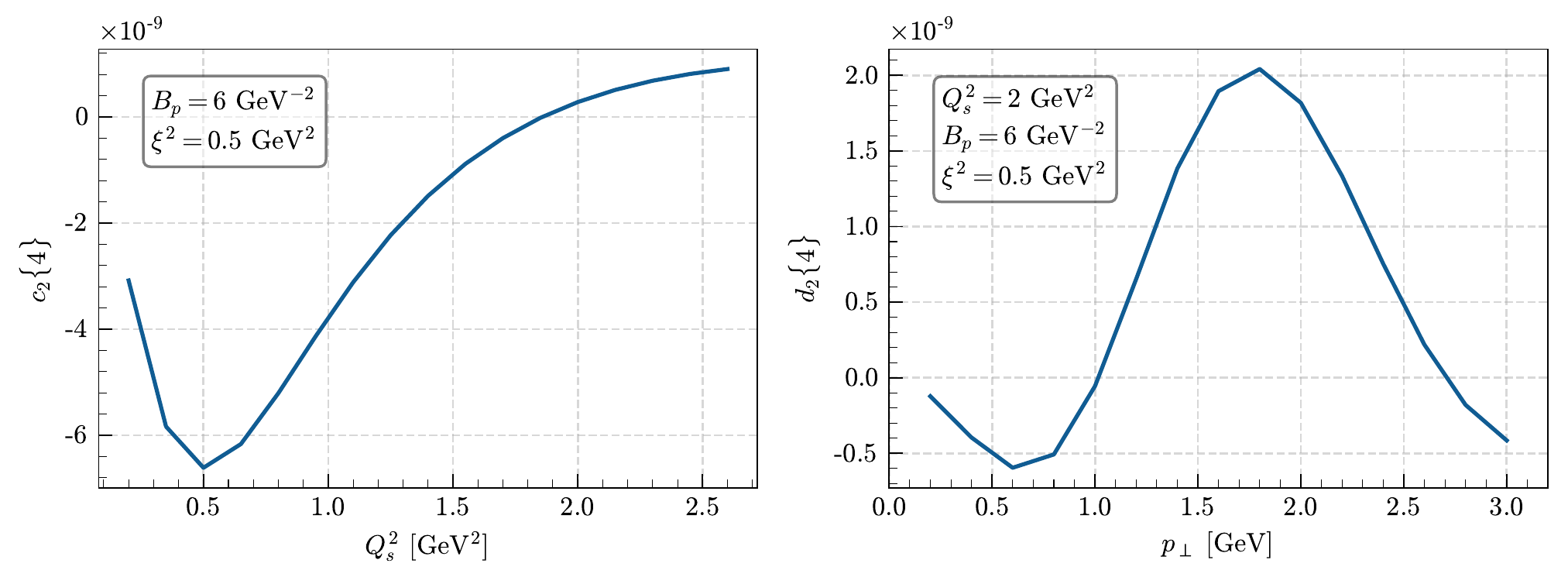}
	\caption{Dependence of the 4-particle integrated cumulants on $Q_s^2$ (left) and of the differential cumulants on $p_\perp$ (right) in the case in which the partons in the projectile wave function are uncorrelated. In these graphs we have used $N_c=3$ and the values of the remaining parameters are indicated on the plots.}
	\label{cn_qs2_uncorr}
\end{figure}

\begin{figure}[h!]
	\centering
	\includegraphics[scale=0.8]{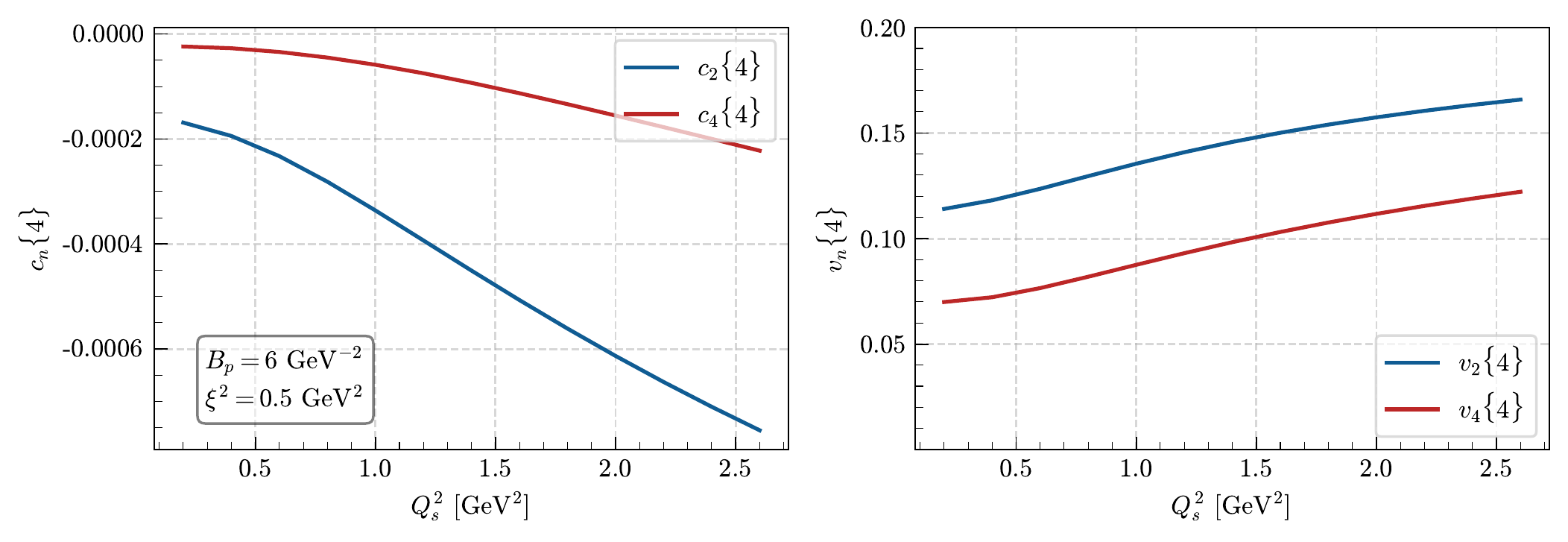}
	\caption{Dependence of the 4-particle cumulant (left) and azimuthal harmonic (right) of second and fourth order with $Q_s^2$. In these graphs we have used $N_c=3$ and the values of the remaining parameters are indicated on the plots.}
	\label{cn_vs_qs2}
\end{figure}

In \cref{cn_qs2_uncorr} left we have plotted our results for \cref{cn42} as a function of $Q_s^2$. The absolute values of the cumulant are very small and it even becomes positive with increasing $Q_s^2$. The reason why it is so comes from the fact that we are not having into account all the contributions that come from the correlation of the partons in the projectile ensemble. Below, see \cref{cn_vs_qs2}, these contributions are taken into account and the values are reasonable and in the ballpark of the ones in experimental data.

The differential 4-particle cumulant in \cref{dn4_def} can be computed in the same fashion but since we are fixing one of the momentums we have to be more careful with the symmetries discussed in the last paragraphs. In \cref{app:4gluon} we show that we can write (again dropping factors 2 and $2\pi$)
\begin{align}
	&\int_{0}^{2\pi} d\phi_1 \int d^2\textbf{k}_3 d^2\textbf{k}_5 d^2\textbf{k}_7 e^{i2n(\phi_1+\phi_3-\phi_5-\phi_7)} 
	\left(\input{n4_14.tikz} + \text{perm}_5 \right) \Bigg \vert_{|\textbf{k}_1|=p_\perp}
	\nonumber \\ 
	=&\int_{0}^{2\pi} d\phi_1 \int d^2\textbf{k}_3 d^2\textbf{k}_5 d^2\textbf{k}_7 e^{i2n(\phi_1+\phi_3-\phi_5-\phi_7)} 
	\left( 32 \input{n4_14.tikz} + 16 \input{fig11.tikz} \right) \Bigg \vert_{|\textbf{k}_1|=p_\perp}
\end{align}
and, therefore,
\begin{align}\label{kdn4ex2}
	\tilde{\kappa}_{2n}\{4\}(p_\perp) = \int_{0}^{2\pi} d\phi_1 \int d^2\textbf{k}_3 d^2\textbf{k}_5 d^2\textbf{k}_7 e^{i2n(\phi_1+\phi_3-\phi_5-\phi_7)} 
	&\Bigg(  32 \input{n4_14.tikz} + 16 \input{fig11.tikz}   \Bigg)\Bigg \vert_{|\textbf{k}_1|=p_\perp} 
	\nonumber \\
	&+2\tilde{\kappa}_n\{2\}(p_\perp)\kappa_n\{2\}.
\end{align}

In order to compute $\tilde{\kappa}_{0}\{4\}(p_\perp)$ we cannot use the same symmetries that we employed for computing $\kappa_{0}\{4\}$ because now one of the momenta is fixed. All the Wick diagrams of \cref{mul4} that are related by a change of variables $\textbf{k}_i \rightarrow - \textbf{k}_i$ still leave the integral invariant but now, since we are fixing $|\textbf{k}_1|=p_\perp$, all the diagrams that are related by a change of variable $\textbf{k}_1 \leftrightarrow \textbf{k}_j$ will give a different value for the integral but the ones that are related by $\textbf{k}_i \leftrightarrow \textbf{k}_j$, with $i,j \ne 1$, still leave the integral invariant. Performing a simple counting of the permutations of \cref{mul4} we can write
\begin{align}\label{kdif0ex}
	\tilde{\kappa}_{0}\{4\}(p_\perp) &= \int_{0}^{2\pi} d\phi_1 \int d^2\textbf{k}_3 d^2\textbf{k}_5 d^2\textbf{k}_7 
	\Bigg[ 
	\input{n4_11.tikz} \Bigg \vert_{|\textbf{k}_1|=p_\perp}
	\nonumber \\ & 
	+\left(6 \input{n4_12.tikz} + 6 (\textbf{k}_1 \leftrightarrow \textbf{k}_5) \right) \Bigg \vert_{|\textbf{k}_1|=p_\perp}
	+\left(8 \input{n4_13.tikz} + 24(\textbf{k}_1 \leftrightarrow \textbf{k}_3) \right) \Bigg \vert_{|\textbf{k}_1|=p_\perp}
	\nonumber \\ & 
	+12 \input{n4_15.tikz} \Bigg \vert_{|\textbf{k}_1|=p_\perp}
	+\left(32 \input{n4_14.tikz} + 16 (\textbf{k}_1 \leftrightarrow \textbf{k}_3) \right)  \Bigg \vert_{|\textbf{k}_1|=p_\perp}
	\Bigg].
\end{align}

Thus, using \cref{kdn4ex2,kdif0ex} the differential 4-particle cumulant can be written in a similar form as \cref{cn42}:
\begin{align}\label{dn42}
	d_{2n}\{4\}(p_\perp) &= \frac{1}{\tilde{\kappa}_0\{4\}(p_\perp)}\int_{0}^{2\pi} d\phi_1 \int d^2\textbf{k}_3 d^2\textbf{k}_5 d^2\textbf{k}_7 e^{i2n(\phi_1+\phi_3-\phi_5-\phi_7)} 
	\Bigg(  32 \input{n4_14.tikz} \nonumber \\
	& \hskip 9cm + 16 \input{fig11.tikz}   \Bigg) \Bigg \vert_{|\textbf{k}_1|=p_\perp}
	\nonumber \\
	&+2 \tilde{\kappa}_{2n}\{2\}(p_\perp) \kappa_{2n}\{2\} \left( \frac{1}{\tilde{\kappa}_0\{4\}(p_\perp)}-\frac{1}{\tilde{\kappa}_0\{2\}(p_\perp)\kappa_0\{2\}} \right),
\end{align}
and the differential 4-particle azimuthal harmonics is
%using \cref{vn4pt}.
defined as
\begin{align}\label{vn4dif}
	v_n\{4\}(p_\perp)=(-d_{2n}\{4\}(p_\perp))^{1/4}.
\end{align}

In \cref{cn_qs2_uncorr} right we have plotted our result for \cref{dn42}. Again, the values are very small and they even become positive with increasing $p_\perp$ because we are not including the diagrams that take into account the correlation of the partons inside the projectile.

With the results of \cref{cn_qs2_uncorr} we have finished our discussion of 4-gluon production in the case in which the partons are not correlated in the projectile wave function. So far, let us recapitulate what we did in this section. First, we wrote the 4-gluon spectrum in terms of the Wick diagrams by classifying them in different topologies and, thus, with a different suppression in powers of $(N_c^2-1)^{-1}$. Then we wrote the diagrams with the same topology as just one plus a bunch of permutations, as in \cref{mul4}. Then we exploited the symmetries of these permutations in order to reduce the number of integrals to be performed in the 4-particle cumulant functions \cref{cn42,dn42}. We also noticed that the contribution of the non vanishing factorizable connected diagrams to $\kappa_n\{4\}$ can be written as $2 \kappa_n\{2\}^2$. Finally, we solved numerically these integrals for given values of $Q_s^2$ and $B_p$.

Now let us jump to the case in which we take into account all the terms of the Wick expansion of the projectile correlator. In this case we have to deal with $(7!!)^2=11025$ terms instead of $7!!=105$. While the calculation becomes more cumbersome, the approach is exactly the same. First, we  group all the Wick diagrams in the 4-gluon spectrum by their topology that defines the power in $(N_c^2-1)^{-1}$, by using the property \ref{prop3} of \cref{sec2c}. Then, we relate the diagrams with the same topology by permutations. Next, in order to compute the 4-particle cumulant we exploit the symmetries of these permutations and reduce as much as possible the number of integrals to be performed. Finally, we  solve numerically each one of these integrals and obtain a result for the azimuthal harmonics. The detailed discussion of this procedure can be found in \cref{app:4gluon}.

\begin{figure}[h!]
	\centering
	\includegraphics[scale=0.8]{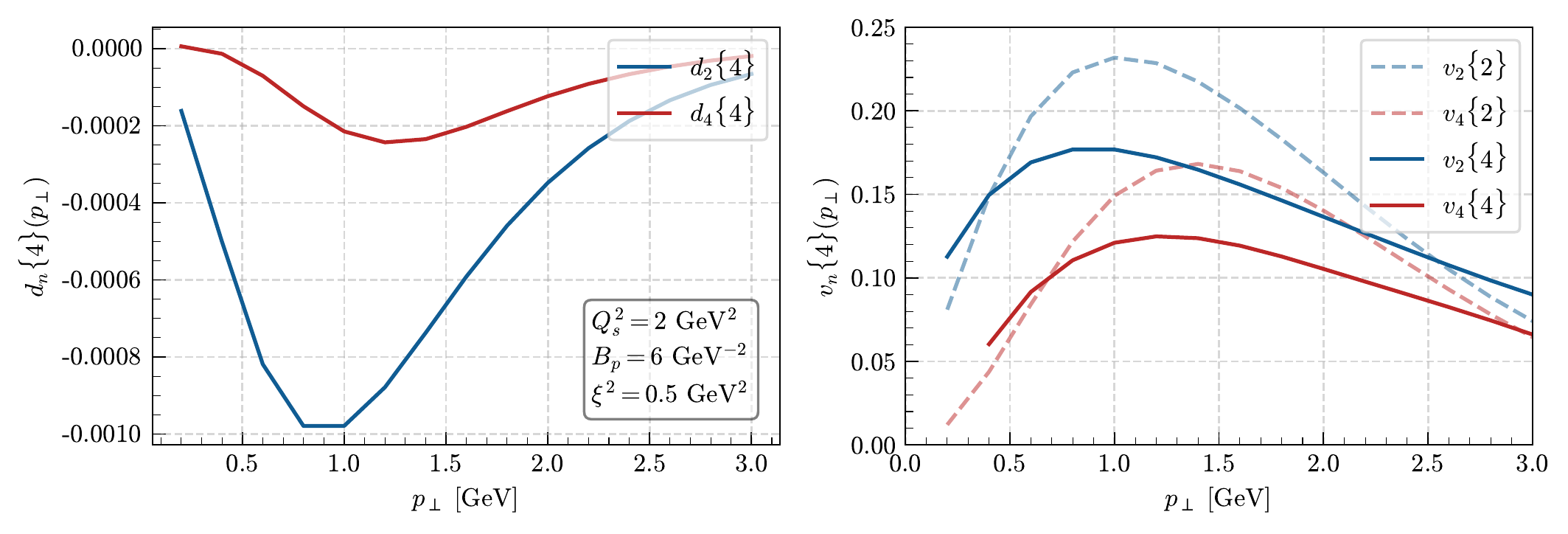}
	\caption{Dependence of the differential 4-particle cumulant (left) and azimuthal harmonic (right) of second and fourth order with $p_\perp$. For the latter we also show the results obtained from 2-particle correlations. In these graphs we have used $N_c=3$ and the values of the remaining parameters are indicated on the plots.}
	\label{dn_pt}
\end{figure}

In \cref{cn_vs_qs2,dn_pt} we show our results for the four gluon cumulants as a function of $Q_s^2$, the differential cumulants as a function of $p_\perp$ and the corresponding azimuthal harmonics for $n=2$ and 4. We use the same parameters that we employed in the two gluon case. Now, in contrast with the case seen above, the values obtained are negative (for the cumulants, thus real for the Fourier coefficients), larger in absolute value and in the ballpark of experimental data. Monte Carlo integration is used, yielding negligible errors except for the smallest $p_\perp$  for $d_4\{4\}$. On the other hand, it is known that when the multiplicity gets low the 4-particle cumulant turns positive~\cite{Abelev:2014mda,Khachatryan:2016txc}. The naive assumption that the multiplicity is proportional to the saturation momentum suggests a change of sign in the cumulant as $Q_s^2 \rightarrow 0$. Indeed, in the glasma graph approach, suitable for dilute-dilute collisions and therefore for lower multiplicities, arguments~\cite{Dumitru:2014yza} suggested that  $c_2\{4\}>0$ -- a result also found in~\cite{Dusling:2017aot} where a transition from positive to negative is found when multiple scattering (that goes beyond glasma graphs) is introduced. This is not seen in \cref{cn_vs_qs2}. A more detailed calculation should be done in this regime of low multiplicities where the transition for the glasma graph approach is expected.

With the results for the azimuthal harmonics in the case of 4-gluon inclusive production we finish our discussion on multi-gluon production. We point out that the procedure that we have developed can be generalised for larger values of $n$. It implies dealing with a large number of diagrams ($[(2n-1)!!]^2$). There is not conceptual problem for doing it since, as we have shown, we can always use the property \ref{prop3} of \cref{sec2c} to group all the diagrams in a systematic way and then exploit the symmetries to reduce the number of integrals to be performed. The remaining issue is dealing with a large number of $2n$-dimensional integrals that must be solved numerically.

We should also note that, although the results shown in this section are consistent with experimental data, no attempt is done to compare with them. We have used the area enhancement argument that should only be valid in the case when the overlap area is large. Furthermore, we have only taken into account the scattering of gluons. For more realistic results, we should at least compute the differential multiplicities for scattering quarks, consider more involved projectile and target averages (e.g. fluctuations) and convolute the results with fragmentation functions.

%The aim of this section was to give a simple way of regularizing the product of Lipatov vertices via \cref{lipatov} and show that using this model and the Wick expansion of the target and projectile correlators it is possible to obtain a result for the azimuthal harmonics of 2 and 4 particle production that is consistent with the experimental data.

\section{Summary}\label{sec4}

In this work we have computed multi-gluon production in the CGC in dilute-dense ($p$A) collisions, extending the work in~\cite{Altinoluk:2018ogz} to four gluons. Our calculation includes the contributions that are leading in the overlap area of the collision~\cite{Altinoluk:2018ogz,Altinoluk:2018hcu,Kovner:2017ssr,Kovner:2018vec}, while keeping all orders in the expansion in the number of colors. We develop a diagrammatic technique to write the numerous color contractions and exploit the symmetries to group the diagrams and simplify the expressions. This technique reduces dramatically the number of integrals needed to compute the multiplicity distributions and integrated and differential cumulants, which results essential for the large number of diagrams, more than 10000, that appears for four gluon production. We use the GBW model~\cite{GolecBiernat:1998js,GolecBiernat:1999qd} for the dipoles that result from the target averages, and the generalised MV model~\cite{McLerran:1993ni,McLerran:1993ka} for projectile averages. In order to proceed analytically as far as possible and simplify the final calculations, we use the Wigner function approach~\cite{Lappi:2015vta,Dusling:2017aot,Davy:2018hsl} that we extend to include quantum correlations in the projectile wave function. The Wigner function approach supposes that the final momenta of gluons is mainly acquired through interaction with the dense target and is thus suitable for a collinear projectile approximation.

Apart from the techniques developed and the discussions on the validity of the area enhancement argument and the Wigner function approach, our main results can be summarised in \cref{vn2_qs2,vn2_pt,cn_vs_qs2,dn_pt}. For two gluon correlations, we provide analytic expressions for integrated and differential cumulants which show smooth dependences on the parameters defining the projectile and target Wigner function and dipole, respectively. For four gluon correlations we find that the second order four particle cumulant $c_2\{4\}<0$ -- thus providing a sensible second order Fourier coefficient $v_2\{4\}$, a result found in~\cite{Dusling:2017aot} (where only quark scattering is considered and partons in the projectile wave function are uncorrelated) and attributed to multiple scattering. We note that the approximation in which gluons in the projectile are uncorrelated gives results for the cumulants that are much smaller in absolute value than when correlations are included, and become positive for some values of $Q_s$ and $p_\perp$. This emphasises the importance of including the full correlations in the projectile.

Our numerical results, due to the Gaussian forms that we employ for the Wigner function and dipole, cannot be considered reliable for $p_\perp$  sizeably larger than $Q_s$. They
lie in the ballpark of experimental data, for values of parameters that look reasonable. But we are aware that further analytic understanding is still required, and several pieces are still missing in our formalism: the contribution from quarks, more involved projectile and target averages, fragmentation functions,$\dots$. All these aspects should be explored before we can establish a model ready for phenomenology.

An immediate outlook of this work that we plan to address in the near future, is exploring the transition to low multiplicities, where the target should behave as a dilute object and the glasma graph approach should be valid. It has been argued that in such approximation $c_2\{4\}>0$~\cite{Dumitru:2014yza,Dusling:2017aot}. It would be most interesting to clarify the origin of such change of behaviour observed in data~\cite{Abelev:2014mda,Khachatryan:2016txc} and implement a framework that consistently goes from the dilute-dilute to the dilute-dense situation to examine the behaviour of the many particle cumulants.

\section*{Acknowledgements}
We thank Fabio Dom\'{\i}nguez, Alex Kovner, Michael Lublinsky and Vladimir Skokov for useful discussions. PA and NA have received financial support from Xunta de Galicia (Centro singular de investigaci\'on de Galicia accreditation 2019-2022), by the European research Council under project
ERC-2018-ADG-835105 YoctoLHC, by European Union ERDF, and by the "Mar\'{\i}a de Maeztu" Units of Excellence program MDM2016-0692 and the Spanish Research State Agency under project FPA2017-83814-P. TA is supported by Grant No.
2018/31/D/ST2/00666 (SONATA 14 - National Science Centre, Poland). PA is supported by the Xunta de Galicia action ”Axudas de apoio a etapa predoutoral”. This
work has been performed in the framework of COST Action CA 15213 "Theory of hot matter and relativistic heavy ion collisions" (THOR), MSCA RISE 823947 "Heavy ion collisions: collectivity and precision in saturation physics"
(HIEIC) and has received funding from the European Union's Horizon 2020 research and innovation programme under
grant agreement No. 824093.

\appendix

\section{On the validity of the area enhancement argument}
\label{app_AE}

In this section we will study the validity of the area enhancement argument, from now on AE model, introduced in Sect.~\ref{secarea}. For the sake of simplicity we will work in the fundamental representation of the Wilson lines instead of in the adjoint representation. Furthermore, we will only consider the expectation value of 4 Wilson lines, i.e. double quark interaction with a target. We expect the discussion presented here to be also valid for any number of Wilson lines or for a different color representation. In our case, the expected value of the double dipole operator is
\begin{align}\label{ae_dd}
	\langle \mathcal{D}(\textbf{x},\textbf{y})  \mathcal{D}(\textbf{u},\textbf{v}) \rangle_T^{\text{AE}} = D(\textbf{x},\textbf{y}) D(\textbf{u},\textbf{v})+\frac{1}{N_c^2} D(\textbf{x},\textbf{v}) D(\textbf{u},\textbf{y}),
\end{align}
where we have introduced the dipole operator $\mathcal{D}(\textbf{x},\textbf{y})=\frac{1}{N_c}\text{Tr}[U(\textbf{x})U^\dagger(\textbf{y})]$ and $D(\textbf{x},\textbf{y})$ is its target average.

As discussed in Sect.~\ref{secarea}, this approximation is only valid after integration over the phase space and at leading order in the transverse size of the interaction region, $B_p$. In order to check the validity of this model we will compare it with the result of~\cite{Dominguez:2008aa,Dominguez:2011wm} that was obtained by assuming multiple coherent scatterings of the quarks within the target. This result was obtained using the MV model and the result is
\begin{align}\label{mv_dd}
	\langle \mathcal{D}(\textbf{x},\textbf{y})  &\mathcal{D}(\textbf{u},\textbf{v}) \rangle_T^{\text{MV}} = D(\textbf{x},\textbf{y}) D(\textbf{u},\textbf{v}) \Bigg[  
	\left( \frac{F(\textbf{x},\textbf{u};\textbf{y},\textbf{v})+\sqrt{\Delta}}{2 \sqrt{\Delta}} - \frac{1}{N_c^2} \frac{F(\textbf{x},\textbf{y};\textbf{u},\textbf{v})}{\sqrt{\Delta}} \right) e^{\frac{N_c}{4} \mu^2 \sqrt{\Delta}}
	\nonumber \\
	&-\left( \frac{F(\textbf{x},\textbf{u};\textbf{y},\textbf{v})-\sqrt{\Delta}}{2 \sqrt{\Delta}} - \frac{1}{N_c^2} \frac{F(\textbf{x},\textbf{y};\textbf{u},\textbf{v})}{\sqrt{\Delta}} \right) e^{-\frac{N_c}{4} \mu^2 \sqrt{\Delta}}
	\Bigg]
	e^{-\frac{N_c}{4}\mu^2 F(\textbf{x},\textbf{u};\textbf{y},\textbf{v})+\frac{1}{2 N_c} \mu^2 F(\textbf{x},\textbf{y};\textbf{u},\textbf{v})},
\end{align}
where
\begin{align}
	\Delta = F(\textbf{x},\textbf{u};\textbf{y},\textbf{v})^2+\frac{4}{N_c^2} F(\textbf{x},\textbf{y};\textbf{u},\textbf{v}) F(\textbf{x},\textbf{v};\textbf{u},\textbf{y})
\end{align}
and  function $F(\textbf{x},\textbf{y};\textbf{u},\textbf{v})$ is defined in~\cite{Dominguez:2008aa}. In the GBW model this function reads simply
\begin{align}
	\mu^2 F(\textbf{x},\textbf{y};\textbf{u},\textbf{v}) = \frac{Q_s^2}{2 C_F} (\textbf{x}-\textbf{y}) \cdot (\textbf{u}-\textbf{v}).
\end{align}

Taking the large-$N_c$ limit we can simplify \cref{mv_dd} drastically to read, in the GBW model,
\begin{align}
	\langle \mathcal{D}(\textbf{x},\textbf{y})  \mathcal{D}(\textbf{u},\textbf{v}) \rangle_T^{\text{MV}} &=
	D(\textbf{x},\textbf{y})D(\textbf{u},\textbf{v}) \\ &
	+\frac{1}{N_c^2} \frac{F(\textbf{x},\textbf{y};\textbf{u},\textbf{v})^2}{F(\textbf{x},\textbf{u};\textbf{y},\textbf{v})^2} 
	\Bigg[  D(\textbf{x},\textbf{v})D(\textbf{u},\textbf{y})
	+D(\textbf{x},\textbf{y})D(\textbf{u},\textbf{v})
	\left( \frac{Q_s^2}{2}(\textbf{u}-\textbf{x}) \cdot (\textbf{v}-\textbf{y})-1\right)  \Bigg] + \mathcal{O}\left(\frac{1}{N_c^3}\right).\nonumber 
\end{align}

Thus, in the large-$N_c$ limit, the ratio between the integral of the double dipole weighted by an arbitrary smooth function  of the coordinates $\Psi(\textbf{x},\textbf{y},\textbf{u},\textbf{v})$ computed in the MV and AE models is
	\begin{align} \label{eq:largenc}
		\frac{\int_{\textbf{x},\textbf{y},\textbf{u},\textbf{v}} \langle \mathcal{D}\mathcal{D} \rangle^{\text{MV}} \Psi(\textbf{x},\textbf{y},\textbf{u},\textbf{v})}{\int_{\textbf{x},\textbf{y},\textbf{u},\textbf{v}} \langle \mathcal{D}\mathcal{D} \rangle^{\text{AE}} \Psi(\textbf{x},\textbf{y},\textbf{u},\textbf{v})}&=1      
		+ \frac{1}{N_c^2} 
		\int_{\textbf{x},\textbf{y},\textbf{u},\textbf{v}}
		\Psi(\textbf{x},\textbf{y},\textbf{u},\textbf{v})  
		\left[
		\left(\frac{F(\textbf{x},\textbf{y};\textbf{u},\textbf{v})^2}{F(\textbf{x},\textbf{u};\textbf{y},\textbf{v})^2} - 1 \right) D(\textbf{x},\textbf{v})D(\textbf{u},\textbf{y}) \right.
		\\
		&\hskip -2cm \left. +  \frac{F(\textbf{x},\textbf{y};\textbf{u},\textbf{v})^2}{F(\textbf{x},\textbf{u};\textbf{y},\textbf{v})^2} \left( \frac{Q_s^2}{2}(\textbf{u}-\textbf{x}) \cdot (\textbf{v}-\textbf{y}) - 1 \right) D(\textbf{x},\textbf{y})D(\textbf{u},\textbf{v})\right]
		{\bigg /} 
		\int_{\textbf{x},\textbf{y},\textbf{u},\textbf{v}} D(\textbf{x},\textbf{y})D(\textbf{u},\textbf{v}) \Psi(\textbf{x},\textbf{y},\textbf{u},\textbf{v}) 	
		\nonumber \\
		+ \mathcal{O}\left(\frac{1}{N_c^3}\right). \nonumber
	\end{align}
	
%\begin{align} \label{eq:largenc}
%	\frac{\int_{\textbf{x},\textbf{y},\textbf{u},\textbf{v}} \langle \mathcal{D}\mathcal{D} \rangle^{\text{MV}} \Psi(\textbf{x},\textbf{y},\textbf{u},\textbf{v})}{\int_{\textbf{x},\textbf{y},\textbf{u},\textbf{v}} \langle \mathcal{D}\mathcal{D} \rangle^{\text{AE}} \Psi(\textbf{x},\textbf{y},\textbf{u},\textbf{v})}&=1      \\
%	&+ \frac{1}{N_c^2} \left[\int_{\textbf{x},\textbf{y},\textbf{u},\textbf{v}}
%	\left(\frac{F(\textbf{x},\textbf{y};\textbf{u},\textbf{v})^2}{F(\textbf{x},\textbf{u};\textbf{y},\textbf{v})^2} - 1 \right) D(\textbf{x},\textbf{v})D(\textbf{u},\textbf{y}) \Psi(\textbf{x},\textbf{y},\textbf{u},\textbf{v})\right.
%	\nonumber \\
%	&\hskip 1.1cm \left. + \int_{\textbf{x},\textbf{y},\textbf{u},\textbf{v}} \frac{F(\textbf{x},\textbf{y};\textbf{u},\textbf{v})^2}{F(\textbf{x},\textbf{u};\textbf{y},\textbf{v})^2} \left( \frac{Q_s^2}{2}(\textbf{u}-\textbf{x}) \cdot (\textbf{v}-\textbf{y}) - 1 \right) D(\textbf{x},\textbf{y})D(\textbf{u},\textbf{v}) \Psi(\textbf{x},\textbf{y},\textbf{u},\textbf{v})\right]\nonumber \\
%	&\hskip 1cm {\bigg /} \int_{\textbf{x},\textbf{y},\textbf{u},\textbf{v}} D(\textbf{x},\textbf{y})D(\textbf{u},\textbf{v}) \Psi(\textbf{x},\textbf{y},\textbf{u},\textbf{v})\nonumber \\
%	 &+ \mathcal{O}\left(\frac{1}{N_c^3}\right).\nonumber
%\end{align}
Using the saddle point approximation, and noting that $F(\textbf{x},\textbf{y};\textbf{u},\textbf{v})\to 0$ when $\textbf{x}\to \textbf{y}$ or $\textbf{u}\to \textbf{v}$ and the fact that the dipole functions are Gaussian functions, it is straightforward to see that, in this approximation, the MV and AE model lead to the same result. For such approximation to hold we must consider the Gaussian functions, with width $\propto 1/Q_s$, to behave $\delta$-like with respect to the integration area. Therefore, corrections must be order $1/Q_s^2$ that, by dimensional reasons, has to be multiplied by an inverse area, with the overlap area, i.e., the size of the proton $B_p$, being the only parameter with such dimensions.

So far the discussion in this section only relies on the dynamics of the target and for this reason $B_p$ does not appear  in the expressions. We will introduce it by defining the phase space measure as
\begin{align}
	d\Omega = d^2\textbf{x} d^2\textbf{y} d^2\textbf{u} d^2\textbf{v} 
	\Theta\left(\sqrt{2 B_p}-|\textbf{x}|\right)
	\Theta\left(\sqrt{2 B_p}-|\textbf{y}|\right)
	\Theta\left(\sqrt{2 B_p}-|\textbf{u}|\right)
	\Theta\left(\sqrt{2 B_p}-|\textbf{v}|\right);
\end{align}
that is, we integrate over a 4-sphere of radius $\sqrt{2 B_p}$ in such a way that the integral over the phase space leads to the expected result
\begin{align}
	\int d\Omega = (2 \pi B_p)^4 = S_\perp^4.
\end{align}

In order to compare the MV and AE models, we perform a Fourier transform over the phase space measure defined as
\begin{align}\label{eq:FT}
	\langle \mathcal{D}(\textbf{q}_1,\textbf{q}_2)  \mathcal{D}(\textbf{q}_3,\textbf{q}_4) \rangle_T	 = \int d\Omega 
	e^{i \textbf{q}_1 \cdot \textbf{x}-i \textbf{q}_2 \cdot \textbf{y}+i \textbf{q}_3 \cdot \textbf{u}-i \textbf{q}_4 \cdot \textbf{v}}
	\langle \mathcal{D}(\textbf{x},\textbf{y})  \mathcal{D}(\textbf{u},\textbf{v}) \rangle_T.
\end{align}

\begin{figure}[h!]
	\centering
	\includegraphics[scale=0.7]{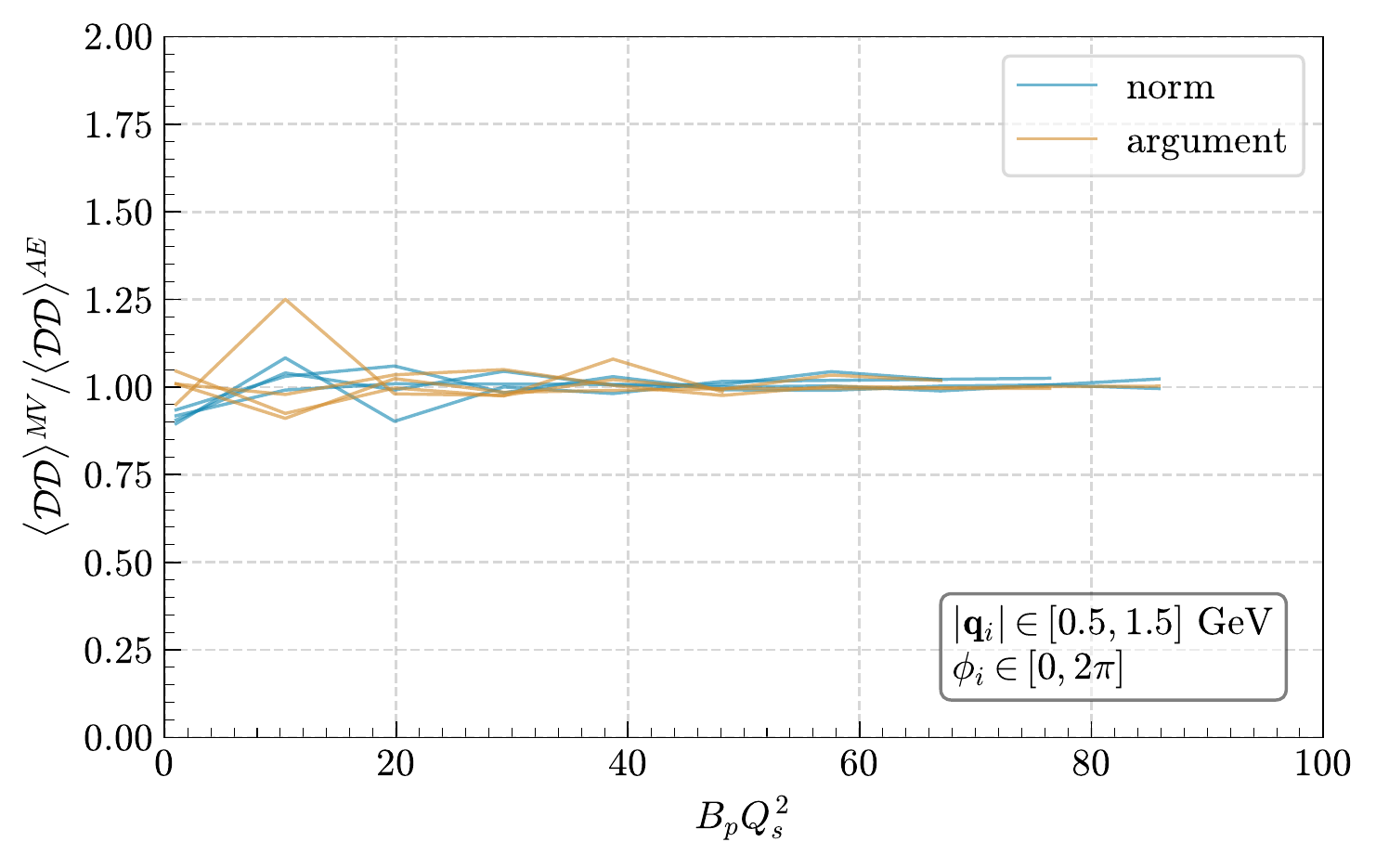}
	\caption{Ratio of the Fourier transforms of \cref{mv_dd} and \cref{ae_dd} at different values of $B_p$. The values of the ratio were computed using four sets of random momenta with moduli between 0.5 GeV and 1.5 GeV. We present both the norm (blue lines) and the argument (yellow lines). We have suppressed the values where the estimated error in the Monte Carlo integration becomes larger than 10 \%.}
	\label{ratio_dd}
\end{figure}

In \cref{ratio_dd} we show the result for the ratio of the Fourier transforms of \cref{mv_dd} and \cref{ae_dd} for different values of $B_p$, taking $Q_s^2=1$ GeV$^2$. The result was generated by using four sets of random momenta with moduli between 0.5 and 1.5 GeV. We see that, as expected, as we increase the value of $B_p Q_s^2$ the results in the AE model tends to those in the MV model, being the difference between both approaches of order a few \% at relatively high $B_p$.

An analogous discussion can be performed to the expected value of the quadrupole operator. In the AE model it can be written as 
\begin{align}\label{ae_q}
	\langle \mathcal{Q}(\textbf{x},\textbf{y},\textbf{u},\textbf{v}) \rangle_T^{\text{AE}} = D(\textbf{x},\textbf{y}) D(\textbf{u},\textbf{v})+ D(\textbf{x},\textbf{v}) D(\textbf{u},\textbf{y}),
\end{align}
where $\mathcal{Q}(\textbf{x},\textbf{y},\textbf{u},\textbf{v})=\frac{1}{N_c}\text{Tr}[U(\textbf{x})U^\dagger(\textbf{y})U(\textbf{u})U^\dagger(\textbf{v})]$. In the MV model it reads~\cite{Dominguez:2011wm}
\begin{align}\label{mv_q}
	\langle \mathcal{Q}(\textbf{x},\textbf{y},\textbf{u},\textbf{v}) \rangle_T^{\text{MV}} &= D(\textbf{x},\textbf{y}) D(\textbf{u},\textbf{v}) \Bigg[  
	\left( \frac{F(\textbf{x},\textbf{u};\textbf{y},\textbf{v})+\sqrt{\Delta}}{2 \sqrt{\Delta}} -  \frac{F(\textbf{x},\textbf{y};\textbf{u},\textbf{v})}{\sqrt{\Delta}} \right) e^{\frac{N_c}{4} \mu^2 \sqrt{\Delta}}
	\nonumber \\
	-&\left( \frac{F(\textbf{x},\textbf{u};\textbf{y},\textbf{v})-\sqrt{\Delta}}{2 \sqrt{\Delta}} -  \frac{F(\textbf{x},\textbf{y};\textbf{u},\textbf{v})}{\sqrt{\Delta}} \right) e^{-\frac{N_c}{4} \mu^2 \sqrt{\Delta}}
	\Bigg]
	e^{-\frac{N_c}{4}\mu^2 F(\textbf{x},\textbf{u};\textbf{y},\textbf{v})+\frac{1}{2 N_c} \mu^2 F(\textbf{x},\textbf{y};\textbf{u},\textbf{v})}.
\end{align}
It turns out that the only difference between the expectation value of the quadrupole and double dipole operators in both models is a factor $1/N_c^2$.

In the large-$N_c$ limit, \cref{mv_q} can be simplified to read
\begin{align}\label{largen_q}
	\langle \mathcal{Q}(\textbf{x},\textbf{y},\textbf{u},\textbf{v}) \rangle_T^{\text{MV}} = D(\textbf{x},\textbf{y})D(\textbf{u},\textbf{v})
	- \frac{F(\textbf{x},\textbf{y};\textbf{u},\textbf{v})}{F(\textbf{x},\textbf{u};\textbf{y},\textbf{v})} 
	\Big[D(\textbf{x},\textbf{y})D(\textbf{u},\textbf{v}) - D(\textbf{x},\textbf{v})D(\textbf{u},\textbf{y})\Big]  + \mathcal{O}\left(\frac{1}{N_c^2}\right).
\end{align}
We can show again that by using the saddle point approximation and the same arguments given below \cref{eq:largenc} that the expected value of the quadrupole is the same in both models. However, in the case of the quadrupole the difference between \cref{ae_q,largen_q} is not suppressed by any power of $1/N_c^2$ and, therefore, we expect a larger discrepancy between both models.

\begin{figure}[h!]
	\centering
	\includegraphics[scale=0.7]{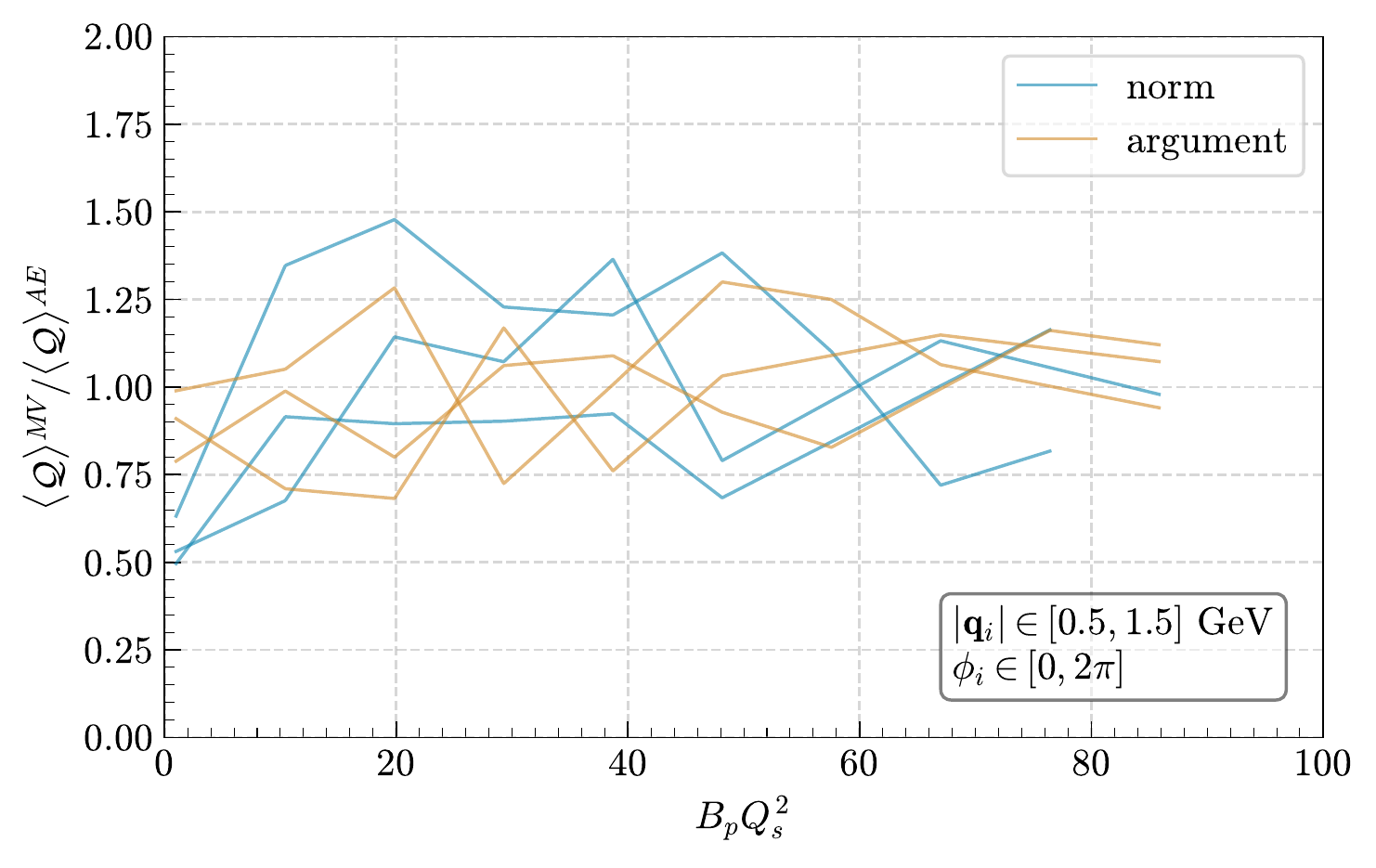}
	\caption{Ratio of the Fourier transform of \cref{mv_q,ae_q} at different values of $B_p$.  The values of the ratio were computed using three sets of random momenta with moduli between 0.5 GeV and 1.5 GeV. We present both the norm (blue lines) and the argument (yellow lines). We have suppressed the values where the estimated error in the Monte Carlo integration becomes larger than 10 \%.}
	\label{ratio_q}
\end{figure}

In \cref{ratio_q} we plot the ratio between the Fourier transform, defined analogous to \cref{eq:FT}, of \cref{mv_q,ae_q} for three sets of random momenta with moduli between 0.5 and 1.5 GeV for different values of $B_p$, taking again $Q_s^2=1$ GeV$^2$. In this case the difference between both models is of order 30\% at relatively high $B_p$, being larger than in \cref{ratio_dd} due to the $1/N_c^2$ suppression present for the double dipole and absent for the quadrupole. It also looks that at high $B_p$ the AE model tends to the MV model but integrals become very time consuming which prevents reaching larger values of $B_p Q_s^2$. Therefore, the tendency is not as clear as in \cref{ratio_dd}.

\section{Table of integrals}\label{app1}

In this section we present all the closed-form solutions of the integrals that have been used in this work. First, let us introduce the well known definitions
\begin{align}
	I_n(x)&=\sum_{k=0}^{\infty} \left(\frac{x}{2}\right)^{2k+n}\frac{1}{k!\Gamma(k+n+1)},
	\label{bessel}
	\\
	\,{}_{p}F_{q}(a_{1},\ldots ,a_{p};b_{1},\ldots ,b_{q};z) &=\sum _{n=0}^{\infty }\frac {(a_{1})_{n}\cdots (a_{p})_{n}}{(b_{1})_{n}\cdots (b_{q})_{n}}\,\frac {z^{n}}{n!},
	\label{hypergeo}
	\\
	(a)_{n}&={\frac {\Gamma (a+n)}{\Gamma (a)}},
	\\
	\Gamma(k)&=\int_0^{\infty} dx e^{-x}x^{k-1},
	\label{gamma}
\end{align}
where $I_n(x)$ is the modified Bessel function of the first kind, $\,{}_{p}F_{q}(a_{1},\ldots ,a_{p};b_{1},\ldots ,b_{q};z)$ is the generalised hypergeometric function, $(a)_{n}$ is the rising factorial (or Pochhammer symbol) and $\Gamma(k)$ is the gamma function. 

We will also introduce for convenience the Jacobi-Anger expansion
\begin{align}\label{JAexp}
	e^{x \cos \phi}=\sum_{n=-\infty}^{\infty}I_n(x)e^{-i n \phi}.
\end{align}

The first integral and the one that will be the most used in this work is the well known Gaussian integral
\begin{align}\label{Gaussintegr}
	\int d^2 \textbf{k}\  e^{-A \textbf{k}^2+\textbf{B} \cdot \textbf{k}}=\frac{\pi}{A}e^{\frac{\textbf{B}^2}{4A}}.
\end{align}

When computing the 2-particle differential cumulant we will have to deal with the following integral
\begin{align}
	&\int_0^{2\pi} d \phi_1 \int d^2 \textbf{k}_2 \ e^{i2n(\phi_1-\phi_2)}e^{-A_1 \textbf{k}_1^2-A_2\textbf{k}_2^2+A_{12}\textbf{k}_1 \cdot \textbf{k}_2} \Big \vert_{|\textbf{k}_1|=p_\perp}
	\nonumber \\ 
	= &\int_0^{2\pi} d \phi_1 d\phi_2\  e^{i2n(\phi_1-\phi_2)} 
	\int_0^{\infty} dk_2 k_2 \ e^{-A_1 p_\perp^2-A_2 k_2^2+A_{12} p_\perp k_2 \cos(\phi_1-\phi_2)}
	\nonumber \\
	=&2 \pi e^{B}       
	\int_0^{\infty} dk_2 k_2 \ e^{-A_2 k_2^2}
	\int_0^{2 \pi} d\phi \ e^{i2n\phi+C k_2 \cos \phi},
 \end{align}
where we have defined $B=-A_1 p_\perp^2$ and $C=A_{12} p_\perp$. Taking into account the Jacobi-Anger expansion \cref{JAexp} and the fact that $n$ is an integer we can write
\begin{align}
	2 \pi e^{B}       
	\int_0^{\infty} dk_2 k_2 \ e^{-A_2 k_2^2}
	\int_0^{2 \pi} d\phi \ e^{i2n\phi+C k_2 \cos \phi}
	&=
	2 \pi e^{B} \sum_{m=-\infty}^{\infty}
	\int_0^{\infty} dk_2 k_2 \ e^{-A_2 k_2^2} I_m(C k_2) 
	\int_0^{2 \pi} d\phi\  e^{i(2n-m)\phi}
	\nonumber \\
	&=
	(2 \pi)^2 e^{B} \int_0^{\infty} dk_2 k_2 \ e^{-A_2 k_2^2} I_{2n}(C k_2).
\end{align}
Using the definition of \cref{bessel} and then making the change of variable $x=A_2 k_2^2$ we have that
\begin{align}
	(2 \pi)^2 e^{B} \int_0^{\infty} dk_2 k_2 \ e^{-A_2 k_2^2} I_{2n}(C k_2) =&(2 \pi)^2 e^{B} \sum_{k=0}^{\infty} \int_0^{\infty} dk_2 k_2 \ e^{-A_2 k_2^2} \left(\frac{C k_2}{2}\right)^{2k+2n}\frac{1}{k!\Gamma(k+2n+1)}
	\nonumber \\
	=&(2 \pi)^2 e^{B} \sum_{k=0}^{\infty} \frac{1}{k!\Gamma(k+2n+1)} \int_0^{\infty} \frac{dx}{2 A_2} e^{-x} \left(\frac{C^2 x}{4 A_2}\right)^{k+n} 
	\nonumber \\
	=&
	\frac{(2 \pi)^2}{2 A_2} e^{B}  \left(\frac{C^2}{4 A_2}\right)^{n}
	\sum_{k=0}^{\infty} \left(\frac{C^2}{4 A_2}\right)^{k} \frac{1}{k!\Gamma(k+2n+1)} \int_0^{\infty} e^{-x} x^{k+n}
	\nonumber \\
	=& \frac{(2 \pi)^2}{2 A_2} e^{B}  \left(\frac{C^2}{4 A_2}\right)^{n}
	\sum_{k=0}^{\infty} \left(\frac{C^2}{4 A_2}\right)^{k} \frac{1}{k!\Gamma(k+2n+1)} \Gamma(k+n+1),
\end{align}
where in the last line we have used the definition of the Gamma function \cref{gamma}. Using the definition of the hypergeometric function \cref{hypergeo} we can write the last sum as
\begin{align}
	\sum_{k=0}^{\infty} \left(\frac{C^2}{4 A_2}\right)^{k} \frac{1}{k!\Gamma(k+2n+1)} \Gamma(k+n+1)
	=
	\frac{\Gamma(n+1)}{\Gamma(2n+1)} {}_1F_1\left(n+1;2n+1;\frac{C^2}{4 A_2}\right)
\end{align}
and, therefore, the result for the integral is 
\begin{align}\label{besselintegr}
	&\int_0^{2\pi} d \phi_1 \int d^2 \textbf{k}_2 \ e^{i2n(\phi_1-\phi_2)}e^{-A_1 \textbf{k}_1^2-A_2\textbf{k}_2^2+A_{12}\textbf{k}_1 \cdot \textbf{k}_2} \Big \vert_{|\textbf{k}_1|=p_\perp}
	\nonumber \\
	=
	&\frac{(2 \pi)^2}{2 A_2} e^{-A_1 p_\perp^2} \left(\frac{A_{12}^2 p_\perp^2}{4 A_2}\right)^{n} \frac{\Gamma(n+1)}{\Gamma(2n+1)} {}_1F_1\left(n+1;2n+1;\frac{A_{12}^2 p_\perp^2}{4 A_2}\right).
\end{align}

The solution of the integrals that we will find when we evaluate the 2-particle cumulant can be obtained in the same fashion and the result is
\begin{align}\label{doublebesselintegr}
	&\int d^2 \textbf{k}_1 d^2 \textbf{k}_2 \ e^{i2n(\phi_1-\phi_2)}e^{-A_1 \textbf{k}_1^2-A_2\textbf{k}_2^2+A_{12}\textbf{k}_1 \cdot \textbf{k}_2}
	\nonumber \\
	&=
	\frac{(2\pi)^2}{4A_1 A_2} \left(\frac{A_{12}^2}{4 A_1 A_2}\right)^{n} \frac{\Gamma(n+1)^2}{\Gamma(2n+1)} \,_2F_1\left(n+1,n+1;2n+1;\frac{A_{12}^2}{4 A_1 A_2}\right).
\end{align}

\section{The Wigner function approach}\label{app:wigner}

The Wigner function approach was used in several works \cite{Lappi:2015vha,Lappi:2015vta,Dusling:2017dqg,Dusling:2017aot,Davy:2018hsl} in order to compute multi-particle production. Here we will follow the arguments in \cite{Lappi:2015vta}. The forward amplitude for a gluon with momentum $\textbf{p}$ scattering on a dense target and leaving with a momentum $\textbf{k}$ is given, at leading order, by
\begin{align}
	\leftidx{_{out}}{\braket{\textbf{k},a | \textbf{p},b}}^{}_{in} = \int_\textbf{x} e^{i(\textbf{k}-\textbf{p}) \cdot \textbf{x}} U^{ab}(\textbf{x}),
\end{align}
where for simplicity we are not taking into account the longitudinal polarization of the gluons. On the other hand the distribution of gluons with momentum $\textbf{k}$ and color $a$ coming from the projectile after the interaction with the target can be written as
\begin{align}
	\frac{dN}{d^2 \textbf{k}}=\leftidx{_{out}}{\bra{\textbf{k},a} \hat{\rho} \ket{\textbf{k},a}}^{}_{out},
\end{align}
where $\hat{\rho}$ is the single gluon density matrix.

Using the completeness relation for the initial state we can write this equation as
\begin{align}
	\frac{dN}{d^2 \textbf{k}}&=\int_{\textbf{q}_1\textbf{q}_2} 
	\leftidx{_{out}}{\braket{\textbf{k},a | \textbf{q}_1,b_1}}^{}_{in}
	\leftidx{_{in}}{\bra{\textbf{q}_1,b_1} \hat{\rho} \ket{\textbf{q}_2,b_2}}^{}_{in}
	\leftidx{_{in}}{\braket{\textbf{q}_2,b_2 | \textbf{k},a}}^{}_{out}
	\nonumber \\
	&=\int_{\textbf{q}_1\textbf{q}_2} \int_{\textbf{x} \bar{\textbf{x}}} e^{-i(\textbf{k}-\textbf{q}_1) \cdot \textbf{x}+i(\textbf{k}-\textbf{q}_2) \cdot \bar{\textbf{x}}} U^{ab_1}(\textbf{x}) U^{b_2a}(\bar{\textbf{x}})^\dagger \leftidx{_{in}}{\bra{\textbf{q}_1,b_1} \hat{\rho} \ket{\textbf{q}_2,b_2}}^{}_{in}.
\end{align}
Doing the change of variables $\textbf{q}_{1,2}=\textbf{p} \pm \textbf{q}/2$ and $\textbf{x},\bar{\textbf{x}}=\textbf{b} \pm \textbf{r}/2$ we get
\begin{align}
	\frac{dN}{d^2 \textbf{k}}=\int_{\textbf{r}\textbf{b}}\int_{\textbf{p}\textbf{q}}
	e^{i \textbf{q} \cdot \textbf{b}}  
	\leftidx{_{in}}{\Bra{\textbf{p}+\frac{\textbf{q}}{2},b_1} \hat{\rho} 
		\Ket{\textbf{p}-\frac{\textbf{q}}{2},b_2}}^{}_{in}
	e^{-i(\textbf{k}-\textbf{p})\cdot\textbf{r}} 
	U^{ab_1}\left(\textbf{b}+\frac{\textbf{r}}{2}\right)
	U^{b_2 a}\left(\textbf{b}-\frac{\textbf{r}}{2}\right)^\dagger.
\end{align}
Defining
\begin{align}
	\Phi^{b_2 b_1}(\textbf{b},\textbf{k}-\textbf{p})=\int_{\textbf{r}} e^{-i(\textbf{k}-\textbf{p})\cdot\textbf{r}} U^{ab_1}\left(\textbf{b}+\frac{\textbf{r}}{2}\right)
	U^{b_2 a}\left(\textbf{b}-\frac{\textbf{r}}{2}\right)^\dagger
\end{align}
and realizing that
\begin{align}
	W^{b_1 b_2}(\textbf{b},\textbf{p})=\int_{\textbf{q}}e^{i \textbf{q} \cdot \textbf{b}}  \leftidx{_{in}}{\Bra{\textbf{p}+\frac{\textbf{q}}{2},b_1} \hat{\rho} 
		\Ket{\textbf{p}-\frac{\textbf{q}}{2},b_2}}^{}_{in}
\end{align}
is the Weyl transform of the density matrix, that is, the Wigner function, we can write the single inclusive gluon spectrum as
\begin{align}\label{Nex2}
	\frac{dN}{d^2 \textbf{k}}=\int_{\textbf{b}}\int_{\textbf{p}}  
	W^{b_1 b_2}(\textbf{b},\textbf{p})
	\Phi^{b_2 b_1}(\textbf{b},\textbf{k}-\textbf{p}).
\end{align}
Since this expression is also dependent of the color charge density of the target we still have to perform the target average.

On the other hand, in the approach used in this work, we can evaluate the single gluon spectrum, before target averaging, using \cref{mul_kfac}:
\begin{align}
	2 (2\pi)^3 \frac{dN}{d^2 \textbf{k}}=\int_{\textbf{q}_1\textbf{q}_2} 
	\Big \langle \rho^{b_1}(\textbf{q}_1)\rho^{b_2}(\textbf{q}_2)^* \Big \rangle_p 
	4 L^i(\textbf{k},\textbf{k}-\textbf{q}_1) L^i(\textbf{k},\textbf{k}-\textbf{q}_2) 
	\int_{\textbf{x} \bar{\textbf{x}}} e^{-i(\textbf{k}-\textbf{q}_1) \cdot \textbf{x}+i(\textbf{k}-\textbf{q}_2) \cdot \bar{\textbf{x}}}
	U^{a b_1}(\textbf{x}) U^{b_2 a}(\bar{\textbf{x}})^\dagger.
\end{align}

Doing the same change of variables that we did before can write this expression as
\begin{align}\label{Nex1}
	\frac{dN}{d^2 \textbf{k}}=\frac{1}{2 (2\pi)^3}
	\int_{\textbf{b}}\int_{\textbf{p}}
	\int_{\textbf{q}} e^{i \textbf{q} \cdot \textbf{b}}
	\Big \langle \rho^{b_1}\left(\textbf{p}+\frac{\textbf{q}}{2}\right)\rho^{b_2}\left(\textbf{p}-\frac{\textbf{q}}{2}\right)^* \Big \rangle_p 
	4 L^i\left(\textbf{k},\textbf{k}-\textbf{p}-\frac{\textbf{q}}{2}\right) L^i\left(\textbf{k},\textbf{k}-\textbf{p}+\frac{\textbf{q}}{2}\right) 
	\Phi^{b_2 b_1}(\textbf{b},\textbf{k}-\textbf{p}).
\end{align}

Thus, comparing \cref{Nex2,Nex1} we see that the single particle Wigner function can be written in terms of the Lipatov vertices and the 2-point correlator of the projectile charge density:
\begin{align}
	W^{b_1 b_2}(\textbf{b},\textbf{p})&=
	\frac{1}{2 (2\pi)^3}
	\int_{\textbf{q}} e^{i \textbf{q} \cdot \textbf{b}}
	\Big \langle \rho^{b_1}\left(\textbf{p}+\frac{\textbf{q}}{2}\right)\rho^{b_2}\left(\textbf{p}-\frac{\textbf{q}}{2}\right)^* \Big \rangle_p 
	4 L^i\left(\textbf{k},\textbf{k}-\textbf{p}-\frac{\textbf{q}}{2}\right) L^i\left(\textbf{k},\textbf{k}-\textbf{p}+\frac{\textbf{q}}{2}\right).
\end{align}

Using the models employed through this work, \cref{lipatov,mu2}, we can write the single particle Wigner function as
\begin{align}
	W^{b_1 b_2}(\textbf{b},\textbf{p})&=\frac{\delta^{b_1 b_2}}{N_c^2-1} \frac{1}{2 (2\pi)^3} \frac{(4 \pi)^2}{\xi^2} e^{-\textbf{p}^2/\xi^2}  \int_{\textbf{q}} e^{i \textbf{q} \cdot \textbf{b}} e^{-\textbf{q}^2/(4 B_p^{-1})}
	\nonumber \\
	&=\frac{\delta^{b_1 b_2}}{N_c^2-1} \frac{1}{\pi^2} \frac{1}{\xi^2 B_p} e^{-\textbf{b}^2/B_p-\textbf{p}^2/\xi^2},
\end{align}
which is the same function found in the literature~\cite{Lappi:2015vta}. We can also check that this function is well normalised by performing the trace and integrating over $\textbf{b}$ and $\textbf{p}$,
\begin{align}
	\int_\textbf{b}\int_\textbf{p} W^{a a}(\textbf{b},\textbf{p}) = 1.
\end{align}

Doing an analogous discussion we can write the 2-particle Wigner function as
\begin{align}
	W^{b_1 b_2 b_3 b_4}&(\textbf{b}_1,\textbf{p}_1,\textbf{b}_2,\textbf{p}_2)=
	\frac{4}{(2\pi)^6}
	\int_{\textbf{q}_1 \textbf{q}_2} e^{i \textbf{q}_1 \cdot \textbf{b}_1+i \textbf{q}_2 \cdot \textbf{b}_2}
	\Big \langle \rho^{b_1}\left(\textbf{p}_1+\frac{\textbf{q}_1}{2}\right)\rho^{b_2}\left(\textbf{p}_1-\frac{\textbf{q}_1}{2}\right)^* 
	\rho^{b_3}\left(\textbf{p}_2+\frac{\textbf{q}_2}{2}\right)\rho^{b_4}\left(\textbf{p}_2-\frac{\textbf{q}_2}{2}\right)^* 
	\Big \rangle_p
	\nonumber \\ 
	&\times 
	L^i\left(\textbf{k}_1,\textbf{k}_1-\textbf{p}_1-\frac{\textbf{q}_1}{2}\right) L^i\left(\textbf{k}_1,\textbf{k}_1-\textbf{p}_1+\frac{\textbf{q}_1}{2}\right) 
	L^i\left(\textbf{k}_2,\textbf{k}_2-\textbf{p}_2-\frac{\textbf{q}_2}{2}\right) L^i\left(\textbf{k}_2,\textbf{k}_2-\textbf{p}_2+\frac{\textbf{q}_2}{2}\right).
\end{align}

Performing the Wick expansion of the projectile correlator and using again \cref{lipatov,mu2} we obtain \cref{wigner2}.
%\begin{align}\label{wigner2}
%	W^{b_1 b_2 b_3 b_4}(\textbf{b}_1,\textbf{p}_1,\textbf{b}_2,\textbf{p}_2)&=\frac{1}{(N_c^2-1)^2} \frac{1}{\pi^4 \xi^4 B_p^2} e^{-(\textbf{p}_1^2+\textbf{p}_2^2)/\xi^2} e^{-(\textbf{b}_1^2+\textbf{b}_2^2)/B_p} \Big[ \delta^{b_1 b_2} \delta^{b_3 b_4}
%	\nonumber \\
%	&
%	+\delta^{b_1 b_3} \delta^{b_2 b_4} 2 \pi B_p \delta^{(2)}(\textbf{b}_1-\textbf{b}_2) e^{-(\textbf{p}_1+\textbf{p}_2)^2/(2 B_p^{-1})}
%	\nonumber \\
%	&
%	+\delta^{b_1 b_4} \delta^{b_2 b_3} 2 \pi B_p \delta^{(2)}(\textbf{b}_1-\textbf{b}_2) e^{-(\textbf{p}_1-\textbf{p}_2)^2/(2 B_p^{-1})}
%	\Big].
%\end{align}
We can check that the quantity defined in that equation is not well normalised:
\begin{align}
	\int_{\textbf{b}_1\textbf{b}_2} \int_{\textbf{p}_1\textbf{p}_2} W^{a a b b}(\textbf{b}_1,\textbf{p}_1,\textbf{b}_2,\textbf{p}_2)=
	1+2\frac{1}{(N_c^2-1)}\frac{1}{1+B_p \xi^2}.
\end{align}
Therefore, in order to have a proper definition of the Wigner function we should normalise \cref{wigner2} by this factor. However, since in correlation studies the overall constants do not contribute to the cumulants, this normalisation factor is not important for us.

We should also note that \cref{wigner2} breaks the factorisation assumption that is used in the literature in which the 2-particle Wigner function factorizes into a product of two single particle Wigner function:
\begin{align}
	W^{b_1 b_2 b_3 b_4}(\textbf{b}_1,\textbf{p}_1,\textbf{b}_2,\textbf{p}_2)=W^{b_1 b_2}(\textbf{b}_1,\textbf{p}_1) W^{b_3 b_4}(\textbf{b}_2,\textbf{p}_2).
\end{align}
The reason for the breaking of this factorisation is that we are including in our approach quantum correlations in the projectile wave function. Thus we can interpret the terms in \cref{wigner2} that break factorisation as Bose enhancement contributions in the projectile wave function.

\section{Calculation of four gluon inclusive production}\label{app:4gluon}

In this section we analyse the four gluon inclusive spectrum by taking into account all the terms in \cref{mul_kfac}. In order to do so we will follow the same arguments that we have used for writing down the triple gluon spectrum in \cref{sec:triple}. First we note that after performing the Wick expansion of either the target or projectile correlators we  have 105 contributions on each side that can be written schematically as
\begin{align}
	&\input{n4_11_2.tikz}
	\label{d41}
	, \\
	\Bigg( &\input{n4_12_2.tikz}+\textbf{k}_7 \rightarrow -\textbf{k}_7   \Bigg)+\textbf{k}_1 \leftrightarrow \textbf{k}_5+\textbf{k}_1 \leftrightarrow \textbf{k}_7+\textbf{k}_3 \leftrightarrow \textbf{k}_5 +\textbf{k}_3 \leftrightarrow \textbf{k}_7  +(\textbf{k}_3 \leftrightarrow \textbf{k}_5)(\textbf{k}_1 \leftrightarrow \textbf{k}_7)
	\label{d42}
	, \\
	\Bigg[\Bigg( &\input{n4_13_2.tikz}
	+\textbf{k}_3 \rightarrow -\textbf{k}_3
	+\textbf{k}_5 \rightarrow -\textbf{k}_5
	+\textbf{k}_7 \rightarrow -\textbf{k}_7  \Bigg)
	+\textbf{k}_3 \leftrightarrow \textbf{k}_7
	\Bigg]
	+\textbf{k}_1 \leftrightarrow \textbf{k}_3
	+\textbf{k}_1 \leftrightarrow \textbf{k}_5
	+\textbf{k}_1 \leftrightarrow \textbf{k}_7 
	\label{d43}
	, \\
	\Bigg( &\input{n4_15_2.tikz}
	+\textbf{k}_1 \rightarrow -\textbf{k}_1
	+\textbf{k}_5 \rightarrow -\textbf{k}_5
	+(\textbf{k}_1 \rightarrow -\textbf{k}_1)(\textbf{k}_5 \rightarrow -\textbf{k}_5)
	\Bigg)
	+\textbf{k}_3 \leftrightarrow \textbf{k}_5
	+\textbf{k}_3 \leftrightarrow \textbf{k}_7
	\label{d44}
	, \\
	\Bigg[ &\input{n4_14_2.tikz}+\textbf{k}_1 \rightarrow -\textbf{k}_1+\textbf{k}_3 \rightarrow -\textbf{k}_3 +\textbf{k}_5 \rightarrow -\textbf{k}_5 +\textbf{k}_7 \rightarrow -\textbf{k}_7 \label{d45}
	 \\ &
	+ \frac{1}{2} \Bigg(  
	(\textbf{k}_1 \rightarrow -\textbf{k}_1)(\textbf{k}_3 \rightarrow -\textbf{k}_3)
	+(\textbf{k}_1 \rightarrow -\textbf{k}_1)(\textbf{k}_5 \rightarrow -\textbf{k}_5)
	+(\textbf{k}_1 \rightarrow -\textbf{k}_1)(\textbf{k}_7 \rightarrow -\textbf{k}_7)
	+(\textbf{k}_3 \rightarrow -\textbf{k}_3)(\textbf{k}_5 \rightarrow -\textbf{k}_5)
	\nonumber \\ &
	+(\textbf{k}_3 \rightarrow -\textbf{k}_3)(\textbf{k}_7 \rightarrow -\textbf{k}_7)
	+(\textbf{k}_5 \rightarrow -\textbf{k}_5)(\textbf{k}_7 \rightarrow -\textbf{k}_7)
	\Bigg) \Bigg] 
	+\textbf{k}_1 \leftrightarrow \textbf{k}_3
	+\textbf{k}_1 \leftrightarrow \textbf{k}_7
	+\textbf{k}_3 \leftrightarrow \textbf{k}_5
	+\textbf{k}_3 \leftrightarrow \textbf{k}_7
	+\textbf{k}_5 \leftrightarrow \textbf{k}_7,
	\nonumber 
\end{align}
where the permutations $\textbf{k}_i \rightarrow -\textbf{k}_i$ and $\textbf{k}_i \leftrightarrow \textbf{k}_j$ are an abuse of notation since we have not contracted the diagrams and thus we cannot apply  properties \ref{prop1} and \ref{prop2} yet. In order to make the notation lighter we write these permutations as
\begin{align}
	&\input{n4_11_2.tikz}
	, \\
	&\input{n4_12_2.tikz} + \text{perm}_2
	, \\
	&\input{n4_13_2.tikz} + \text{perm}_3
	, \\
	&\input{n4_15_2.tikz} + \text{perm}_4
	, \\
	&\input{n4_14_2.tikz} + \text{perm}_5.
\end{align}

Now we  generate the Wick diagrams in such a way that they are grouped by their powers of $(N_c^2-1)^{-1}$. In order to do so we  exploit  property \ref{prop3} of \cref{sec2c}. In this case the suppression of a given diagram is given by $(N_c^2-1)^{n_p+n_T-8}$, with the values of $n_p$ and $n_T$ fixing the topology of the diagram. 

It is straightforward to realise that all the diagrams with $n_T=4$ will have the configuration of \cref{d41} on the right side. All the diagrams with $n_T=3$ will have one of the 12 configurations of \cref{d42} on the right side. All the diagrams with $n_T=2$ will have one of the 32 configurations of \cref{d43} or one of the 12 configurations of \cref{d44} on the right side and all the diagrams with $n_T=1$ will have one of the 48 configurations of \cref{d45} on the right side. Therefore the value of $n_T$ is fixed by the configuration that we have on the right side of the diagram.

The value of $n_p$, on the other hand, will depend on the configuration that we have on both sides. It is determined by the number of disconnected pieces that we obtain after drawing the right configuration of the diagram on top of the left one. Thus, the only way of obtaining $n_p=4$ is by having a configuration on the left that has the same links as the one on the right. The only way of obtaining $n_p=3$ is by having a configuration on the left that has just two links that are equal to the ones on the right. The way of obtaining $n_p=2$ is by having a configuration on the left that has only one link that is equal to the right configuration or by having all the links different but in such a way that, after the projection, we obtain two disconnected pieces. Finally, the only way of obtaining $n_p=1$ is by having a configuration on the left side that has all the links different to the right configuration in such a way that, after the projection, we have a fully connected piece. The number of possibilities for $n_p=4$ is 1, for $n_p=3$ is 12, for $n_p=2$ is 32 and 12, respectively, and for $n_p=1$ is 48.

Having this into account we can find all the diagrams with a given suppression in powers of $(N_c^2-1)^{-1}$. As an example, let us see which are the diagrams with power suppression $(N_c^2-1)^{-3}$. In this case we have  $n_p+n_T=5$ and we will have 4 different topologies that are fixed by this constraint: $n_T=4$ and $n_p=1$ ; $n_T=3$ and $n_p=2$ ; $n_T=2$ and $n_p=3$ ; $n_T=1$ and $n_p=4$. Let us study this situation case by case:
\begin{enumerate}[label=\roman*)]
	\item $n_T=4$ and $n_p=1$. In this case we will have the configuration of \cref{d41} on the right side of the diagram and on the left side we will have all the diagrams that have zero links in common with the one on the right in such a way that after the projection we just have one connected piece. Thus we will have $1\times48$ possibilities that are
	\begin{align}\label{Nc31}
		&\input{nt4np1_1.tikz}+\input{nt4np1_2.tikz}+\input{nt4np1_3.tikz}+\input{nt4np1_4.tikz}+\input{nt4np1_5.tikz}
		\nonumber \\ +&
		\input{nt4np1_6.tikz} +\input{nt4np1_7.tikz}+\input{nt4np1_8.tikz}+\input{nt4np1_9.tikz}+\input{nt4np1_10.tikz}
		\nonumber \\ +&
		\input{nt4np1_11.tikz}+\input{nt4np1_12.tikz}+\input{nt4np1_13.tikz}+\input{nt4np1_14.tikz}+\input{nt4np1_15.tikz}
		\nonumber \\ +&
		\input{nt4np1_16.tikz}+\input{nt4np1_17.tikz}+\input{nt4np1_18.tikz}+\input{nt4np1_19.tikz}+\input{nt4np1_20.tikz}
		\nonumber \\ +&
		\input{nt4np1_21.tikz}+\input{nt4np1_22.tikz}+\input{nt4np1_23.tikz}+\input{nt4np1_24.tikz}+\input{nt4np1_25.tikz}
		\nonumber \\ +&
		\input{nt4np1_26.tikz}+\input{nt4np1_27.tikz}+\input{nt4np1_28.tikz}+\input{nt4np1_29.tikz}+\input{nt4np1_30.tikz}
		\nonumber \\ +&
		\input{nt4np1_31.tikz}+\input{nt4np1_32.tikz}+\input{nt4np1_33.tikz}+\input{nt4np1_34.tikz}+\input{nt4np1_35.tikz}
		\nonumber \\ +&
		\input{nt4np1_36.tikz}+\input{nt4np1_37.tikz}+\input{nt4np1_38.tikz}+\input{nt4np1_39.tikz}+\input{nt4np1_40.tikz}
		\nonumber \\ +&
		\input{nt4np1_41.tikz}+\input{nt4np1_42.tikz}+\input{nt4np1_43.tikz}+\input{nt4np1_44.tikz}+\input{nt4np1_45.tikz}
		\nonumber \\ +&
		\input{nt4np1_46.tikz}+\input{nt4np1_47.tikz}+\input{nt4np1_48.tikz}
		\nonumber \\
		= &\input{nt4np1_14.tikz} + \text{perm}_5,
	\end{align}
	where in the last line we have used the fact that since on the right side of the diagram we have a fully disconnected piece we can write the sum of these $48$ diagrams as just one plus perm$_5$.
	
	\item \label{nT3_np2} $n_T=3$ and $n_p=2$. In this case we will have the configurations of \cref{d42} on the right side of the diagram and on the left side we will have all the diagrams that have just one link in common with the right one. This gives a total of $12\times32$ possibilities
	\begin{align}\label{Nc32}
		&\input{nt3np21_1.tikz}+\input{nt3np21_2.tikz}+\input{nt3np21_3.tikz}+\input{nt3np21_4.tikz}+\input{nt3np21_5.tikz}
		\nonumber \\ +&
		\input{nt3np21_6.tikz} +\input{nt3np21_7.tikz}+\input{nt3np21_8.tikz}+\input{nt3np21_9.tikz}+\input{nt3np21_10.tikz}
		\nonumber \\ +&
		\input{nt3np21_11.tikz}+\input{nt3np21_12.tikz}+\input{nt3np21_13.tikz}+\input{nt3np21_14.tikz}+\input{nt3np21_15.tikz}
		\nonumber \\ +&
		\input{nt3np21_16.tikz}+\input{nt3np21_17.tikz}+\input{nt3np21_18.tikz}+\input{nt3np21_19.tikz}+\input{nt3np21_20.tikz}
		\nonumber \\ +&
		\input{nt3np21_21.tikz}+\input{nt3np21_22.tikz}+\input{nt3np21_23.tikz}+\input{nt3np21_24.tikz}+\input{nt3np21_25.tikz}
		\nonumber \\ +&
		\input{nt3np21_26.tikz}+\input{nt3np21_27.tikz}+\input{nt3np21_28.tikz}+\input{nt3np21_29.tikz}+\input{nt3np21_30.tikz}
		\nonumber \\ +&
		\input{nt3np21_31.tikz}+\input{nt3np21_32.tikz}+\text{perm}_2,
	\end{align}
or we can have on the left side the configurations that have no links in common with the one on the right in such a way that, after the projection, we have two connected pieces. This gives a total of $12 \times 12$ possibilities:
	\begin{align}\label{Nc33}
		&\input{nt3np22_1.tikz}+\input{nt3np22_2.tikz}+\input{nt3np22_3.tikz}+\input{nt3np22_4.tikz}+\input{nt3np22_5.tikz}
		\nonumber \\ +&
		\input{nt3np22_6.tikz} +\input{nt3np22_7.tikz}+\input{nt3np22_8.tikz}+\input{nt3np22_9.tikz}+\input{nt3np22_10.tikz}
		\nonumber \\ +&
		\input{nt3np22_11.tikz}+\input{nt3np22_12.tikz}+\text{perm}_2.
	\end{align}
	
	\item $n_T=2$ and $n_p=3$. This implies that we will have the configurations of \cref{d43} or \cref{d44} on the right side of the diagram and on the left side we will have the configurations that have two links equal to the one on the right. This gives a total of $32 \times 12$ possibilities for the first case,
	\begin{align}\label{Nc34}
		&\input{nt2np11_1.tikz}+\input{nt2np11_2.tikz}+\input{nt2np11_3.tikz}+\input{nt2np11_4.tikz}+\input{nt2np11_5.tikz}
		\nonumber \\ +&
		\input{nt2np11_6.tikz} +\input{nt2np11_7.tikz}+\input{nt2np11_8.tikz}+\input{nt2np11_9.tikz}+\input{nt2np11_10.tikz}
		\nonumber \\ +&
		\input{nt2np11_11.tikz}+\input{nt2np11_12.tikz}+\text{perm}_3,
	\end{align}
	and $12 \times 12$ possibilities for the second case,
	\begin{align}\label{Nc35}
		&\input{nt2np12_1.tikz}+\input{nt2np12_2.tikz}+\input{nt2np12_3.tikz}+\input{nt2np12_4.tikz}+\input{nt2np12_5.tikz}
		\nonumber \\ +&
		\input{nt2np12_6.tikz} +\input{nt2np12_7.tikz}+\input{nt2np12_8.tikz}+\input{nt2np12_9.tikz}+\input{nt2np12_10.tikz}
		\nonumber \\ +&
		\input{nt2np12_11.tikz}+\input{nt2np12_12.tikz}+\text{perm}_4.
	\end{align}
	
	\item $n_T=1$ and $n_p=4$. This implies that we have the configurations of \cref{d45} on the right side of the diagram and on the left side we have the configuration that have the same links with respect to the right one. This gives a total of $48 \times 1$ possibilities:
	\begin{align}\label{Nc36}
		\input{nt1np4.tikz}+\text{perm}_5.
	\end{align}
\end{enumerate}

With \crefrange{Nc31}{Nc36} we have found all the 1152 diagrams with a suppression of $(N_c^2-1)^{-3}$ and written them as a bunch of diagrams plus permutations -- which was our goal. We should also note that some of the diagrams that are drawn in these equations can also be related by symmetries $\textbf{k}_i \rightarrow -\textbf{k}_i$ or $\textbf{k}_i \leftrightarrow \textbf{k}_j$, which could lead to a better optimisation of the calculation but we have not found any systematic way of finding these symmetries. Therefore, we have decided to not include them in the calculation since we see not advantage in doing this by hand. 

We can find the other diagrams with a different suppression in the same fashion obtaining 1 diagram with a suppression of $(N_c^2-1)^{0}$, 24 diagrams with a suppression of $(N_c^2-1)^{-1}$, 232 diagrams with a suppression of $(N_c^2-1)^{-2}$, 3088 diagrams with a suppression of $(N_c^2-1)^{-4}$, 4224 diagrams with a suppression of $(N_c^2-1)^{-5}$ and 2304 diagrams with a suppression of $(N_c^2-1)^{-6}$.

The next step is to exploit the symmetries encoded within the permutations in order to evaluate the cumulants through \cref{kappa_def}. We will do  as an example the calculation only for the terms that contribute, again, with a power $(N_c^2-1)^{-3}$. Let us introduce the shorthand notation $\hat{\mathcal{D}}_{n_p}$ as the sum of all the diagrams that satisfy the topology given by $n_p$ with a given configuration on the right side. Then we can write the contribution of order $(N_c^2-1)^{-3}$ to the 4-gluon spectrum as
\begin{align}\label{mulnc3}
	N^{(3)}=&\left(\input{nt4np1_14.tikz} + \text{perm}_5\right) 
	+\left(\input{nt3np21_ex.tikz}+\input{nt3np22_ex.tikz} + \text{perm}_2 \right)
	\nonumber \\
	+&\left(\input{nt2np11_ex.tikz} + \text{perm}_3\right)
	+\left(\input{nt2np12_ex.tikz} + \text{perm}_4\right)
	+\left(\input{nt1np4.tikz} + \text{perm}_5\right),
\end{align}
with $\hat{\mathcal{D}}_2^{(1)}$ and $\hat{\mathcal{D}}_2^{(2)}$ referring to the first and second contributions to \cref{nT3_np2} discussed above, respectively.

In order to evaluate $\kappa_0\{4\}$ we can use the fact that all the permutations, perm$_i$, of \crefrange{d41}{d42} will give the same result since we are integrating over all the momentum $\textbf{k}_i$. Thus we can write
\begin{align}\label{k0nc3}
	\kappa_0^{(3)}\{4\}=\int_{\textbf{k}_1 \textbf{k}_3 \textbf{k}_5 \textbf{k}_7} 
	&\Bigg[48 \input{nt4np1_14.tikz} + 12 \input{nt3np21_ex.tikz} + 12 \input{nt3np22_ex.tikz} 
	\nonumber \\ 
	& + 32 \input{nt2np11_ex.tikz} + 12 \input{nt2np12_ex.tikz} + 48 \input{nt1np4.tikz}\Bigg].
\end{align}

When we evaluate $\kappa_n\{4\}$ with $n \ne 0$ we have to integrate the spectrum times $e^{in(\phi_1+\phi_3-\phi_5-\phi_7)}$ which will break some of the symmetries encoded in the permutations, perm$_i$. In order to check how we can simplify the integration let us start with the permutations of the $n_T=3$ case with a generic $n_p$. In this case we can define the sum of the diagrams without the permutations as
\begin{align}
	\input{nt3_ex.tikz} \equiv f_2 (\textbf{k}_1,\textbf{k}_3,\textbf{k}_5,\textbf{k}_7).
\end{align}
By using the properties \ref{prop1} and \ref{prop2} of \cref{sec2c} we can check that this sum has the following symmetry
\begin{align}\label{sym1}
	f_2(\textbf{k}_1,\textbf{k}_3,\textbf{k}_5,\textbf{k}_7)=f_2(\textbf{k}_1,\textbf{k}_3,\textbf{k}_7,\textbf{k}_5).
\end{align}
Thus, the contribution of the $n_T=3$ diagrams to the $\kappa$-function can be written as
\begin{align}
	\int_{\textbf{k}_1 \textbf{k}_3 \textbf{k}_5 \textbf{k}_7} e^{in(\phi_1+\phi_3-\phi_5-\phi_7)} \Bigg[ &\input{nt3_ex.tikz} + \text{perm}_2 \Bigg]
	\nonumber \\
	=2 \int_{\textbf{k}_1 \textbf{k}_3 \textbf{k}_5 \textbf{k}_7} e^{in(\phi_1+\phi_3-\phi_5-\phi_7)} 
	\Big[ &f_2(\textbf{k}_1,\textbf{k}_3,\textbf{k}_5,\textbf{k}_7)+f_2(\textbf{k}_5,\textbf{k}_3,\textbf{k}_1,\textbf{k}_7)
	+f_2(\textbf{k}_7,\textbf{k}_3,\textbf{k}_5,\textbf{k}_1)
	\nonumber \\
	+&f_2(\textbf{k}_1,\textbf{k}_5,\textbf{k}_3,\textbf{k}_7)+f_2(\textbf{k}_1,\textbf{k}_7,\textbf{k}_5,\textbf{k}_3)
	+f_2(\textbf{k}_7,\textbf{k}_5,\textbf{k}_3,\textbf{k}_1) \Big]
	\nonumber \\
	=2 \int_{\textbf{k}_1 \textbf{k}_3 \textbf{k}_5 \textbf{k}_7} e^{in(\phi_1+\phi_3-\phi_5-\phi_7)} 
	\Big[ &2 f_2(\textbf{k}_1,\textbf{k}_3,\textbf{k}_5,\textbf{k}_7)+ 4 f_2(\textbf{k}_1,\textbf{k}_5,\textbf{k}_3,\textbf{k}_7)
	\Big],
\end{align}
where the factor 2 comes from exploiting the symmetries $\textbf{k}_i \rightarrow -\textbf{k}_i$ that are in \cref{d42} and in the last equality we have used \cref{sym1} and relabelled the variables. Therefore, we can write
\begin{align}\label{symcumu1}
	&\int_{\textbf{k}_1 \textbf{k}_3 \textbf{k}_5 \textbf{k}_7} e^{in(\phi_1+\phi_3-\phi_5-\phi_7)} \Bigg[ \input{nt3_ex.tikz} + \text{perm}_2 \Bigg] 
	\nonumber \\
	= &\int_{\textbf{k}_1 \textbf{k}_3 \textbf{k}_5 \textbf{k}_7} e^{in(\phi_1+\phi_3-\phi_5-\phi_7)} \Bigg[ 4 \input{nt3_ex.tikz} + 8 \input{nt3_ex2.tikz} \Bigg].
\end{align}

For the $n_T=2$ diagrams that are defined by \cref{d43} we can follow the same arguments by defining
\begin{align}
	\input{nt21_ex.tikz} \equiv f_3 (\textbf{k}_1,\textbf{k}_3,\textbf{k}_5,\textbf{k}_7).
\end{align}
We can check that this function has the following symmetries
\begin{align}
	f_3(\textbf{k}_1,\textbf{k}_3,\textbf{k}_5,\textbf{k}_7)=f_3(-\textbf{k}_1,-\textbf{k}_7,-\textbf{k}_5,-\textbf{k}_3)
	=f_3(-\textbf{k}_1,-\textbf{k}_3,-\textbf{k}_7,-\textbf{k}_5)=f_3(-\textbf{k}_1,-\textbf{k}_5,-\textbf{k}_3,-\textbf{k}_7).
\end{align}
Thus, exploiting this symmetries we can write the contribution of the diagrams that have the configuration of \cref{d43} on the right side with generic $n_p$ to the cumulant as
\begin{align}\label{symcumu2}
	\int_{\textbf{k}_1 \textbf{k}_3 \textbf{k}_5 \textbf{k}_7} e^{in(\phi_1+\phi_3-\phi_5-\phi_7)} \Bigg[ \input{nt21_ex.tikz} + \text{perm}_3 \Bigg]
	= 32 \int_{\textbf{k}_1 \textbf{k}_3 \textbf{k}_5 \textbf{k}_7} e^{in(\phi_1+\phi_3-\phi_5-\phi_7)} \input{nt21_ex.tikz}.
\end{align}

For the $n_T=2$ diagrams that are defined by \cref{d44} we define
\begin{align}
	\input{nt22_ex.tikz} \equiv f_4 (\textbf{k}_1,\textbf{k}_3,\textbf{k}_5,\textbf{k}_7),
\end{align}
which has the following symmetries
\begin{align}
	f_4(\textbf{k}_1,\textbf{k}_3,\textbf{k}_5,\textbf{k}_7)=f_4(\textbf{k}_3,\textbf{k}_1,\textbf{k}_5,\textbf{k}_7)=f_4(\textbf{k}_1,\textbf{k}_3,\textbf{k}_7,\textbf{k}_5).
\end{align}
Thus, the contribution of the diagrams that have the configuration of \cref{d44} on the right side with generic $n_p$ to the cumulant is
\begin{align}\label{symcumu3}
	&\int_{\textbf{k}_1 \textbf{k}_3 \textbf{k}_5 \textbf{k}_7} e^{in(\phi_1+\phi_3-\phi_5-\phi_7)} \Bigg[ \input{nt22_ex.tikz} + \text{perm}_4 \Bigg] 
	\nonumber \\
	= &\int_{\textbf{k}_1 \textbf{k}_3 \textbf{k}_5 \textbf{k}_7} e^{in(\phi_1+\phi_3-\phi_5-\phi_7)} \Bigg[ 4 \input{nt22_ex.tikz} + 8 \input{nt22_ex2.tikz} \Bigg].
\end{align}

For the $n_T=1$ diagrams that are defined by \cref{d45} we define
\begin{align}
	\input{nt1_ex.tikz} \equiv f_5 (\textbf{k}_1,\textbf{k}_3,\textbf{k}_5,\textbf{k}_7).
\end{align}
This function has the following symmetries
\begin{align}
	f_5(\textbf{k}_1,\textbf{k}_3,\textbf{k}_5,\textbf{k}_7)=f_5(-\textbf{k}_5,-\textbf{k}_3,-\textbf{k}_1,-\textbf{k}_7)
	=f_5(-\textbf{k}_1,-\textbf{k}_7,-\textbf{k}_5,-\textbf{k}_3)=f_5(-\textbf{k}_3,-\textbf{k}_1,-\textbf{k}_7,-\textbf{k}_5).
\end{align}
Therefore, the contribution of the diagrams that have the configuration of \cref{d45} on the right side with generic $n_p$ to the cumulant is
\begin{align}\label{symcumu4}
	&\int_{\textbf{k}_1 \textbf{k}_3 \textbf{k}_5 \textbf{k}_7} e^{in(\phi_1+\phi_3-\phi_5-\phi_7)} \Bigg[ \input{nt1_ex.tikz} + \text{perm}_5 \Bigg] 
	\nonumber \\
	= &\int_{\textbf{k}_1 \textbf{k}_3 \textbf{k}_5 \textbf{k}_7} e^{in(\phi_1+\phi_3-\phi_5-\phi_7)} \Bigg[ 32 \input{nt1_ex.tikz} + 16 \input{nt1_ex2.tikz} \Bigg].
\end{align}

All in all, using \cref{mulnc3} and the simplifications in \cref{symcumu1,symcumu2,symcumu3,symcumu4}, we can reduce the number of integral when computing $\kappa_n\{4\}$ at order $(N_c^2-1)^{-3}$ down to
\begin{align}\label{knnc3}
	\kappa_n^{(3)}\{4\}=&\int_{\textbf{k}_1 \textbf{k}_3 \textbf{k}_5 \textbf{k}_7} e^{in(\phi_1+\phi_3-\phi_5-\phi_7)} \Bigg[32 \input{nt4np1_14.tikz} +16 \input{nt4np1_30.tikz} + 4 \input{nt3np21_ex.tikz} 
	\nonumber \\ 
	& + 8 \input{nt3np21_ex2.tikz}  + 4 \input{nt3np22_ex.tikz}+ 8 \input{nt3np22_ex2.tikz} + 32 \input{nt2np11_ex.tikz} + 4 \input{nt2np12_ex.tikz}
	\nonumber \\ 
	&+ 8 \input{nt2np12_ex2.tikz} + 32 \input{nt1np4.tikz} + 16 \input{nt1np42.tikz}\Bigg].
\end{align}

We compute the contribution at different orders of $(N_c^2-1)^{-1}$ in the same way. Finally, we just have to solve numerically \cref{knnc3,k0nc3} and the equivalent ones at different order. By doing that we are able to obtain \cref{cn_vs_qs2}.

\bibliography{mybib}

\end{document}